\documentclass[journal=jctc,manuscript=article,layout=twocolumn]{achemso}
\usepackage{amsmath}
\usepackage{graphicx}
\usepackage{tikz,multirow,tabularx}
\usetikzlibrary{arrows.meta}
\usepackage{footmisc}
\usepackage{booktabs, afterpage}
\usepackage{algorithm}
\usepackage{algorithmic}

\newcommand{\Nwalkers}{N_{\mathrm{repl}}} 
\newcommand{\Nopt}{N_{\mathrm{opt}}} 
\newcommand{\Cinf}{C^{\infty}} 
\newcommand{\Vb}{V_{\mathrm{bias}}} 
\newcommand{\biasprop}{\alpha_{b}} 

\usepackage{dcolumn}

\newcommand{\resstructarr}[2]{\draw [stealth-stealth, line width=0.6mm] (#1) -- (#2)}

\newcommand{\mutarrend}{Triangle[blue,fill=blue,scale=.5]}
\newcommand{\mutrevarr}[2]{\draw [blue, arrows={\mutarrend-\mutarrend}, line width=1.5mm] (#1) -- (#2)}

\usepackage{enumitem}
\newcommand{\mutationlabel}[1]{M#1}
\newcommand{\dipole}{D}
\newcommand{\dEsolv}{\Delta G_{\mathrm{solv}}}
\newcommand{\gap}{\Delta \epsilon}
\newcommand{\Pacc}{P_{\mathrm{acc}}}
\newcommand{\Pprop}{P_{\mathrm{prop}}}
 \usepackage{hyperref}

\newcommand{\candidate}[2]{$\mathrm{C}_{#1}^{\mathrm{#2}}$}

\newcommand{\mutationfigureblock}[4]{
\vspace{#3ex}

\hspace{0.0ex}\begin{tikzpicture}
        \node[anchor=south west,inner sep=0] (image) at (0.55,0) {\includegraphics[width=0.18\textwidth]{figures/elementary_mutations/M#1/forward.png}};
        \node[anchor=south west,inner sep=0] (image) at (5.45,0) {\includegraphics[width=0.18\textwidth]{figures/elementary_mutations/M#1/backward.png}};
        \mutrevarr{4.1,1.1}{5.4,1.1};
\end{tikzpicture}
\vspace{-#2ex}
\begin{itemize}
\item[\textbf{\mutationlabel{#1}:}] #4
\end{itemize}
}

\newcommand{\resstructexblock}[2]{
\hspace{-3.0ex}

\begin{tikzpicture}
        \node[anchor=south west,inner sep=0] (image) at (0.0,0) {\includegraphics[width=0.22\textwidth]{figures/resonance_structures/rot_resonance_#1_0.png}};
        \node[anchor=south west,inner sep=0] (image) at (4.5,0) {\includegraphics[width=0.22\textwidth]{figures/resonance_structures/rot_resonance_#1_1.png}};
        \resstructarr{3.7,1.3}{4.7,1.3};
        \node at (-0.2,1.7) {(#2)};
\end{tikzpicture}

}

\SectionNumbersOn

\title{Evolutionary Monte Carlo of QM properties in chemical space: Electrolyte design}

\author{Konstantin Karandashev}
\email{konstantin.karandashev@univie.ac.at}
\affiliation{University of Vienna, Faculty of Physics, Kolingasse 14-16, AT-1090 Wien, Austria}
\author{Jan Weinreich}
\affiliation{University of Vienna, Faculty of Physics, Kolingasse 14-16, AT-1090 Wien, Austria}
\author{Stefan Heinen}
\affiliation{Vector Institute for Artificial Intelligence, Toronto, ON, M5S 1M1, Canada}
\author{Daniel Jose Arismendi Arrieta}
\affiliation{Department of Chemistry-{\AA}ngstr\"om Laboratory, Uppsala University, Box 538, SE-75121 Uppsala, Sweden}
\author{Guido Falk von Rudorff}
\affiliation{University Kassel, Department of Chemistry, Heinrich-Plett-Str.40, 34132 Kassel, Germany}
\alsoaffiliation{Center for Interdisciplinary Nanostructure Science and Technology (CINSaT), Heinrich-Plett-Straße 40, 34132 Kassel}
\author{Kersti Hermansson}
\affiliation{Department of Chemistry-{\AA}ngstr\"om Laboratory, Uppsala University, Box 538, SE-75121 Uppsala, Sweden}
\author{O. Anatole von Lilienfeld}
\affiliation{Vector Institute for Artificial Intelligence, Toronto, ON, M5S 1M1, Canada}
\alsoaffiliation{Departments of Chemistry, Materials Science and Engineering, and Physics, University of Toronto, St. George Campus, Toronto, ON, Canada}
\alsoaffiliation{Machine Learning Group, Technische Universit\"at Berlin and Institute for the Foundations of Learning and Data, 10587 Berlin, Germany}

\begin{tocentry}
\includegraphics[width=8.3cm]{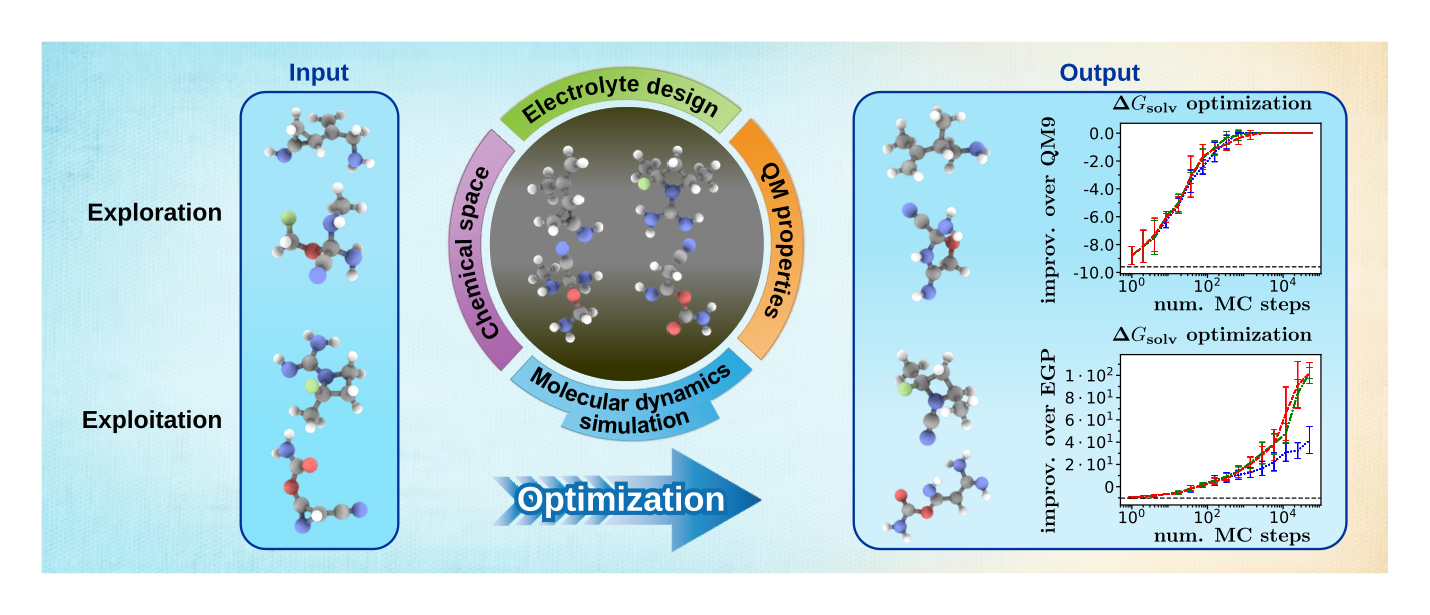}
\end{tocentry}

\begin{document}

\begin{abstract}
Optimizing a target function over the space of organic molecules is an important problem appearing in many fields of applied science, but also a very difficult one due to the vast number of possible molecular systems. We propose an Evolutionary Monte Carlo algorithm for solving such problems which is capable of straightforwardly tuning both exploration and exploitation characteristics of an optimization procedure while retaining favourable properties of genetic algorithms.  The method, dubbed MOSAiCS (\textbf{M}etropolis \textbf{O}ptimization by \textbf{S}ampling \textbf{A}daptively \textbf{i}n \textbf{C}hemical \textbf{S}pace), is tested on problems related to optimizing components of battery electrolytes, namely minimizing solvation energy in water or maximizing dipole moment while enforcing a lower bound on the HOMO-LUMO gap; optimization was done over sets of molecular graphs inspired by QM9 and Electrolyte Genome Project (EGP) datasets. MOSAiCS reliably generated molecular candidates with good target quantity values, which were in most cases better than the ones found in QM9 or EGP. While the optimization results presented in this work sometimes required up to $10^{6}$ QM calculations and were thus only feasible thanks to computationally efficient \emph{ab initio} approximations of properties of interest, we discuss possible strategies for accelerating MOSAiCS using machine learning approaches.
\end{abstract}

\maketitle

\section{Introduction}
\label{sec:introduction}

Increasing efficiency and longevity of energy storage systems is critical for improving economic sustainability of lowering greenhouse gas emissions.\cite{Jafari_Apurba:2022} One aspect of this problem is searching chemical compound space for organic molecules optimal for a target application, such as lithium battery electrolyte component\cite{Korth:2014,Cheng_Curtiss:2015,Borodin_Knap:2015,Qu_Persson:2015,Lian_Wu:2019} or electroactive molecules for redox flow batteries.\cite{Agarwal_Assary:2021,Sorkun_Er:2022} In this work we focused on the former, more specifically on finding electrochemically stable organic molecules that are good solvents for alkali salts. While such searches can be aided with high-throughput screening,\cite{Korth:2014,Cheng_Curtiss:2015,Borodin_Knap:2015} there has been a surge of ways to go beyond by increasing efficiency of compound property evaluations, \emph{e.g.} with machine learning\cite{Huang_Lilienfeld:2023} or quantum alchemy,\cite{Chang_Lilienfeld:2018,Griego_Keith:2021,Eikey_Keith:2022} and by sampling chemical space more efficiently. In the context of optimizing small organic molecules, most methods of the latter category can be classified as those based on Markov decision processes,\cite{You_Leskovec:2018,Zhou_Riley:2019,Stahl_Bostrom:2019,Khemchandani_Kell:2020,Horwood_Noutahi:2020,Pereira_Arrais:2021} recurrent neural networks,\cite{Gupta_Schneider:2018,Popova_Isayev:2019} genetic algorithms,\cite{Globus_Wipke:1999,Brown_Gasteiger:2004,Virshup_Beratan:2013,Jensen:2019,Nigam_Aspuru-Guzik:2022,Laplaza_Corminboeuf:2022} and variational autoencoders.\cite{Gomez-Bombarelli_Aspuru-Guzik:2018,Oliveira_Quiles:2022}

\begin{figure*}
\center
\includegraphics[width=0.8\textwidth]{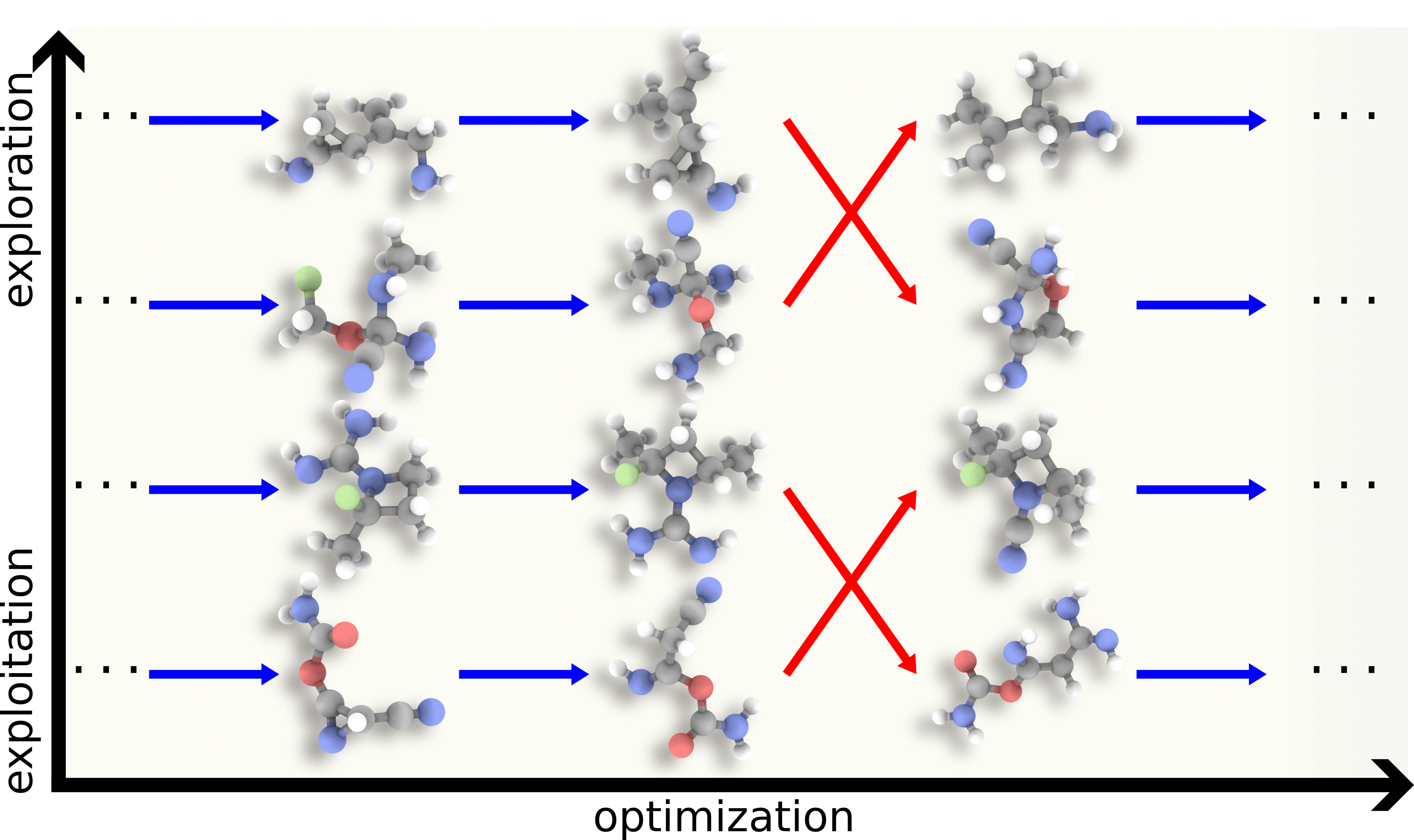}
          \caption{Scheme of an Evolutionary Monte Carlo workflow for molecular optimization that introduces several \emph{replicas} with varying degree of focus on \emph{exploitation} or \emph{exploration} aspects of the calculation and evolves them by applying \emph{elementary mutations} (blue) to single replicas and \emph{crossover moves} (red) to pairs of replicas in a way satisfying detailed balance.
          }
     \label{fig:workflow_illustration} 
 \end{figure*}

While several variants of Markov chain Monte Carlo\cite{Levin_Peres:2017} sampling have also been applied to molecule optimization problems,\cite{Fu_Sun:2021,Xie_Lei:2021} one intriguing variant, namely Evolutionary Monte Carlo,\cite{Liang_Wong:2000,Hu_Tsui:2010,Spezia:2020} has been overlooked so far. The approach combines two philosophies that have demonstrated reliable performance for a range of optimization problems: parallel tempering\cite{Hukushima_Nemoto:1996,Sambridge:2014,Angelini_Ricci-Tersenghi:2019} and genetic algorithms.\cite{Holland:1975,Johannesson_Norskov:2002,Sharma_Balint-Kurti:2010} As illustrated in Figure~\ref{fig:workflow_illustration}, Evolutionary Monte Carlo involves running several Markov chain Monte Carlo simulations that focus on \emph{exploitation} (\emph{i.e.} refining already known molecules via incremental changes) or \emph{exploration} (\emph{i.e.} finding promising regions of chemical space), which interact by swapping configurations analogously to parallel tempering or by creating ``child configurations'' similarly to genetic algorithms in a way that observes detailed balance condition.\cite{Hastings:1970} As is the case for genetic algorithms, increasing the number of replicas yields more opportunities for creating "child configurations," thus accelerating exploration of chemical space. Unlike genetic algorithms though, Evolutionary Monte Carlo allows straightforward control of its exploration and exploitation aspects while guaranteeing to \emph{eventually} find the global minimum due to the properties of Markov chain Monte Carlo. Evolutionary Monte Carlo can also potentially be combined with nested Monte Carlo techniques\cite{Iftimie_Schofield:2000,Gelb:2003,Jadrich_Leiding:2020} to utilize multiple optimized quantity evaluation methods at once, \emph{e.g.} when laboratory experiments are used alongside theoretical and machine learning approaches,\cite{Zhang_Henkelman:2013,Anderson_Crooks:2015,Shields_Doyle:2021,Park_Jung:2023} an advantage particularly relevant for high-throughput automated laboratory workflows.\cite{Rahmanian_Stein:2022,Stein_Schroeder:2022,Manzano_Cronin:2022,Park_Jung:2023}

With these reasons in mind, we implemented an Evolutionary Monte Carlo algorithm inspired by a family of genetic algorithms for optimization in the space of molecular graphs.\cite{Globus_Wipke:1999,Brown_Gasteiger:2004,Virshup_Beratan:2013,Jensen:2019} While some recently proposed methods for molecular optimization operate in string representations,\cite{Gupta_Schneider:2018,Pereira_Arrais:2021,Nigam_Aspuru-Guzik:2021,Nigam_Aspuru-Guzik:2022,Born_Manica:2023} we performed all procedures directly on chemical graphs to facilitate ensuring validity of generated molecules and maintaining detailed balance, as well as provide a more direct connection between the molecules considered and graph-based representations that have proven efficient in machine learning applications.\cite{Lemm_Lilienfeld:2021,Weinreich_Lilienfeld:2022} Lastly, we implemented a simple Wang-Landau biasing potential\cite{Wang_Landau:2001} as a \emph{curiosity reward}\cite{Thiede_Aspuru-Guzik:2022} increasing exploration aspect of the algorithm by ``pushing'' our Markov chain Monte Carlo simulations out of previously occupied graphs. The resulting method is named MOSAiCS (\textbf{M}etropolis \textbf{O}ptimization by \textbf{S}ampling \textbf{A}daptively \textbf{i}n \textbf{C}hemical \textbf{S}pace). While we were mainly designing our approach with battery applications in mind, we think it should be useful for other molecular optimization problems, such as those arising in drug design.\cite{Gupta_Schneider:2018,Reker:2019,Stahl_Bostrom:2019,Xie_Lei:2021,Horwood_Noutahi:2020, Fu_Sun:2022,Carter_Jorgensen:2023}

The rest of the paper is organized as follows. Section~\ref{sec:theory} presents the main ideas behind our approach in Subsecs.~\ref{subsec:base_definitions}-\ref{subsec:Monte_Carlo_moves}, following up with description of the optimization problem on which we test it in Subsec.~\ref{subsec:minfunc_choice} and details of our Monte Carlo simulations in Subsec.~\ref{subsec:target_min_comp_details}. Section~\ref{sec:experiment} discusses our experimental results, Section~\ref{sec:conclusions_outlook} concludes the paper with a results summary and outline of possible strategies to improve our approach. Some technical details of our method's implementation, experimental setup, and results were left for Supporting Information.

\section{Theory}
\label{sec:theory}

\subsection{Chemical space definition}
\label{subsec:base_definitions}

\begin{figure}
          \centering           
\begin{tabularx}{\textwidth}{c}
\resstructexblock{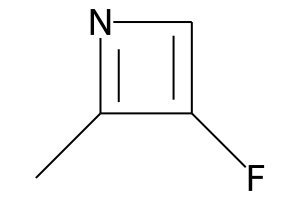}{a}
        \\
\resstructexblock{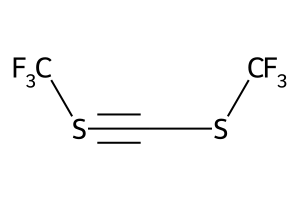}{b}
\end{tabularx}
          \caption{Examples of molecules that can be written in terms of several resonance structures that differ in (a) covalent bond orders and (b) both covalent bond orders and heavy atom valences.}
     \label{fig:resonance_structure_examples} 
 \end{figure}

We aim to minimize a loss function $F$ over a set of molecules, the latter represented by their \emph{chemical graphs}. We define a chemical graph as an undirected graph whose \emph{nodes} correspond to heavy atoms, along with, where present, hydrogen atoms covalently connected to them, and whose \emph{edges} connect a pair of nodes if their heavy atoms share a covalent bond. For a chemical graph we also define a \emph{resonance structure} as a set of \emph{valences} of nodes' heavy atoms and \emph{orders of covalent bonds} connecting these heavy atoms, both quantities taking integer values. The sum of covalent bond orders connecting a heavy atom to other atoms equals its valence, with the orders of bonds between a heavy atom and a hydrogen atom counted as one. Valence numbers are chosen to be chemically reasonable (\emph{e.g.} IV for C, II, or IV, or VI for S) and we require their sum to be the minimum needed to build a set of covalent bond orders. We also forbid a covalent bond order to be larger than three. The reasons for not including valences and bond orders in the definition of a chemical graph, but rather enumerating their possible values separately, are illustrated in Figure~\ref{fig:resonance_structure_examples} demonstrating examples of molecules for which several resonance structures differing in bond orders or bond orders and heavy atom valences can be defined. We emphasize that while this definition of bond orders and valences is loosely based on valence structure theory, it was designed not to reflect actual electronic structure of a molecule, but to allow convenient definitions of changes of chemical graphs that are illustrated in Figures~\ref{fig:elementary_mutations} and~\ref{fig:cross_coupling_moves}, as will be discussed in detail in Subsec.~\ref{subsec:Monte_Carlo_moves}.

Our definition of chemical graph \emph{a priori} prevents us from differentiating between conformers or stereoisomers and we will assume our optimization problem to be unaffected by this, \emph{e.g.} if for a given chemical graph we are interested only in the most stable stereoisomer and we optimize a Boltzmann average. We also did not implement support for molecules where valid Lewis structures can only be generated by assigning charges to atoms, \emph{e.g.} compounds with nitro groups, hence they were ignored during all calculations done in this work.

\subsection{Monte Carlo sampling}
\label{subsec:extended_ensemble}

We perform optimization by running a Markov chain Monte Carlo simulation (referred to as just ``simulation'' from now on) of unnormalized probability density similar to the one used for parallel tempering
\begin{equation}
    \begin{split}
    P(\mathbf{X})= &  \exp\left\{-\sum_{i=1}^{\Nopt}\Cinf\left[F\left(X^{(i)}\right)\right] \right. \\ &\left.-\sum_{i=\Nopt+1}^{\Nwalkers}\beta^{(i)}\left[F\left(X^{(i)}\right)+\Vb^{(i)}\left(X^{(i)}\right)\right]\right\},
    \end{split}
    \label{eq:sample_prob}
\end{equation}
where $\mathbf{X}$ is a set of $\Nwalkers$ chemical graphs (also referred to as \emph{replicas}) $X^{(i)}$ ($i=1,\ldots,\Nwalkers$), $\beta^{(i)}$ ($i=\Nopt+1,\ldots,\Nwalkers$) are temperature parameters, $\Vb$ is biasing potential, $\Cinf$ is the ``infinitely convex function'' defined to be such that for arbitrary sets of numbers $x^{(j)}$ and $y^{(j)}$ ($j=1,\ldots,\Nopt$)
\begin{equation}
\begin{split}
    \min_{j=1,\ldots,\Nopt}x^{(j)}-\min_{j=1,\ldots,\Nopt}y^{(j)}<0
    \Leftrightarrow\\
    \sum_{j=1}^{\Nopt}\left[\Cinf(x^{(j)})-\Cinf(y^{(j)})\right]\rightarrow+\infty,
\end{split}
\label{eq:Cinf_defined}
\end{equation}
and $\Nopt$ is the number of replicas that, as will become clear later, effectively undergo greedy stochastic minimization and are referred to as \emph{greedy replicas}, with the other replicas, referred to as \emph{exploration replicas}, providing a less restricted exploration of chemical space and preventing greedy replicas from getting stuck in a local minimum of $F$. The history-dependent biasing potential $\Vb^{(i)}$ is defined as\cite{Wang_Landau:2001}
\begin{equation}
    \Vb^{(i)}\left(X\right)=\frac{\biasprop}{\beta^{(i)}}\sum_{j=\Nopt+1}^{\Nwalkers}\rho^{(j)}(X),
    \label{eq:biasing_potential}
\end{equation}
where $\rho^{(j)}(X)$ is the number of times $X$ has been visited during the simulation by replica with index $j$ (details on how it was evaluated are left for Subsec.~\ref{subsec:target_min_comp_details}), $\biasprop$ is the user-defined bias proportionality coefficient. Setting a non-zero $\biasprop$ makes sampling $\mathbf{X}$ non-Markovian; as a result our certainty that in this regime a global minimum of $F$ w.r.t. $X$ is eventually found is based not on properties of Markov chain Monte Carlo, but on heuristic expectation that the biasing potential would make probability distribution of each exploration replica approach uniformity, leading to at least one replica coming across the global minimum over a finite number of simulation steps.

\subsection{Monte Carlo moves}
\label{subsec:Monte_Carlo_moves}

A simulation consists of taking a sequence of \emph{moves} in a way outlined in Algorithm~\ref{alg:MC_outline}. If the current set of replicas is in configuration $\mathbf{X}_{1}$, a move involves randomly generating parameters $\mathbf{w}$ of a change and deciding to replace $\mathbf{X}_{1}$ with the change's outcome (or \emph{trial configuration}) $\mathbf{X}_{2}$ with an acceptance probability similar to the standard Metropolis-Hastings expression\cite{Hastings:1970}
\begin{equation}
\begin{split}
    \Pacc\left(\mathbf{X}_{1},\mathbf{w},\mathbf{X}_{2}\right)=&\mathrm{min}\left[1,
    \vphantom{\frac{\Pprop\left(\mathbf{X}_{\mathrm{2}},\mathbf{w}^{-1}\right)P\left(\mathbf{X}_{2}\right)}
    {\Pprop(\mathbf{X}_{1},\mathbf{w})P\left(\mathbf{X}_{1}\right)}}\right.\\
    &\left.\frac{\Pprop(\mathbf{X}_{1},\mathbf{w})P\left(\mathbf{X}_{2}\right)}
    {\Pprop\left(\mathbf{X}_{\mathrm{2}},\mathbf{w}^{-1}\right)P\left(\mathbf{X}_{1}\right)}\right],
\end{split}
    \label{eq:acceptance_probability}
\end{equation}
where $\Pprop(\mathbf{X}_{1},\mathbf{w})$ is the probability that $\mathbf{w}$ is proposed given that $\mathbf{X}_{1}$ is the initial configuration and $\mathbf{w}^{-1}$ are parameters of a random change yielding $\mathbf{X}_{1}$ when applied to $\mathbf{X}_{2}$ and corresponding to a unique $\mathbf{w}$. The latter property ensures that detailed balance still holds in situations when several $\mathbf{w}$ yield the same trial configuration $\mathbf{X}_{2}$. Trial configurations decreasing the minimal value of $F$ among greedy replicas compared to initial configurations is accepted automatically due to our definition of $\Cinf$~(\ref{eq:Cinf_defined}).

\begin{algorithm}
\caption{Sampling $P(\mathbf{X})$~(\ref{eq:sample_prob}).}
\begin{algorithmic} 
\REQUIRE{Initial configuration: $\mathbf{X}_{1}$;}
\LOOP
\STATE{Randomly choose change parameters $\mathbf{w}$;}
\STATE{Use $\mathbf{w}$ on $\mathbf{X}_{1}$ to generate $\mathbf{X}_{2}$;}
\STATE{Randomly sample $r$ from uniform distribution in $[0,1]$;}
\STATE{$r_{\mathrm{acc.}}\leftarrow P_{\mathrm{acc.}}(\mathbf{X}_{1},\mathbf{w},\mathbf{X}_{2})$ (see Eq.~(\ref{eq:acceptance_probability});}
\IF{$r<r_{\mathrm{acc.}}$}
\STATE{$\mathbf{X}_{1}\leftarrow\mathbf{X}_{2}$;}
\ENDIF
\ENDLOOP
\end{algorithmic}
\label{alg:MC_outline}
\end{algorithm}

\begin{figure*}
          \centering           
\begin{tabularx}{\textwidth}{p{0.48\textwidth}p{0.48\textwidth}}
\mutationfigureblock{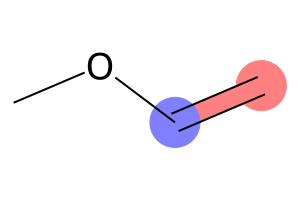}{3.0}{0.0}{Addition/removal of a node (red) connected to a single neighbor (blue).}

&
\mutationfigureblock{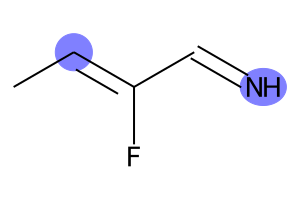}{3.0}{0.0}{Changing order of a covalent bond (green) and number of hydrogen atoms in the nodes it connects (blue).}
    \\

\mutationfigureblock{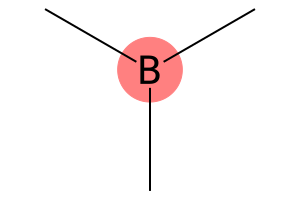}{3.0}{-5.2}{Replacing a heavy atom (red) in a node without changing its connections to other nodes.}

&

\mutationfigureblock{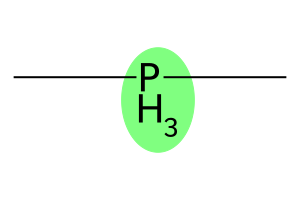}{3.0}{-5.2}{Change valence of a heavy atom (green) by adding or removing hydrogen atoms connected to it.}

\\

\mutationfigureblock{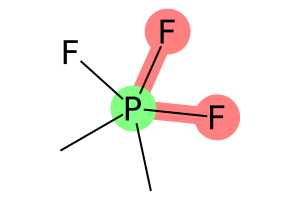}{3.0}{-5.2}{Change valence of a heavy atom (green) by adding or removing nodes (red) connected to it.}

&
\mutationfigureblock{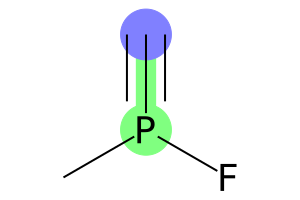}{3.0}{-5.2}{Change valence of a heavy atom (green) by changing order of a covalent bond (green) with another node whose hydrogen number is changed (blue).}

\end{tabularx}
          \caption{
          Definitions and examples of elementary mutations. A node is colored green if its heavy atom changes valence, otherwise it is colored red if it is destroyed or created, or colored blue if it changes the number of hydrogen atoms connected to the heavy atom. A bond is colored red if it is destroyed or created or green if it, in general, only changes bond order, though the latter may involve changing the order to and from 0. The valences in \mutationlabel{5} and \mutationlabel{6} always change to adjacent values (\emph{e.g.} for S valence can change from II to IV, but not from II to VI); nodes created and destroyed in \mutationlabel{1}, \mutationlabel{3}, and \mutationlabel{5} contain heavy atoms in their smallest valence state (\emph{e.g.} valence II for S).}
     \label{fig:elementary_mutations} 
 \end{figure*}

We use three types of moves to propose the trial configurations $\mathbf{X}_{2}$; we will only discuss the general idea behind them here with implementation details left for Supporting Information. The first type, referred to as \emph{elementary moves}, applies an \emph{elementary mutation} outlined in Figure~\ref{fig:elementary_mutations} to a single replica; such moves correspond to incremental exploration of chemical space. To accelerate greedy optimization of molecules, we additionally introduced the ``no reconsiderations condition'': if change parameters $\mathbf{w}$ corresponding to an elementary move have been rejected for a greedy replica they are not considered again. The second type of moves is \emph{tempering swap} moves that are analogous to the swap moves in conventional parallel tempering techniques and involve randomly choosing replicas with indices $i$ and $j$ in such a way that at least one of them is an exploration replica, considering a swap of the corresponding chemical graphs, and accepting it with acceptance probability~(\ref{eq:acceptance_probability}). These moves allow greedy replicas stuck in a local minimum of $F$ to get to chemical graphs with lower values of $F$ if the latter are discovered by an exploration replica.

\begin{figure}
          \centering           
\begin{tabularx}{\textwidth}{c}
\begin{tikzpicture}
        \node[anchor=south west,inner sep=0] (image) at (0.5,0) {\includegraphics[width=0.175\textwidth]{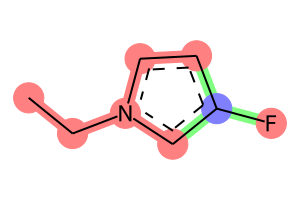}};
        \node[anchor=south west,inner sep=0] (image) at (4.4,0) {\includegraphics[width=0.175\textwidth]{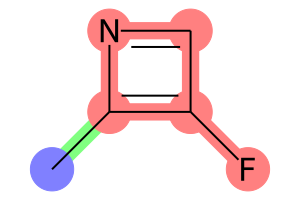}};
        \node at (0,1.4) {\textbf{I}};
\end{tikzpicture}
    \\

\begin{tikzpicture}
        \node[anchor=south west,inner sep=0] (image) at (0.5,0) {\includegraphics[width=0.175\textwidth]{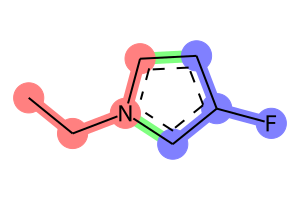}};
        \node[anchor=south west,inner sep=0] (image) at (4.4,0) {\includegraphics[width=0.175\textwidth]{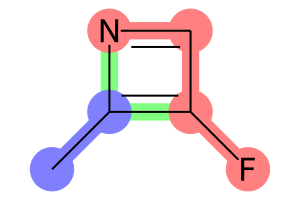}};
        \node at (0,1.4) {\textbf{II}};
\end{tikzpicture}

\\

\begin{tikzpicture}
        \node[anchor=south west,inner sep=0] (image) at (0.5,0) {\includegraphics[width=0.175\textwidth]{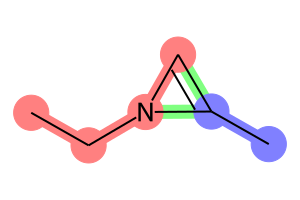}};
        \node[anchor=south west,inner sep=0] (image) at (4.4,0) {\includegraphics[width=0.175\textwidth]{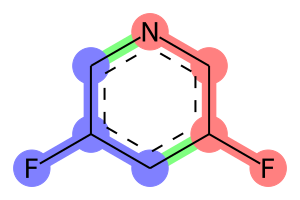}};
        \node at (0,1.4) {\textbf{III}};
\end{tikzpicture}

\\

\end{tabularx}
          \caption{An example of a crossover move. I - selection of two ``blue'' nodes in two molecules; II - selecting the ``blue'' neighborhoods of the two blue nodes, coloring the rest ``red,'' with the bonds connecting blue and red fragments colored ``green''; III - exchanging the blue fragments between molecules and reconnecting them with red fragments by exchanging connected nodes between pairs of green bonds.}
     \label{fig:cross_coupling_moves} 
 \end{figure}

The third type of moves are \emph{crossover moves} inspired by the procedure developed in Ref.~\citenum{Globus_Wipke:1999}, which are introduced to allow drastic changes of chemical graphs occupied by replicas. The general idea is illustrated in Figure~\ref{fig:cross_coupling_moves}: a pair of nodes is randomly chosen in two chemical graphs and the neighborhoods of these two nodes are exchanged to create two new chemical graphs. Thus defined crossover moves are more restrictive than the ones of Ref.~\citenum{Globus_Wipke:1999} as they do not allow exchanging fragments of arbitrary shape and connectivity. These restrictions, however, make it straightforward to ensure that the resulting chemical graphs satisfy constraints on the number of nodes, are connected, and correspond to a change for which the $\Pprop(\mathbf{X}_{1},\mathbf{w})/\Pprop(\mathbf{X}_{2},\mathbf{w}^{-1})$ ratio in $\Pacc$~(\ref{eq:acceptance_probability}) can be easily calculated.

Setting $F$ to be infinitely large for chemical graphs violating certain user-defined constraints is a general way to enforce the latter on the optimization result. However, it is in general preferable to maintain a given constraint as early as during the proposition of trial configuration $\mathbf{X}_{2}$ to increase average acceptance probability and the resulting speed of chemical space exploration. We implemented the corresponding algorithms for maintaining constraints on the number of heavy atoms in a molecule and the kinds of atoms that can share a covalent bond since they are simple to maintain, yet quite important for our applications. Lastly, the question of the moves' sufficiency to access the chemical space and sets of molecules considered in Section~\ref{sec:experiment} in their entirety is discussed in Supporting Information.

\subsection{Minimization problems}
\label{subsec:minfunc_choice}

A good battery electrolyte is a good solvent for lithium salts and is electrochemically stable. We approximated the former property with polarity; maximizing a molecule's polarity was in turn approximated by either maximizing the dipole moment $\dipole$ or minimizing the free energy of solvation in water $\dEsolv$. We approximated the electrochemical stability requirement with a lower bound on the HOMO-LUMO $\gap$, with which we approximated the width of the compound's electrochemical stability window.\cite{Korth:2014} While the latter relation is not actually practical for battery design,\cite{Borodin:2019} we still opted for a $\gap$-based electrochemical stability criterion to connect our work with other compound optimization problems where $\gap$ can be used.\cite{Teunissen_De_Vleeschouwer:2017,De_Lile_Lee:2020} For both $\dipole$ and $\dEsolv$ optimization we constrained the molecules' $\gap$ to be larger than either benzene (\emph{strong $\gap$ constraint}) or octa-1,3,5,7-tetraene (\emph{weak $\gap$ constraint}), resulting in four minimization problems of differing difficulty. While in this work we focused on testing performance of MOSAiCS against these single objective optimization problems, our approach can also be used to optimize several properties at once via a suitable multiobjective loss function.\cite{Fromer_Coley:2023}

We aimed to estimate $\dEsolv$, $\dipole$, and $\gap$ as computationally cheaply as possible while being qualitatively correct over a wide range of chemical compounds; the resulting protocol is explained in detail in Supporting Information. Here we just mention that for a given chemical graph we used the MMFF94 forcefield\cite{Halgren:1996_I,*Halgren:1996_II,*Halgren:1996_III,*Halgren_Nachbar:1996_IV,*Halgren:1996_V,*Halgren:1999_VI,*Halgren:1999_VII} to generate molecular conformers, for which we performed GFN2-xTB\cite{Bannwarth_Grimme:2019} calculations with analytical linearized
Poisson-Boltzmann model\cite{Ehlert_Grimme:2021} simulating presence of water. The root mean square error (RMSE) that is presented for calculated quantities corresponds to statistical error from randomness of conformer generation. We used two sets of parameters for our protocol: ``converged'' that produced reasonable RMSEs for a wide variety of compounds, but was relatively computationally expensive, and ``cheap'' that was used during our simulations. From now on, $\dEsolv$, $\dipole$, and $\gap$ will denote estimates of these quantities obtained with the ``converged'' protocol, while estimates obtained with the ``cheap'' protocol will be marked with addition of ``cheap'' superscript.

Each of the four minimization problems was solved in two sets of molecules based on QM9\cite{Ruddigkeit_Reymond:2012,*Ramakrishnan_Lilienfeld:2014} and the Electrolyte Genome Project\cite{Qu_Persson:2015} (EGP) datasets. The QM9 dataset consists of 134k molecules containing up to 9 heavy atoms (C, O, N, and F). We defined the ``QM9*'' set to consist of molecules (not necessarily in QM9) that also contain up to 9 heavy atoms of the same elements as QM9, but are additionally constrained by not allowing bonds between N, O, and F atoms, as well as O-H and H-F bonds, since these covalent bonds are typically associated with increased chemical reactivity. The EGP dataset was generated with the Materials Project\cite{Jain_Persson:2013} workflows in an effort to facilitate discovery of novel battery electrolyte molecules; the version currently hosted on the Materials Project website contains 24.5k species in total; neutral species for which MMFF94 coordinates could be generated included 19.7k individual chemical graphs containing up to 92 heavy atoms. These characteristics of the EGP dataset were the basis for defining the ``EGP*'' set, whose molecules (not necessarily in EGP) contain up to 15 heavy atoms (B, C, N, O, F, Si, P, S, Cl, and Br, which are elements present in organic molecules of EGP) and, for the sake of chemical stability, do not contain covalent bonds between N, O, F, Cl, and Br, between H and B, O, F, Si, P, S, Cl, or Br, as well as S-S and P-P bonds.

We chose 15 as the maximum number of heavy atoms allowed in EGP* molecules because this size restriction is obeyed by 87.0\% and 97.0\% of EGP's chemical graphs satisfying weak and strong $\gap$ constraints. When choosing which elements can not share a covalent bond in QM9* and EGP* molecules we mainly aimed for excluding weak bonds, although we also forbade some relatively strong bonds whose presence can signify molecular reactivity. Since we only consider molecules whose valid Lewis structures can be generated without assigning charges to atoms, H-F and double O-O bonds can only be encountered in hydrogen fluoride and oxygen, which we excluded from consideration due to their corrosive properties. Creating N-N bonds inside an organic compound risks making it prone to releasing nitrogen on excitation, adding functional groups containing double N-O bonds to a molecule risks making the latter prone to self-oxidation, and hydroxyl groups engage relatively easily in reactions involving oxidation or nucleophilic attacks.\cite{Clayden_Warren:2012} We note that in practice, managing this kind of reactive behavior would require additional use of more sophisticated compound stability measures.

While both QM9* and EGP* are well defined and finite sets of chemical graphs, their huge size makes evaluating any of their properties exactly, \emph{i.e.} through exact enumeration of all their chemical graphs, unfeasible. However, we do summarize properties of intersections of QM9* and QM9, as well as EGP* and EGP, in Supporting Information.

\subsection{Simulation details}
\label{subsec:target_min_comp_details}

During a simulation we used $\gap^{\mathrm{cheap}}$ to estimate whether a molecule satisfies the constraint on $\gap$; dimensionless loss functions corresponding to $\dipole$ and $\dEsolv$ were defined as
\begin{align}
F_{\mathrm{solv.}}(X)=&\frac{\dEsolv^{\mathrm{cheap}}(X)}{\mathrm{STD}_\mathrm{dataset}(\dEsolv)},\label{eq:F_solv}\\
F_{\mathrm{dipole}}(X)=&-\frac{\dipole^{\mathrm{cheap}}(X)}{\mathrm{STD}_\mathrm{dataset}(\dipole)},\label{eq:F_dipole}
\end{align}
where $\mathrm{STD}_\mathrm{dataset}$ refers to standard deviation of a quantity over molecules at the intersection of chemical graph set of interest and the reference dataset (QM9 for QM9* and EGP for EGP*) which satisfy the $\gap$ constraint of interest. We chose 1000 ``pre-final'' molecules exhibiting the smallest value of loss function out of the molecules visited during the simulation and evaluated converged estimates of the quantities of interest for them; the molecule with the best $\dipole$ or $\dEsolv$ value among pre-final molecules satisfying the $\gap$ constraint is the one considered the \emph{candidate} molecule proposed by the simulation.

We used $\Nwalkers=36$ with $\Nopt=4$ (cf. definitions in Subsection~\ref{subsec:extended_ensemble}); virtual temperature parameters $\beta^{(i)}$ appearing in $P$~(\ref{eq:sample_prob}) were defined in such a way that the smallest and largest $\beta^{(i)}$ were 1 and 8, and the other $\beta^{(i)}$ formed a geometric progression between the two extrema values, the latter being a simple recipe taken from applications of parallel tempering to configuration space sampling.\cite{Rathore_Pablo:2005,Kone_Kofke:2005} A simulation consisted of 50000 ``global'' steps, out of which 60\% were ``simple'' steps applying an elementary move to each replica, 20\% were ``tempering'' steps making tempering swap moves on 128 randomly chosen pairs of replicas, and another 20\% were ``crossover'' steps making  crossover moves on 32 randomly chosen pairs of replicas. $\rho^{(j)}(X)$ appearing in $\Vb^{(i)}$~(\ref{eq:biasing_potential}) was counted as the number of times replica $j$ was found in $X$ after a global step had been completed. For elementary moves we additionally set that: the nodes added or removed during \mutationlabel{1} mutation could be connected to the molecule with bonds of order from 1 to 3; bonds changed with \mutationlabel{2} and \mutationlabel{6} mutations could have their order increased or decreased by 1 and 2 respectively; nodes added or removed with \mutationlabel{5} mutation could be connected to the molecule with bonds of order~1 or~2.

We set $\biasprop$ to $0.0$, $0.2$, or $0.4$; for each of the resulting 12 combinations of $\biasprop$ and optimization problem we ran 8 simulations with different random number generator seeds. For all simulations all replicas initially occupied the chemical graph of methane. While it would be natural to assign each replica a randomly chosen molecule from the intersection of QM9 and QM9* or EGP and EGP*, we went with the intentional handicap of using methane as the starting molecule to demonstrate that MOSAiCS is capable of constructing all the candidate molecules presented in this Section from scratch. The effect of choice of initial conditions on the final result is briefly addressed in Supporting Information.

\section{Results and discussion}
\label{sec:experiment}

In this Section we describe the main results of our numerical experiments. The more technical aspects, such as full information on generated candidates and influence of biasing potential on search efficiency, are left for Supporting Information.

\begin{figure}
          \centering           
\hspace{-2ex}
\begin{tikzpicture}
        \node[anchor=south west,inner sep=0] (image) at (0,0) {\includegraphics[width=0.235\textwidth]{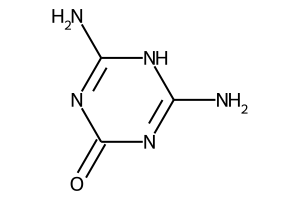}};
        \node[anchor=south west,inner sep=0] (image) at (4,0) {\includegraphics[width=0.235\textwidth]{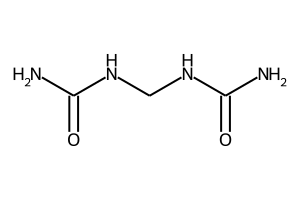}};
        \node[anchor=south west,inner sep=0] (image) at (0,-2.85) {\includegraphics[width=0.235\textwidth]{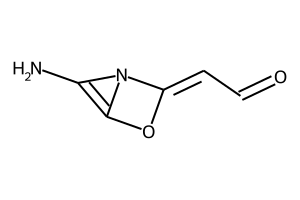}};
        \node[anchor=south west,inner sep=0] (image) at (4,-2.85) {\includegraphics[width=0.235\textwidth]{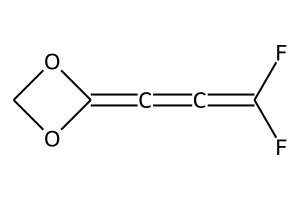}};
        \node at (2.0,3.2) {\vphantom{$\sum$}weak $\gap$ constr.};
        \node at (6.0,3.2) {\vphantom{$\sum$}strong $\gap$ constr.};
        \node[rotate=90] at (-0.1,1.3) {min. $\dEsolv$};
        \node[rotate=90] at (-0.1,-1.55) {max. $\dipole$};
        \node at (2.0,2.5) {\candidate{1}{QM9}};
        \node at (6.0,2.4) {\candidate{2}{QM9}};
        \node at (2.0,-0.55) {\candidate{4}{QM9}};
        \node at (6.0,-0.55) {\candidate{6}{QM9}};
\end{tikzpicture}
          \caption{QM9* candidates that exhibited smallest free energy of solvation $\dEsolv$ or largest dipole $\dipole$ under weak or strong constraint on the HOMO-LUMO gap $\gap$. See Supporting Information for more information about them.}
     \label{fig:qm9_best_compounds} 
 \end{figure}

\begin{figure}
          \centering           
\includegraphics[width=0.45\textwidth]{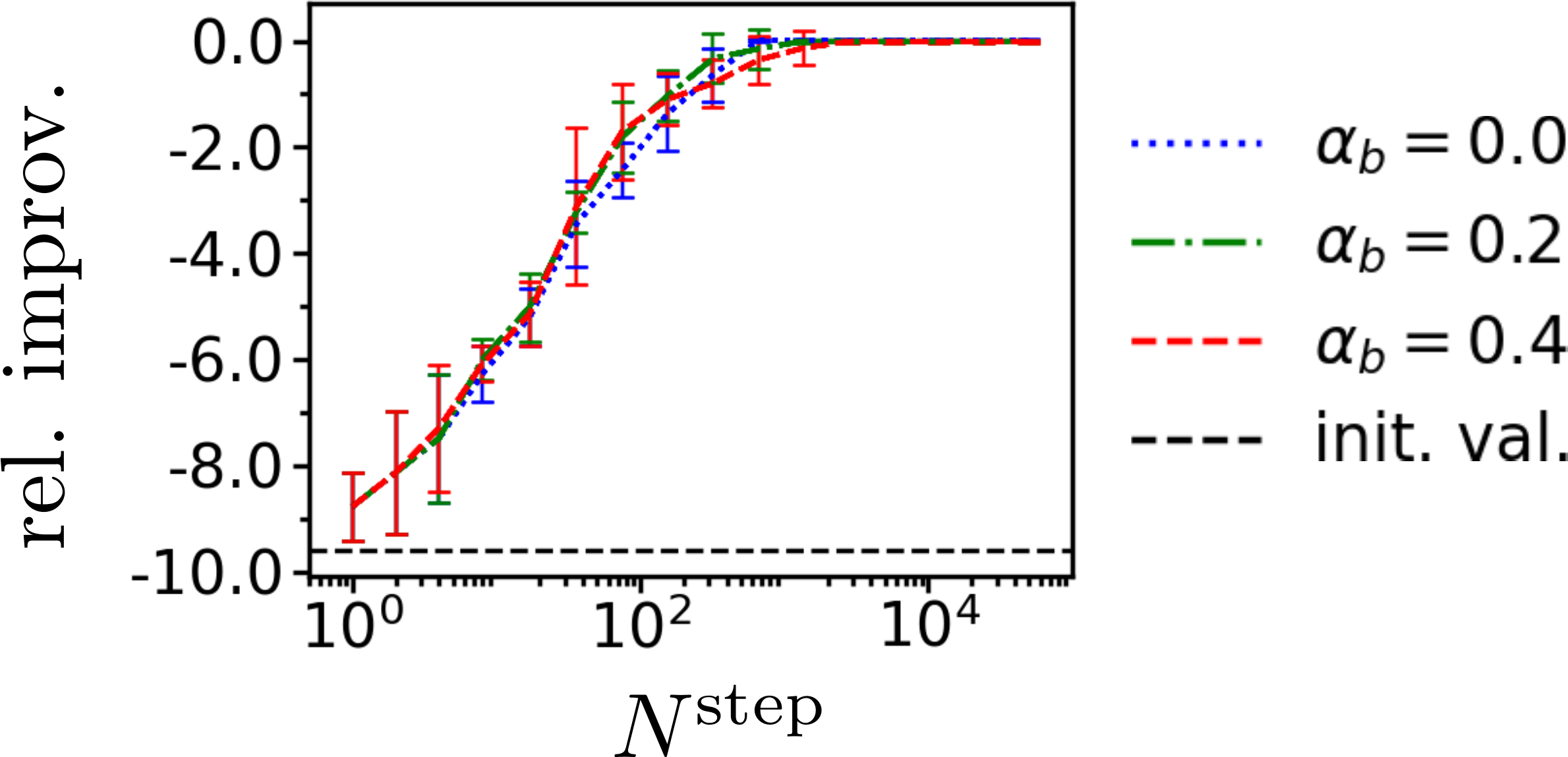}
          \caption{Relative improvement (as defined in Subsec.~\ref{sec:experiment} and estimated from $\dEsolv^{\mathrm{cheap}}$) observed at $N^{\mathrm{step}}$ global Monte Carlo steps for QM9* simulations optimizing $\dEsolv$ with weak $\gap$ constraint. For each bias proportionality coefficient $\biasprop$ we plot the mean over different random generator seeds, the error bars corresponding to standard deviation. "init. val." is the relative improvement of the simulations' starting molecule, \emph{i.e.} methane.}
     \label{fig:optimization_log_QM9} 
 \end{figure}

\begin{figure*}
          \centering           
          \includegraphics[width=1.0\linewidth]{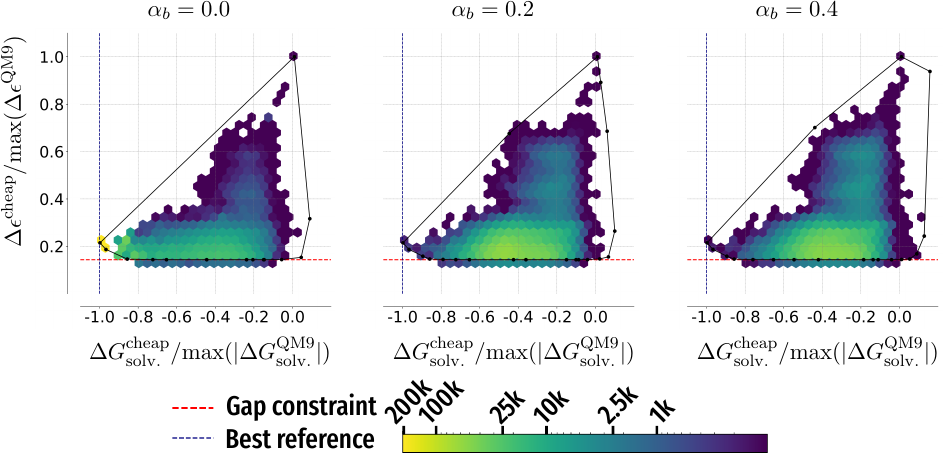}
          \caption{Densities of molecules encountered by example simulations minimizing free energy of solvation $\dEsolv$ with weak HOMO-LUMO gap $\gap$ constraint in QM9* for different values of bias proportionality coefficient $\biasprop$.}
     \label{fig:qm9_pareto_front_solvation_weak} 
 \end{figure*}

While we ran in total 96 simulations in QM9*, or 24 simulations with different random generator seed and $\biasprop$ values for each optimization problem, they agreed remarkably often on candidates proposed, only yielding 10 candidates in total. Table~\ref{tab:qm9_summary} summarizes the best and worst values of optimized quantities of candidates proposed by MOSAiCS along with the corresponding relative improvement, which we define as absolute difference between a candidate’s optimized quantity value and the corresponding value for the best molecule for the optimization problem taken from the reference dataset (cf. Table~S2 in Supporting Information), divided by the corresponding $\mathrm{STD}_{\mathrm{dataset}}$. For optimization of $\dEsolv$ with weak $\gap$ constraint all trajectories proposed the minimum of $\dEsolv$ already present in QM9, while for all other optimization problems all trajectories proposed candidates that improved significantly on molecules in QM9. Best candidates proposed for a given optimization problem are shown in Figure~\ref{fig:qm9_best_compounds}; note that to facilitate discussion of candidates' properties in Supporting Information, each candidate is referred by a capital $C$ with a unique index subscript and a superscript denoting the reference dataset. We see how MOSAiCS successfully constructed complex conjugated bond structures facilitating charge transfer which, in turn, led to smaller, \emph{i.e.} more negative, $\dEsolv$  or larger $\dipole$. Figure~\ref{fig:optimization_log_QM9} displays optimization progress with number of global Monte Carlo steps for different $\biasprop$ values for minimizing $\dEsolv$ with weak $\gap$ constraint. We observe convergence of the optimized property with a rate not significantly affected by changing $\biasprop$; the same is true with varying degree for other optimization problems as discussed in Supporting Information. To visualize how simulations explored chemical space for different optimization problems and values of $\biasprop$, for each such combination we took a simulation that had produced the best candidate and plotted the density of molecules it encountered with respect to the optimized quantity and $\gap$. Figure~\ref{fig:qm9_pareto_front_solvation_weak} presents such plots for minimizing $\dEsolv$ with weak $\gap$ constraint, with plots for other optimization problems presented and discussed in Supporting Information. We see that increasing $\biasprop$ tended to increase diversity of molecules encountered during the simulations, but this was mainly done by considering more molecules in regions of chemical space with larger values of $\dEsolv$.

 \begin{table*}
 \centering
\begin{tabular}{lllllll}
\toprule
                     \multicolumn{1}{c}{optimized} & \multicolumn{1}{c}{$\Delta\epsilon$} & \multicolumn{2}{c}{optimized quantity value} & \phantom{.} & \multicolumn{2}{c}{relative improvement} \\
\cline{3-4}\cline{6-7}\multicolumn{1}{c}{quantity} &       \multicolumn{1}{c}{constraint} &                 \multicolumn{1}{c}{best} &                \multicolumn{1}{c}{worst} & \phantom{.} & \multicolumn{1}{c}{best} & \multicolumn{1}{c}{worst} \\
\midrule
              \multirow{2}{*}{$\dEsolv$} &                                 weak &             $-94.79 \pm 0.06\phantom{0}$ &                   \multicolumn{1}{c}{\_} &             &        $0.004 \pm 0.009$ &    \multicolumn{1}{c}{\_} \\
                            \phantom{\_} &                               strong &             $-68.19 \pm 0.60\phantom{0}$ &             $-67.27 \pm 0.00\phantom{0}$ &             &        $1.665 \pm 0.099$ &         $1.537 \pm 0.052$ \\
   \cline{1-2}\multirow{2}{*}{$\dipole$} &                                 weak & $\phantom{\pm}15.73 \pm 0.10\phantom{0}$ & $\phantom{\pm}15.23 \pm 0.00\phantom{0}$ &             &        $1.287 \pm 0.052$ &         $1.013 \pm 0.000$ \\
                            \phantom{\_} &                               strong & $\phantom{\pm}11.14 \pm 0.00\phantom{0}$ & $\phantom{\pm}10.00 \pm 0.00\phantom{0}$ &             &        $1.924 \pm 0.030$ &         $1.087 \pm 0.030$ \\
\bottomrule
\end{tabular}

\caption{Best and worst QM9* candidates proposed during minimization of free energy of solvation $\dEsolv$ or maximization of dipole $\dipole$ with weak or strong constraint on the HOMO-LUMO gap $\gap$, along with their optimized quantity values and the relative improvement compared to QM9 dataset as defined in Sec.~\ref{sec:experiment}. $\dEsolv$ and $\dipole$ values are in kJ/mol and debye. The full list of candidate molecules can be found in Supporting Information.}
\label{tab:qm9_summary}
\end{table*}

\begin{figure}
          \centering           
\hspace{-2ex}
\begin{tikzpicture}
        \node[anchor=south west,inner sep=0] (image) at (0,0) {\includegraphics[width=0.235\textwidth]{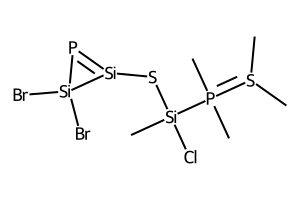}};
        \node[anchor=south west,inner sep=0] (image) at (4,0) {\includegraphics[width=0.235\textwidth]{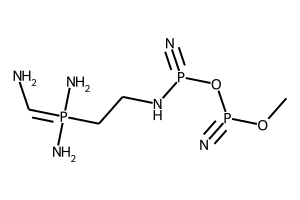}};
        \node[anchor=south west,inner sep=0] (image) at (0,-2.85) {\includegraphics[width=0.235\textwidth]{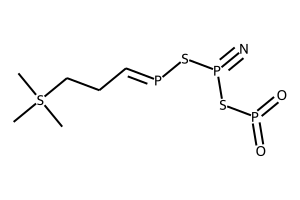}};
        \node[anchor=south west,inner sep=0] (image) at (4,-2.85) {\includegraphics[width=0.235\textwidth]{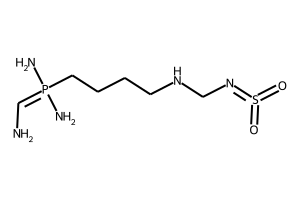}};
        \node at (2.0,3.2) {\vphantom{$\sum$}weak $\gap$ constr.};
        \node at (6.0,3.2) {\vphantom{$\sum$}strong $\gap$ constr.};
        \node[rotate=90] at (-0.1,1.3) {min. $\dEsolv$};
        \node[rotate=90] at (-0.1,-1.55) {max. $\dipole$};
        \node at (2.0,2.5) {\candidate{1}{EGP}};
        \node at (6.0,2.4) {\candidate{21}{EGP}};
        \node at (2.0,-0.55) {\candidate{41}{EGP}};
        \node at (6.0,-0.55) {\candidate{65}{EGP}};
\end{tikzpicture}
          \caption{EGP* candidates that exhibited smallest free energy of solvation $\dEsolv$ or largest dipole $\dipole$ under weak or strong constraint on the HOMO-LUMO gap $\gap$. See the Supporting Information for more information about them.}
     \label{fig:egp_best_candidates} 
 \end{figure}

\begin{figure}
          \centering           
\includegraphics[width=0.45\textwidth]{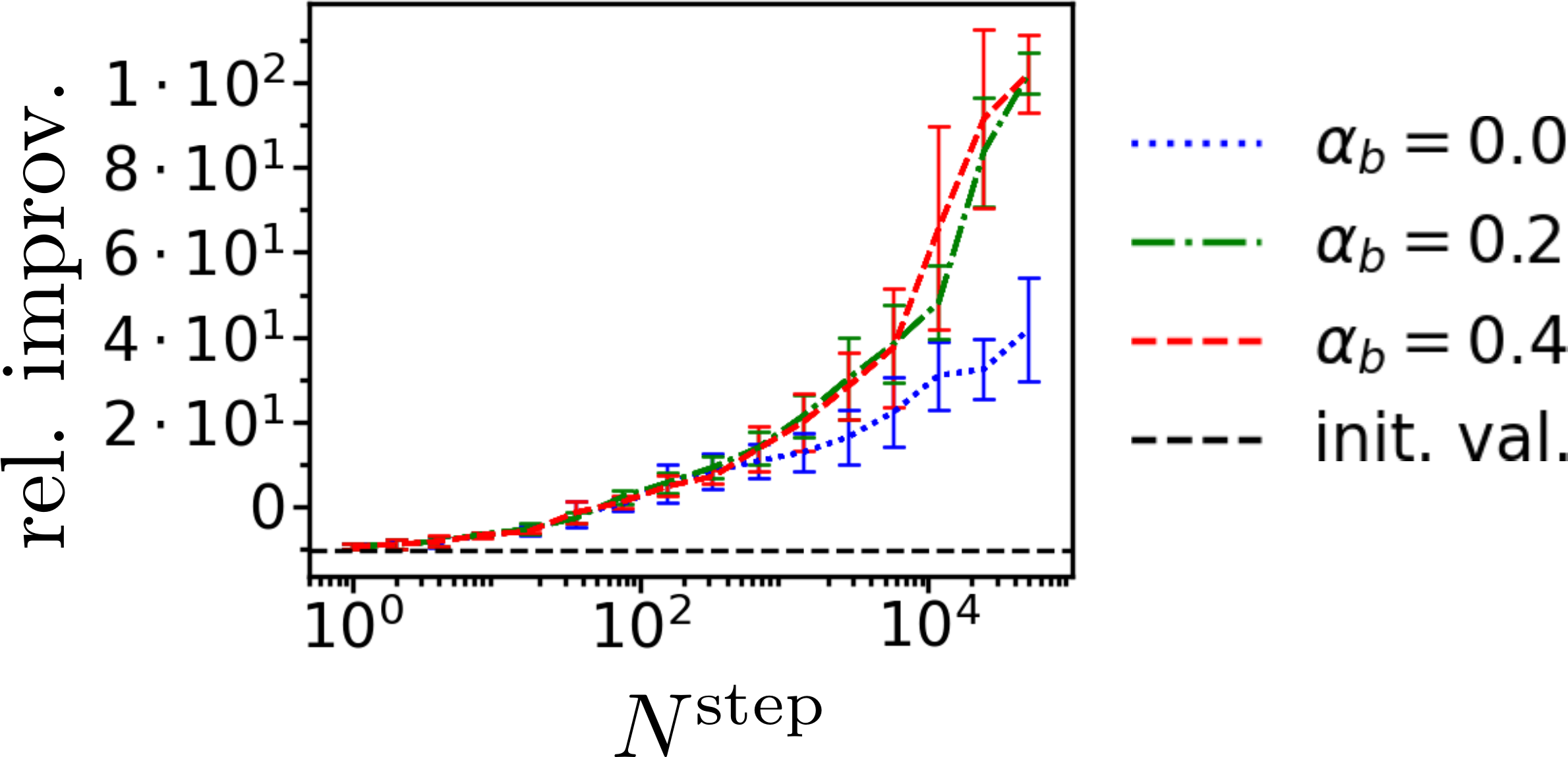}
          \caption{Relative improvement observed at $N^{\mathrm{step}}$ global Monte Carlo steps for EGP* simulations optimizing $\dEsolv$ with weak $\gap$ constraint; results are organized analogously to Figure~\ref{fig:optimization_log_QM9}.}
     \label{fig:optimization_log_EGP} 
 \end{figure}

\begin{figure*}
          \centering
          \includegraphics[width=1.0\linewidth]{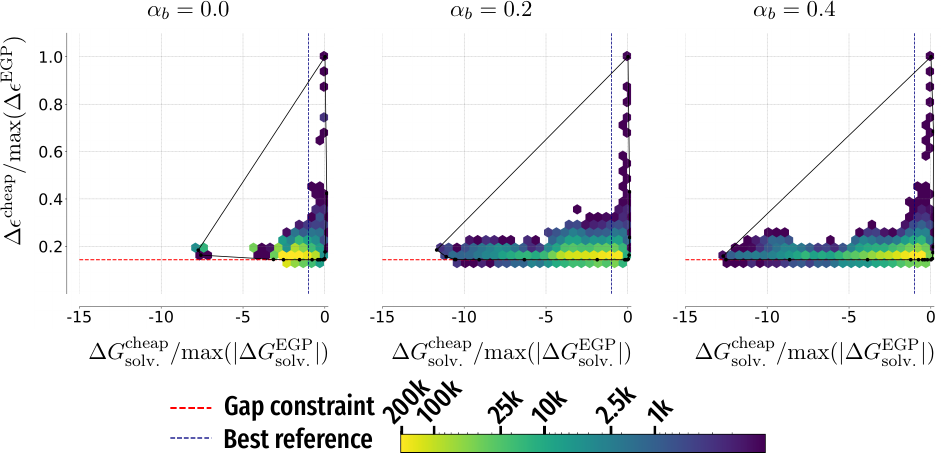}
          \caption{Densities of molecules encountered by simulations minimizing free energy of solvation $\dEsolv$ with weak HOMO-LUMO gap $\gap$ constraint in EGP* that produced the candidate with smallest $\dEsolv$ for a given value of bias proportionality coefficient $\biasprop$.}
     \label{fig:egp_pareto_front_solvation_weak} 
 \end{figure*}

Optimization in EGP* was harder than in QM9* due to larger size of the former set of molecules, resulting in our protocol generating underconverged simulations that rarely agreed on candidates, producing 83 candidates in total. However, as summarized in Table~\ref{tab:egp_summary}, we still observed significant improvement of optimized quantities compared to EGP, though the improvements' impressive magnitudes are largely due to EGP containing a much less representative portion of EGP* compared to the case of QM9 and QM9*. Best EGP* candidates for each optimization problem are presented in Figure~\ref{fig:egp_best_candidates}; unlike QM9* candidates, no chemical intuition is seen in how they were constructed beyond adding as many polar covalent bonds as possible, which may be due to underconvergence of our EGP* simulations. The underconvergence can also be observed on  our optimization progress plots, the plot for minimizing $\dEsolv$ with weak $\gap$ constraint presented in  Figure~\ref{fig:optimization_log_EGP} and the rest found in Supporting Information. Figure~\ref{fig:optimization_log_EGP} also demonstrates how adding $\biasprop$ can accelerate optimization as a function of global Monte Carlo steps, though we need to note that simulations with larger $\biasprop$ on average process more chemical graphs per global Monte Carlo steps as discussed in Supporting Information. Densities of molecules encountered during simulations minimizing $\dEsolv$ with weak $\gap$ constraint that produced the best candidates are presented in Figure~\ref{fig:egp_pareto_front_solvation_weak}; unlike the case of QM9*, increasing $\biasprop$ helped simulations explore parts of chemical space with smaller values of $\dEsolv$. Analogous plots for other optimization problems in EGP* are presented in Supporting Information.

 \begin{table*}
 \centering
\begin{tabular}{lllllll}
\toprule
                     \multicolumn{1}{c}{optimized} & \multicolumn{1}{c}{$\Delta\epsilon$} & \multicolumn{2}{c}{optimized quantity value} & \phantom{.} & \multicolumn{2}{c}{relative improvement} \\
\cline{3-4}\cline{6-7}\multicolumn{1}{c}{quantity} &       \multicolumn{1}{c}{constraint} &                                         \multicolumn{1}{c}{best} &                              \multicolumn{1}{c}{worst} & \phantom{.} &              \multicolumn{1}{c}{best} &     \multicolumn{1}{c}{worst} \\
\midrule
                        \multirow{2}{*}{$\dEsolv$} &                                 weak &     $-1194\phantom{00}\phantom{.} \pm 7\phantom{000}\phantom{.}$ &   $\phantom{0}$$-382.2\phantom{0} \pm 1.2\phantom{00}$ &             & $118.0\phantom{0} \pm 0.7\phantom{0}$ & $\phantom{0}$$30.84 \pm 0.13$ \\
                                      \phantom{\_} &                               strong &             $\phantom{0}$$-269.8\phantom{0} \pm 2.1\phantom{00}$ &   $\phantom{0}$$-207.6\phantom{0} \pm 0.9\phantom{00}$ &             &         $\phantom{0}$$23.19 \pm 0.25$ & $\phantom{0}$$15.65 \pm 0.11$ \\
             \cline{1-2}\multirow{2}{*}{$\dipole$} &                                 weak & $\phantom{0}$$\phantom{\pm}109.5\phantom{0} \pm 0.9\phantom{00}$ & $\phantom{00}$$\phantom{\pm}53.69 \pm 0.12\phantom{0}$ &             &         $\phantom{0}$$51.16 \pm 0.48$ & $\phantom{0}$$21.49 \pm 0.07$ \\
                                      \phantom{\_} &                               strong &           $\phantom{00}$$\phantom{\pm}59.81 \pm 1.14\phantom{0}$ & $\phantom{00}$$\phantom{\pm}25.88 \pm 0.17\phantom{0}$ &             &         $\phantom{0}$$34.70 \pm 0.76$ & $\phantom{0}$$11.95 \pm 0.12$ \\
\bottomrule
\end{tabular}

\caption{Best and worst EGP* candidates, data notation is analogous to Table~\ref{tab:qm9_summary}. Full list of candidate molecules can be found in Supporting Information.}
\label{tab:egp_summary}
\end{table*}

\section{Conclusions and outlook}
\label{sec:conclusions_outlook}

We have proposed an effective algorithm for optimization in chemical space, dubbed MOSAiCS, and successfully applied it to several test optimization problems connected to lithium battery electrolyte design.  In the current implementation, it is only feasible to optimize estimates of quantities of interest that can be evaluated with little computational effort due to the large number of evaluations of loss function made during a simulation (see Supporting Information); given successes in using active learning for optimization problems in both configuration\cite{Gastegger_Marquetand:2017,Podryabinkin_Shapeev:2017,Schaaf_Csanyi:2023,Vandermause_Kozinsky:2023} and chemical\cite{Hernandez-Lobato_Aspuru-Guzik:2017,Smith_Roitberg:2018,Reker:2019} space, our first priority is to combine MOSAiCS with a similar protocol to decrease the number of loss function evaluations done during the simulations. Successful use of Markov Decision Process formalism to accelerate genetic algorithms in chemical space\cite{Fu_Sun:2022} suggests MOSAiCS might similarly be improved with a smarter policy for choosing elementary mutations and crossover moves. On a more general note, any method for generating chemical graphs that can also provide corresponding proposition probability $\Pprop$ needed for $\Pacc$~(\ref{eq:acceptance_probability}) can be integrated into MOSAiCS framework directly.

While we aimed to propose an approach that would be agnostic to how much is known about the chemical graph set of interest, we still relied on QM9 and EGP to get reasonably rescaled loss functions $ F_{\mathrm{solv.}} $~(\ref{eq:F_solv}) and $F_{\mathrm{dipole}}$~(\ref{eq:F_dipole}) that were then used during simulations in QM9* and EGP*. This dependence on previously published data should become avoidable by implementing more sophisticated schemes\cite{Vousden_Mandel:2015} for adjusting temperature parameters $\beta^{(i)}$ based on trajectory history. Also, while we used heavy atoms with connected hydrogens as nodes of chemical graphs to maximize chemical diversity of generated compounds, it is possible to expand the algorithm to using larger compound fragments as nodes instead. If applicable to the optimization problem at hand, this modification should simplify the search by both decreasing effective dimensionality of the graphs considered\cite{Xie_Lei:2021,Laplaza_Corminboeuf:2022} and improving scalability of machine learning models for the molecules of interest.\cite{Huang_Lilienfeld:2020,Huang_Benali:2023}

\section{Code availability}

Python implementation of MOSAiCS is available at \url{https://github.com/chemspacelab/mosaics}.

\begin{suppinfo}
Details of MOSAiCS implementation and quantity of interest evaluation, discussion of accessibility of QM9* and EGP* by our simulations, additional experimental results.
\end{suppinfo}

\begin{acknowledgement}
This project has received funding from the European Union’s Horizon 2020 research and innovation programme under grant agreement No~957189 (BIG-MAP) and  No~957213 (BATTERY 2030+). O.A.v.L. has received funding from the European Research Council (ERC) under the European Union’s Horizon 2020 research and innovation programme (grant agreement No.~772834). O.A.v.L. has
received support as the Ed Clark Chair of Advanced Materials and as a Canada CIFAR AI Chair. O.A.v.L. acknowledges that this research is part of the University of Toronto’s Acceleration Consortium, which receives funding from the Canada First Research Excellence Fund (CFREF). Obtaining the presented computational results has been facilitated using the queueing system implemented at \href{https://leruli.com}{http://leruli.com}. The project has been supported by the Swedish Research Council (Vetenskapsrådet), and the Swedish National Strategic e-Science program eSSENCE as well as by computing resources from the Swedish National Infrastructure for Computing (SNIC/NAISS).
\end{acknowledgement}

\bibliography{references}

\providecommand{\latin}[1]{#1}
\makeatletter
\providecommand{\doi}
  {\begingroup\let\do\@makeother\dospecials
  \catcode`\{=1 \catcode`\}=2 \doi@aux}
\providecommand{\doi@aux}[1]{\endgroup\texttt{#1}}
\makeatother
\providecommand*\mcitethebibliography{\thebibliography}
\csname @ifundefined\endcsname{endmcitethebibliography}
  {\let\endmcitethebibliography\endthebibliography}{}
\begin{mcitethebibliography}{89}
\providecommand*\natexlab[1]{#1}
\providecommand*\mciteSetBstSublistMode[1]{}
\providecommand*\mciteSetBstMaxWidthForm[2]{}
\providecommand*\mciteBstWouldAddEndPuncttrue
  {\def\EndOfBibitem{\unskip.}}
\providecommand*\mciteBstWouldAddEndPunctfalse
  {\let\EndOfBibitem\relax}
\providecommand*\mciteSetBstMidEndSepPunct[3]{}
\providecommand*\mciteSetBstSublistLabelBeginEnd[3]{}
\providecommand*\EndOfBibitem{}
\mciteSetBstSublistMode{f}
\mciteSetBstMaxWidthForm{subitem}{(\alph{mcitesubitemcount})}
\mciteSetBstSublistLabelBeginEnd
  {\mcitemaxwidthsubitemform\space}
  {\relax}
  {\relax}

\bibitem[Jafari \latin{et~al.}(2022)Jafari, Botterud, and
  Sakti]{Jafari_Apurba:2022}
Jafari,~M.; Botterud,~A.; Sakti,~A. Decarbonizing power systems: A critical
  review of the role of energy storage. \emph{Renewable Sustainable Energy
  Rev.} \textbf{2022}, \emph{158}, 112077\relax
\mciteBstWouldAddEndPuncttrue
\mciteSetBstMidEndSepPunct{\mcitedefaultmidpunct}
{\mcitedefaultendpunct}{\mcitedefaultseppunct}\relax
\EndOfBibitem
\bibitem[Korth(2014)]{Korth:2014}
Korth,~M. Large-scale virtual high-throughput screening for the identification
  of new battery electrolyte solvents: evaluation of electronic structure
  theory methods. \emph{Phys. Chem. Chem. Phys.} \textbf{2014}, \emph{16},
  7919--7926\relax
\mciteBstWouldAddEndPuncttrue
\mciteSetBstMidEndSepPunct{\mcitedefaultmidpunct}
{\mcitedefaultendpunct}{\mcitedefaultseppunct}\relax
\EndOfBibitem
\bibitem[Cheng \latin{et~al.}(2015)Cheng, Assary, Qu, Jain, Ong, Rajput,
  Persson, and Curtiss]{Cheng_Curtiss:2015}
Cheng,~L.; Assary,~R.~S.; Qu,~X.; Jain,~A.; Ong,~S.~P.; Rajput,~N.~N.;
  Persson,~K.; Curtiss,~L.~A. Accelerating Electrolyte Discovery for Energy
  Storage with High-Throughput Screening. \emph{J. Phys. Chem. Lett.}
  \textbf{2015}, \emph{6}, 283–291\relax
\mciteBstWouldAddEndPuncttrue
\mciteSetBstMidEndSepPunct{\mcitedefaultmidpunct}
{\mcitedefaultendpunct}{\mcitedefaultseppunct}\relax
\EndOfBibitem
\bibitem[Borodin \latin{et~al.}(2015)Borodin, Olguin, Spear, Leiter, and
  Knap]{Borodin_Knap:2015}
Borodin,~O.; Olguin,~M.; Spear,~C.~E.; Leiter,~K.~W.; Knap,~J. Towards high
  throughput screening of electrochemical stability of battery electrolytes.
  \emph{Nanotechnology} \textbf{2015}, \emph{26}, 354003\relax
\mciteBstWouldAddEndPuncttrue
\mciteSetBstMidEndSepPunct{\mcitedefaultmidpunct}
{\mcitedefaultendpunct}{\mcitedefaultseppunct}\relax
\EndOfBibitem
\bibitem[Qu \latin{et~al.}(2015)Qu, Jain, Rajput, Cheng, Zhang, Ong, Brafman,
  Maginn, Curtiss, and Persson]{Qu_Persson:2015}
Qu,~X.; Jain,~A.; Rajput,~N.~N.; Cheng,~L.; Zhang,~Y.; Ong,~S.~P.; Brafman,~M.;
  Maginn,~E.; Curtiss,~L.~A.; Persson,~K.~A. The Electrolyte Genome project: A
  big data approach in battery materials discovery. \emph{Comput. Mater. Sci.}
  \textbf{2015}, \emph{103}, 56--67\relax
\mciteBstWouldAddEndPuncttrue
\mciteSetBstMidEndSepPunct{\mcitedefaultmidpunct}
{\mcitedefaultendpunct}{\mcitedefaultseppunct}\relax
\EndOfBibitem
\bibitem[Lian \latin{et~al.}(2019)Lian, Liu, Li, and Wu]{Lian_Wu:2019}
Lian,~C.; Liu,~H.; Li,~C.; Wu,~J. Hunting ionic liquids with large
  electrochemical potential windows. \emph{AIChE J.} \textbf{2019}, \emph{65},
  804--810\relax
\mciteBstWouldAddEndPuncttrue
\mciteSetBstMidEndSepPunct{\mcitedefaultmidpunct}
{\mcitedefaultendpunct}{\mcitedefaultseppunct}\relax
\EndOfBibitem
\bibitem[Agarwal \latin{et~al.}(2021)Agarwal, Doan, Robertson, Zhang, and
  Assary]{Agarwal_Assary:2021}
Agarwal,~G.; Doan,~H.~A.; Robertson,~L.~A.; Zhang,~L.; Assary,~R.~S. Discovery
  of Energy Storage Molecular Materials Using Quantum Chemistry-Guided
  Multiobjective Bayesian Optimization. \emph{Chem. Mater.} \textbf{2021},
  \emph{33}, 8133–8144\relax
\mciteBstWouldAddEndPuncttrue
\mciteSetBstMidEndSepPunct{\mcitedefaultmidpunct}
{\mcitedefaultendpunct}{\mcitedefaultseppunct}\relax
\EndOfBibitem
\bibitem[Sorkun \latin{et~al.}(2019)Sorkun, Zhang, Khetan, Sorkun, and
  Er]{Sorkun_Er:2022}
Sorkun,~E.; Zhang,~Q.; Khetan,~A.; Sorkun,~M.~C.; Er,~S. RedDB, a computational
  database of electroactive molecules for aqueous redox flow batteries.
  \emph{Sci. Data} \textbf{2019}, \emph{9}, 718\relax
\mciteBstWouldAddEndPuncttrue
\mciteSetBstMidEndSepPunct{\mcitedefaultmidpunct}
{\mcitedefaultendpunct}{\mcitedefaultseppunct}\relax
\EndOfBibitem
\bibitem[Huang \latin{et~al.}(2023)Huang, {von Rudorff}, and von
  Lilienfeld]{Huang_Lilienfeld:2023}
Huang,~B.; {von Rudorff},~G.~F.; von Lilienfeld,~O.~A. The central role of
  density functional theory in the AI age. \emph{Science} \textbf{2023},
  \emph{381}, 170--175\relax
\mciteBstWouldAddEndPuncttrue
\mciteSetBstMidEndSepPunct{\mcitedefaultmidpunct}
{\mcitedefaultendpunct}{\mcitedefaultseppunct}\relax
\EndOfBibitem
\bibitem[Chang and {von Lilienfeld}(2018)Chang, and {von
  Lilienfeld}]{Chang_Lilienfeld:2018}
Chang,~K. Y.~S.; {von Lilienfeld},~O.~A.
  ${\mathrm{Al}}_{x}{\mathrm{Ga}}_{1\ensuremath{-}x}\text{As}$ crystals with
  direct 2 eV band gaps from computational alchemy. \emph{Phys. Rev. Mater.}
  \textbf{2018}, \emph{2}, 073802\relax
\mciteBstWouldAddEndPuncttrue
\mciteSetBstMidEndSepPunct{\mcitedefaultmidpunct}
{\mcitedefaultendpunct}{\mcitedefaultseppunct}\relax
\EndOfBibitem
\bibitem[Griego \latin{et~al.}(2021)Griego, Kitchin, and
  Keith]{Griego_Keith:2021}
Griego,~C.~D.; Kitchin,~J.~R.; Keith,~J.~A. Acceleration of catalyst discovery
  with easy, fast, and reproducible computational alchemy. \emph{Int. J.
  Quantum Chem.} \textbf{2021}, \emph{121}, e26380\relax
\mciteBstWouldAddEndPuncttrue
\mciteSetBstMidEndSepPunct{\mcitedefaultmidpunct}
{\mcitedefaultendpunct}{\mcitedefaultseppunct}\relax
\EndOfBibitem
\bibitem[Eikey \latin{et~al.}(2022)Eikey, Maldonado, Griego, {von Rudorff}, and
  Keith]{Eikey_Keith:2022}
Eikey,~E.~A.; Maldonado,~A.~M.; Griego,~C.~D.; {von Rudorff},~G.~F.;
  Keith,~J.~A. Evaluating quantum alchemy of atoms with thermodynamic cycles:
  Beyond ground electronic states. \emph{J. Chem. Phys.} \textbf{2022},
  \emph{156}, 064106\relax
\mciteBstWouldAddEndPuncttrue
\mciteSetBstMidEndSepPunct{\mcitedefaultmidpunct}
{\mcitedefaultendpunct}{\mcitedefaultseppunct}\relax
\EndOfBibitem
\bibitem[You \latin{et~al.}(2018)You, Liu, Ying, Pande, and
  Leskovec]{You_Leskovec:2018}
You,~J.; Liu,~B.; Ying,~R.; Pande,~V.; Leskovec,~J. Graph Convolutional Policy
  Network for Goal-Directed Molecular Graph Generation. \emph{arXiv}
  \textbf{2018}, 1806.02473\relax
\mciteBstWouldAddEndPuncttrue
\mciteSetBstMidEndSepPunct{\mcitedefaultmidpunct}
{\mcitedefaultendpunct}{\mcitedefaultseppunct}\relax
\EndOfBibitem
\bibitem[Zhou \latin{et~al.}(2019)Zhou, Kearnes, Li, Zare, and
  Riley]{Zhou_Riley:2019}
Zhou,~Z.; Kearnes,~S.; Li,~L.; Zare,~R.~N.; Riley,~P. Optimization of Molecules
  via Deep Reinforcement Learning. \emph{Sci. Rep.} \textbf{2019}, \emph{9},
  10752\relax
\mciteBstWouldAddEndPuncttrue
\mciteSetBstMidEndSepPunct{\mcitedefaultmidpunct}
{\mcitedefaultendpunct}{\mcitedefaultseppunct}\relax
\EndOfBibitem
\bibitem[St{\aa}hl \latin{et~al.}(2019)St{\aa}hl, Falkman, Karlsson, Mathiason,
  and Bostr\"{o}m]{Stahl_Bostrom:2019}
St{\aa}hl,~N.; Falkman,~G.; Karlsson,~A.; Mathiason,~G.; Bostr\"{o}m,~J. Deep
  Reinforcement Learning for Multiparameter Optimization in \emph{de novo} Drug
  Design. \emph{J. Chem. Inf. Model.} \textbf{2019}, \emph{59},
  3166--3176\relax
\mciteBstWouldAddEndPuncttrue
\mciteSetBstMidEndSepPunct{\mcitedefaultmidpunct}
{\mcitedefaultendpunct}{\mcitedefaultseppunct}\relax
\EndOfBibitem
\bibitem[Khemchandani \latin{et~al.}(2020)Khemchandani, O'Hagan, Samanta,
  Swainston, Roberts, Bollegala, and Kell]{Khemchandani_Kell:2020}
Khemchandani,~Y.; O'Hagan,~S.; Samanta,~S.; Swainston,~N.; Roberts,~T.~J.;
  Bollegala,~D.; Kell,~D.~B. DeepGraphMolGen, a multi-objective, computational
  strategy for generating molecules with desirable properties: a graph
  convolution and reinforcement learning approach. \emph{J. Cheminform.}
  \textbf{2020}, \emph{12}, 53\relax
\mciteBstWouldAddEndPuncttrue
\mciteSetBstMidEndSepPunct{\mcitedefaultmidpunct}
{\mcitedefaultendpunct}{\mcitedefaultseppunct}\relax
\EndOfBibitem
\bibitem[Horwood and Noutahi(2020)Horwood, and Noutahi]{Horwood_Noutahi:2020}
Horwood,~J.; Noutahi,~E. Molecular Design in Synthetically Accessible Chemical
  Space via Deep Reinforcement Learning. \emph{{ACS} Omega} \textbf{2020},
  \emph{5}, 32984--32994\relax
\mciteBstWouldAddEndPuncttrue
\mciteSetBstMidEndSepPunct{\mcitedefaultmidpunct}
{\mcitedefaultendpunct}{\mcitedefaultseppunct}\relax
\EndOfBibitem
\bibitem[Pereira \latin{et~al.}(2021)Pereira, Abbasi, Ribeiro, and
  Arrais]{Pereira_Arrais:2021}
Pereira,~T.; Abbasi,~M.; Ribeiro,~B.; Arrais,~J.~P. Diversity oriented Deep
  Reinforcement Learning for targeted molecule generation. \emph{J.
  Cheminform.} \textbf{2021}, \emph{13}, 21\relax
\mciteBstWouldAddEndPuncttrue
\mciteSetBstMidEndSepPunct{\mcitedefaultmidpunct}
{\mcitedefaultendpunct}{\mcitedefaultseppunct}\relax
\EndOfBibitem
\bibitem[Gupta \latin{et~al.}(2018)Gupta, M\"{u}ller, Huisman, Fuchs,
  Schneider, and Schneider]{Gupta_Schneider:2018}
Gupta,~A.; M\"{u}ller,~A.~T.; Huisman,~B. J.~H.; Fuchs,~J.~A.; Schneider,~P.;
  Schneider,~G. Generative Recurrent Networks for De Novo Drug Design.
  \emph{Mol. Inform.} \textbf{2018}, \emph{37}, 1700111\relax
\mciteBstWouldAddEndPuncttrue
\mciteSetBstMidEndSepPunct{\mcitedefaultmidpunct}
{\mcitedefaultendpunct}{\mcitedefaultseppunct}\relax
\EndOfBibitem
\bibitem[Popova \latin{et~al.}(2019)Popova, Shvets, Oliva, and
  Isayev]{Popova_Isayev:2019}
Popova,~M.; Shvets,~M.; Oliva,~J.; Isayev,~O. MolecularRNN: Generating
  realistic molecular graphs with optimized properties. \emph{arXiv}
  \textbf{2019}, 1905.13372\relax
\mciteBstWouldAddEndPuncttrue
\mciteSetBstMidEndSepPunct{\mcitedefaultmidpunct}
{\mcitedefaultendpunct}{\mcitedefaultseppunct}\relax
\EndOfBibitem
\bibitem[Globus \latin{et~al.}(1999)Globus, Lawton, and
  Wipke]{Globus_Wipke:1999}
Globus,~A.; Lawton,~J.; Wipke,~T. Automatic molecular design using evolutionary
  techniques. \emph{Nanotechnology} \textbf{1999}, \emph{10}, 290\relax
\mciteBstWouldAddEndPuncttrue
\mciteSetBstMidEndSepPunct{\mcitedefaultmidpunct}
{\mcitedefaultendpunct}{\mcitedefaultseppunct}\relax
\EndOfBibitem
\bibitem[Brown \latin{et~al.}(2004)Brown, McKay, Gilardoni, and
  Gasteiger]{Brown_Gasteiger:2004}
Brown,~N.; McKay,~B.; Gilardoni,~F.; Gasteiger,~J. A Graph-Based Genetic
  Algorithm and Its Application to the Multiobjective Evolution of Median
  Molecules. \emph{J. Chem. Inf. Comput. Sci.} \textbf{2004}, \emph{44},
  1079–1087\relax
\mciteBstWouldAddEndPuncttrue
\mciteSetBstMidEndSepPunct{\mcitedefaultmidpunct}
{\mcitedefaultendpunct}{\mcitedefaultseppunct}\relax
\EndOfBibitem
\bibitem[Virshup \latin{et~al.}(2013)Virshup, Contreras-Garc\'{i}a, Wipf, Yang,
  and Beratan]{Virshup_Beratan:2013}
Virshup,~A.~M.; Contreras-Garc\'{i}a,~J.; Wipf,~P.; Yang,~W.; Beratan,~D.~N.
  Stochastic Voyages into Uncharted Chemical Space Produce a Representative
  Library of All Possible Drug-Like Compounds. \emph{J. Am. Chem. Soc.}
  \textbf{2013}, \emph{135}, 7296–7303\relax
\mciteBstWouldAddEndPuncttrue
\mciteSetBstMidEndSepPunct{\mcitedefaultmidpunct}
{\mcitedefaultendpunct}{\mcitedefaultseppunct}\relax
\EndOfBibitem
\bibitem[Jensen(2019)]{Jensen:2019}
Jensen,~J.~H. A graph-based genetic algorithm and generative model/Monte Carlo
  tree search for the exploration of chemical space. \emph{Chem. Sci.}
  \textbf{2019}, \emph{10}, 3567–3572\relax
\mciteBstWouldAddEndPuncttrue
\mciteSetBstMidEndSepPunct{\mcitedefaultmidpunct}
{\mcitedefaultendpunct}{\mcitedefaultseppunct}\relax
\EndOfBibitem
\bibitem[Nigam \latin{et~al.}(2022)Nigam, Pollice, and
  Aspuru-Guzik]{Nigam_Aspuru-Guzik:2022}
Nigam,~A.; Pollice,~R.; Aspuru-Guzik,~A. Parallel tempered genetic algorithm
  guided by deep neural networks for inverse molecular design. \emph{Digital
  Discovery} \textbf{2022}, \emph{1}, 390--404\relax
\mciteBstWouldAddEndPuncttrue
\mciteSetBstMidEndSepPunct{\mcitedefaultmidpunct}
{\mcitedefaultendpunct}{\mcitedefaultseppunct}\relax
\EndOfBibitem
\bibitem[Laplaza \latin{et~al.}(2022)Laplaza, Gallarati, and
  Corminboeuf]{Laplaza_Corminboeuf:2022}
Laplaza,~R.; Gallarati,~S.; Corminboeuf,~C. Genetic Optimization of Homogeneous
  Catalysts. \emph{Chem. Methods} \textbf{2022}, \emph{2}, e202100107\relax
\mciteBstWouldAddEndPuncttrue
\mciteSetBstMidEndSepPunct{\mcitedefaultmidpunct}
{\mcitedefaultendpunct}{\mcitedefaultseppunct}\relax
\EndOfBibitem
\bibitem[G\'{o}mez-Bombarelli \latin{et~al.}(2018)G\'{o}mez-Bombarelli, Wei,
  Duvenaud, Hern\'{a}ndez-Lobato, S\'{a}nchez-Lengeling, Sheberla,
  Aguilera-Iparraguirre, Hirzel, Adams, and
  Aspuru-Guzik]{Gomez-Bombarelli_Aspuru-Guzik:2018}
G\'{o}mez-Bombarelli,~R.; Wei,~J.~N.; Duvenaud,~D.;
  Hern\'{a}ndez-Lobato,~J.~M.; S\'{a}nchez-Lengeling,~B.; Sheberla,~D.;
  Aguilera-Iparraguirre,~J.; Hirzel,~T.~D.; Adams,~R.~P.; Aspuru-Guzik,~A.
  Automatic Chemical Design Using a Data-Driven Continuous Representation of
  Molecules. \emph{ACS Cent. Sci.} \textbf{2018}, \emph{4}, 268–276\relax
\mciteBstWouldAddEndPuncttrue
\mciteSetBstMidEndSepPunct{\mcitedefaultmidpunct}
{\mcitedefaultendpunct}{\mcitedefaultseppunct}\relax
\EndOfBibitem
\bibitem[Oliveira \latin{et~al.}(2022)Oliveira, {Da Silva}, and
  Quiles]{Oliveira_Quiles:2022}
Oliveira,~A.~F.; {Da Silva},~J. L.~F.; Quiles,~M.~G. Molecular Property
  Prediction and Molecular Design Using a Supervised Grammar Variational
  Autoencoder. \emph{J. Chem. Inf. Model.} \textbf{2022}, \emph{62},
  817--828\relax
\mciteBstWouldAddEndPuncttrue
\mciteSetBstMidEndSepPunct{\mcitedefaultmidpunct}
{\mcitedefaultendpunct}{\mcitedefaultseppunct}\relax
\EndOfBibitem
\bibitem[Levin and Peres(2017)Levin, and Peres]{Levin_Peres:2017}
Levin,~D.~A.; Peres,~Y. \emph{Markov Chains and Mixing Times: Second Edition};
  American Mathematical Society, 2017\relax
\mciteBstWouldAddEndPuncttrue
\mciteSetBstMidEndSepPunct{\mcitedefaultmidpunct}
{\mcitedefaultendpunct}{\mcitedefaultseppunct}\relax
\EndOfBibitem
\bibitem[Fu \latin{et~al.}(2021)Fu, Xiao, Li, Glass, and Sun]{Fu_Sun:2021}
Fu,~T.; Xiao,~C.; Li,~X.; Glass,~L.~M.; Sun,~J. MIMOSA: Multi-constraint
  Molecule Sampling for Molecule Optimization. \emph{Proc. Conf. AAAI Artif.
  Intell.} \textbf{2021}, \emph{35}, 125--133\relax
\mciteBstWouldAddEndPuncttrue
\mciteSetBstMidEndSepPunct{\mcitedefaultmidpunct}
{\mcitedefaultendpunct}{\mcitedefaultseppunct}\relax
\EndOfBibitem
\bibitem[Xie \latin{et~al.}(2021)Xie, Shi, Zhou, Yang, Zhang, Yu, and
  Lei]{Xie_Lei:2021}
Xie,~Y.; Shi,~C.; Zhou,~H.; Yang,~Y.; Zhang,~W.; Yu,~Y.; Lei,~L. MARS: Markov
  Molecular Sampling for Multi-objective Drug Discovery. \emph{arXiv}
  \textbf{2021}, 2103.10432\relax
\mciteBstWouldAddEndPuncttrue
\mciteSetBstMidEndSepPunct{\mcitedefaultmidpunct}
{\mcitedefaultendpunct}{\mcitedefaultseppunct}\relax
\EndOfBibitem
\bibitem[Liang and Wong(2000)Liang, and Wong]{Liang_Wong:2000}
Liang,~F.; Wong,~W.~H. Evolutionary Monte Carlo: applications to C-p model
  sampling and change point problem. \emph{Stat. Sin.} \textbf{2000},
  \emph{10}, 317--342\relax
\mciteBstWouldAddEndPuncttrue
\mciteSetBstMidEndSepPunct{\mcitedefaultmidpunct}
{\mcitedefaultendpunct}{\mcitedefaultseppunct}\relax
\EndOfBibitem
\bibitem[Hu and Tsui(2010)Hu, and Tsui]{Hu_Tsui:2010}
Hu,~B.; Tsui,~K.-W. Distributed evolutionary Monte Carlo for Bayesian
  computing. \emph{Comput. Stat. Data Anal.} \textbf{2010}, \emph{54},
  688--697\relax
\mciteBstWouldAddEndPuncttrue
\mciteSetBstMidEndSepPunct{\mcitedefaultmidpunct}
{\mcitedefaultendpunct}{\mcitedefaultseppunct}\relax
\EndOfBibitem
\bibitem[Spezia(2020)]{Spezia:2020}
Spezia,~L. Bayesian variable selection in non-homogeneous hidden Markov models
  through an evolutionary Monte Carlo method. \emph{Comput. Stat. Data. Anal.}
  \textbf{2020}, \emph{143}, 106840\relax
\mciteBstWouldAddEndPuncttrue
\mciteSetBstMidEndSepPunct{\mcitedefaultmidpunct}
{\mcitedefaultendpunct}{\mcitedefaultseppunct}\relax
\EndOfBibitem
\bibitem[Hukushima and Nemoto(1996)Hukushima, and
  Nemoto]{Hukushima_Nemoto:1996}
Hukushima,~K.; Nemoto,~K. Exchange Monte Carlo Method and Application to Spin
  Glass Simulations. \emph{J. Phys. Soc. Jpn.} \textbf{1996}, \emph{65},
  1604--1608\relax
\mciteBstWouldAddEndPuncttrue
\mciteSetBstMidEndSepPunct{\mcitedefaultmidpunct}
{\mcitedefaultendpunct}{\mcitedefaultseppunct}\relax
\EndOfBibitem
\bibitem[Sambridge(2014)]{Sambridge:2014}
Sambridge,~M. A Parallel Tempering algorithm for probabilistic sampling and
  multimodal optimization. \emph{Geophys. J. Int.} \textbf{2014}, \emph{196},
  357–374\relax
\mciteBstWouldAddEndPuncttrue
\mciteSetBstMidEndSepPunct{\mcitedefaultmidpunct}
{\mcitedefaultendpunct}{\mcitedefaultseppunct}\relax
\EndOfBibitem
\bibitem[Angelini and Ricci-Tersenghi(2019)Angelini, and
  Ricci-Tersenghi]{Angelini_Ricci-Tersenghi:2019}
Angelini,~M.~C.; Ricci-Tersenghi,~F. Monte Carlo algorithms are very effective
  in finding the largest independent set in sparse random graphs. \emph{Phys.
  Rev. E} \textbf{2019}, \emph{100}, 013302\relax
\mciteBstWouldAddEndPuncttrue
\mciteSetBstMidEndSepPunct{\mcitedefaultmidpunct}
{\mcitedefaultendpunct}{\mcitedefaultseppunct}\relax
\EndOfBibitem
\bibitem[Holland(1975)]{Holland:1975}
Holland,~J.~H. \emph{Adaptation in Natural and Artificial Systems}; University
  of Michigan Press, Ann Arbor, 1975\relax
\mciteBstWouldAddEndPuncttrue
\mciteSetBstMidEndSepPunct{\mcitedefaultmidpunct}
{\mcitedefaultendpunct}{\mcitedefaultseppunct}\relax
\EndOfBibitem
\bibitem[J\'{o}hannesson \latin{et~al.}(2002)J\'{o}hannesson, Bligaard, Ruban,
  Skriver, Jacobsen, and N\o{o}rskov]{Johannesson_Norskov:2002}
J\'{o}hannesson,~G.~H.; Bligaard,~T.; Ruban,~A.~V.; Skriver,~H.~L.;
  Jacobsen,~K.~W.; N\o{o}rskov,~J.~K. Combined Electronic Structure and
  Evolutionary Search Approach to Materials Design. \emph{Phys. Rev. Lett.}
  \textbf{2002}, \emph{88}, 255506\relax
\mciteBstWouldAddEndPuncttrue
\mciteSetBstMidEndSepPunct{\mcitedefaultmidpunct}
{\mcitedefaultendpunct}{\mcitedefaultseppunct}\relax
\EndOfBibitem
\bibitem[Sharma \latin{et~al.}(2010)Sharma, Singh, and
  Balint-Kurti]{Sharma_Balint-Kurti:2010}
Sharma,~S.; Singh,~H.; Balint-Kurti,~G.~G. Genetic algorithm optimization of
  laser pulses for molecular quantum state excitation. \emph{J. Chem. Phys.}
  \textbf{2010}, \emph{132}, 064108\relax
\mciteBstWouldAddEndPuncttrue
\mciteSetBstMidEndSepPunct{\mcitedefaultmidpunct}
{\mcitedefaultendpunct}{\mcitedefaultseppunct}\relax
\EndOfBibitem
\bibitem[Hastings(1970)]{Hastings:1970}
Hastings,~W.~K. Monte Carlo sampling methods using Markov chains and their
  applications. \emph{Biometrika} \textbf{1970}, \emph{57}, 97--109\relax
\mciteBstWouldAddEndPuncttrue
\mciteSetBstMidEndSepPunct{\mcitedefaultmidpunct}
{\mcitedefaultendpunct}{\mcitedefaultseppunct}\relax
\EndOfBibitem
\bibitem[Iftimie \latin{et~al.}(2000)Iftimie, Salahub, Wei, and
  Schofield]{Iftimie_Schofield:2000}
Iftimie,~R.; Salahub,~D.; Wei,~D.; Schofield,~J. Using a classical potential as
  an efficient importance function for sampling from an ab initio potential.
  \emph{J. Chem. Phys.} \textbf{2000}, \emph{113}, 4852--4862\relax
\mciteBstWouldAddEndPuncttrue
\mciteSetBstMidEndSepPunct{\mcitedefaultmidpunct}
{\mcitedefaultendpunct}{\mcitedefaultseppunct}\relax
\EndOfBibitem
\bibitem[Gelb(2003)]{Gelb:2003}
Gelb,~L.~D. Monte Carlo simulations using sampling from an approximate
  potential. \emph{J. Chem. Phys.} \textbf{2003}, \emph{118}, 7747--7750\relax
\mciteBstWouldAddEndPuncttrue
\mciteSetBstMidEndSepPunct{\mcitedefaultmidpunct}
{\mcitedefaultendpunct}{\mcitedefaultseppunct}\relax
\EndOfBibitem
\bibitem[Jadrich and Leiding(2020)Jadrich, and Leiding]{Jadrich_Leiding:2020}
Jadrich,~R.~B.; Leiding,~J.~A. Accelerating Ab Initio Simulation via Nested
  Monte Carlo and Machine Learned Reference Potentials. \emph{J. Phys. Chem. B}
  \textbf{2020}, \emph{124}, 5488--5497\relax
\mciteBstWouldAddEndPuncttrue
\mciteSetBstMidEndSepPunct{\mcitedefaultmidpunct}
{\mcitedefaultendpunct}{\mcitedefaultseppunct}\relax
\EndOfBibitem
\bibitem[Zhang \latin{et~al.}(2013)Zhang, Iyyamperumal, Yancey, Crooks, and
  Henkelman]{Zhang_Henkelman:2013}
Zhang,~L.; Iyyamperumal,~R.; Yancey,~D.~F.; Crooks,~R.~M.; Henkelman,~G. Design
  of Pt-Shell Nanoparticles with Alloy Cores for the Oxygen Reduction Reaction.
  \emph{ACS Nano} \textbf{2013}, \emph{7}, 9168--9172\relax
\mciteBstWouldAddEndPuncttrue
\mciteSetBstMidEndSepPunct{\mcitedefaultmidpunct}
{\mcitedefaultendpunct}{\mcitedefaultseppunct}\relax
\EndOfBibitem
\bibitem[Anderson \latin{et~al.}(2015)Anderson, Yancey, Zhang, Chill,
  Henkelman, and Crooks]{Anderson_Crooks:2015}
Anderson,~R.~M.; Yancey,~D.~F.; Zhang,~L.; Chill,~S.~T.; Henkelman,~G.;
  Crooks,~R.~M. A Theoretical and Experimental Approach for Correlating
  Nanoparticle Structure and Electrocatalytic Activity. \emph{Acc. Chem. Res.}
  \textbf{2015}, \emph{48}, 1351--1357\relax
\mciteBstWouldAddEndPuncttrue
\mciteSetBstMidEndSepPunct{\mcitedefaultmidpunct}
{\mcitedefaultendpunct}{\mcitedefaultseppunct}\relax
\EndOfBibitem
\bibitem[Shields \latin{et~al.}(2021)Shields, Stevens, Li, Parasram, Damani,
  Alvarado, Janey, Adams, and Doyle]{Shields_Doyle:2021}
Shields,~B.~J.; Stevens,~J.; Li,~J.; Parasram,~M.; Damani,~F.; Alvarado,~J.
  I.~M.; Janey,~J.~M.; Adams,~R.~P.; Doyle,~A.~G. Bayesian reaction
  optimization as a tool for chemical synthesis. \emph{Nature} \textbf{2021},
  \emph{590}, 89--96\relax
\mciteBstWouldAddEndPuncttrue
\mciteSetBstMidEndSepPunct{\mcitedefaultmidpunct}
{\mcitedefaultendpunct}{\mcitedefaultseppunct}\relax
\EndOfBibitem
\bibitem[Park \latin{et~al.}(2023)Park, Kim, Hong, Han, Nam, and
  Jung]{Park_Jung:2023}
Park,~J.; Kim,~Y.~M.; Hong,~S.; Han,~B.; Nam,~K.~T.; Jung,~Y. Closed-loop
  optimization of nanoparticle synthesis enabled by robotics and machine
  learning. \emph{Matter} \textbf{2023}, \emph{6}, 677--690\relax
\mciteBstWouldAddEndPuncttrue
\mciteSetBstMidEndSepPunct{\mcitedefaultmidpunct}
{\mcitedefaultendpunct}{\mcitedefaultseppunct}\relax
\EndOfBibitem
\bibitem[Rahmanian \latin{et~al.}(2022)Rahmanian, Flowers, Guevarra, Richter,
  Fichtner, Donnely, Gregoire, and Stein]{Rahmanian_Stein:2022}
Rahmanian,~F.; Flowers,~J.; Guevarra,~D.; Richter,~M.; Fichtner,~M.;
  Donnely,~P.; Gregoire,~J.~M.; Stein,~H.~S. Enabling Modular Autonomous
  Feedback-Loops in Materials Science through Hierarchical Experimental
  Laboratory Automation and Orchestration. \emph{Adv. Mater. Interfaces}
  \textbf{2022}, \emph{9}, 2101987\relax
\mciteBstWouldAddEndPuncttrue
\mciteSetBstMidEndSepPunct{\mcitedefaultmidpunct}
{\mcitedefaultendpunct}{\mcitedefaultseppunct}\relax
\EndOfBibitem
\bibitem[Stein \latin{et~al.}(2022)Stein, Sanin, Rahmanian, Zhang, Vogler,
  Flowers, Fischer, Fuchs, Choudhary, and Schroeder]{Stein_Schroeder:2022}
Stein,~H.~S.; Sanin,~A.; Rahmanian,~F.; Zhang,~B.; Vogler,~M.; Flowers,~J.~K.;
  Fischer,~L.; Fuchs,~S.; Choudhary,~N.; Schroeder,~L. From materials discovery
  to system optimization by integrating combinatorial electrochemistry and data
  science. \emph{Curr. Opin. Electrochem.} \textbf{2022}, \emph{35},
  101053\relax
\mciteBstWouldAddEndPuncttrue
\mciteSetBstMidEndSepPunct{\mcitedefaultmidpunct}
{\mcitedefaultendpunct}{\mcitedefaultseppunct}\relax
\EndOfBibitem
\bibitem[Manzano \latin{et~al.}(2022)Manzano, Hou, Zalesskiy, Frei, Wang,
  Kitson, and Cronin]{Manzano_Cronin:2022}
Manzano,~J.~S.; Hou,~W.; Zalesskiy,~S.~S.; Frei,~P.; Wang,~H.; Kitson,~P.~J.;
  Cronin,~L. An autonomous portable platform for universal chemical synthesis.
  \emph{Nat. Chem.} \textbf{2022}, \emph{14}, 1311--1318\relax
\mciteBstWouldAddEndPuncttrue
\mciteSetBstMidEndSepPunct{\mcitedefaultmidpunct}
{\mcitedefaultendpunct}{\mcitedefaultseppunct}\relax
\EndOfBibitem
\bibitem[Nigam \latin{et~al.}(2021)Nigam, Pollice, Krenn, dos Passos~Gomes, and
  Aspuru-Guzik]{Nigam_Aspuru-Guzik:2021}
Nigam,~A.; Pollice,~R.; Krenn,~M.; dos Passos~Gomes,~G.; Aspuru-Guzik,~A.
  Beyond generative models: superfast traversal, optimization, novelty,
  exploration and discovery (STONED) algorithm for molecules using SELFIES.
  \emph{Chem. Sci.} \textbf{2021}, \emph{12}, 7079--7090\relax
\mciteBstWouldAddEndPuncttrue
\mciteSetBstMidEndSepPunct{\mcitedefaultmidpunct}
{\mcitedefaultendpunct}{\mcitedefaultseppunct}\relax
\EndOfBibitem
\bibitem[Born and Manica(2023)Born, and Manica]{Born_Manica:2023}
Born,~J.; Manica,~M. Regression Transformer enables concurrent sequence
  regression and generation for molecular language modelling. \emph{Nat. Mach.
  Intell.} \textbf{2023}, \emph{5}, 432--444\relax
\mciteBstWouldAddEndPuncttrue
\mciteSetBstMidEndSepPunct{\mcitedefaultmidpunct}
{\mcitedefaultendpunct}{\mcitedefaultseppunct}\relax
\EndOfBibitem
\bibitem[Lemm and {von Lilienfeld}(2021)Lemm, and {von
  Lilienfeld}]{Lemm_Lilienfeld:2021}
Lemm,~D.; {von Lilienfeld},~G. F. v. O.~A. Machine learning based energy-free
  structure predictions of molecules, transition states, and solids. \emph{Nat.
  Commun.} \textbf{2021}, \emph{12}, 4468\relax
\mciteBstWouldAddEndPuncttrue
\mciteSetBstMidEndSepPunct{\mcitedefaultmidpunct}
{\mcitedefaultendpunct}{\mcitedefaultseppunct}\relax
\EndOfBibitem
\bibitem[Weinreich \latin{et~al.}(2022)Weinreich, Lemm, {von Rudorff}, and von
  Lilienfeld]{Weinreich_Lilienfeld:2022}
Weinreich,~J.; Lemm,~D.; {von Rudorff},~G.~F.; von Lilienfeld,~O. Ab initio
  machine learning of phase space averages. \emph{J. Chem. Phys.}
  \textbf{2022}, \emph{157}, 024303\relax
\mciteBstWouldAddEndPuncttrue
\mciteSetBstMidEndSepPunct{\mcitedefaultmidpunct}
{\mcitedefaultendpunct}{\mcitedefaultseppunct}\relax
\EndOfBibitem
\bibitem[Wang and Landau(2001)Wang, and Landau]{Wang_Landau:2001}
Wang,~F.; Landau,~D.~P. Efficient, Multiple-Range Random Walk Algorithm to
  Calculate the Density of States. \emph{Phys. Rev. Lett.} \textbf{2001},
  \emph{86}, 2050--2053\relax
\mciteBstWouldAddEndPuncttrue
\mciteSetBstMidEndSepPunct{\mcitedefaultmidpunct}
{\mcitedefaultendpunct}{\mcitedefaultseppunct}\relax
\EndOfBibitem
\bibitem[Thiede \latin{et~al.}(2022)Thiede, Krenn, Nigam, and
  Aspuru-Guzik]{Thiede_Aspuru-Guzik:2022}
Thiede,~L.~A.; Krenn,~M.; Nigam,~A.~K.; Aspuru-Guzik,~A. Curiosity in exploring
  chemical spaces: intrinsic rewards for molecular reinforcement learning.
  \emph{Mach. Learn.: Sci. Technol.} \textbf{2022}, \emph{3}, 035008\relax
\mciteBstWouldAddEndPuncttrue
\mciteSetBstMidEndSepPunct{\mcitedefaultmidpunct}
{\mcitedefaultendpunct}{\mcitedefaultseppunct}\relax
\EndOfBibitem
\bibitem[Reker(2019)]{Reker:2019}
Reker,~D. Practical considerations for active machine learning in drug
  discovery. \emph{Drug Discov. Today Technol.} \textbf{2019}, \emph{32-33},
  73--79\relax
\mciteBstWouldAddEndPuncttrue
\mciteSetBstMidEndSepPunct{\mcitedefaultmidpunct}
{\mcitedefaultendpunct}{\mcitedefaultseppunct}\relax
\EndOfBibitem
\bibitem[Fu \latin{et~al.}(2022)Fu, Gao, Coley, and Sun]{Fu_Sun:2022}
Fu,~T.; Gao,~W.; Coley,~C.; Sun,~J. Reinforced Genetic Algorithm for
  Structure-based Drug Design. Advances in Neural Information Processing
  Systems. 2022; pp 12325--12338\relax
\mciteBstWouldAddEndPuncttrue
\mciteSetBstMidEndSepPunct{\mcitedefaultmidpunct}
{\mcitedefaultendpunct}{\mcitedefaultseppunct}\relax
\EndOfBibitem
\bibitem[Carter \latin{et~al.}(2023)Carter, Hollander, Spasov, Anderson, and
  Jorgensen]{Carter_Jorgensen:2023}
Carter,~Z.~J.; Hollander,~K.; Spasov,~K.~A.; Anderson,~K.~S.; Jorgensen,~W.~L.
  Design, synthesis, and biological testing of biphenylmethyloxazole inhibitors
  targeting HIV-1 reverse transcriptase. \emph{Bioorg. Med. Chem. Lett.}
  \textbf{2023}, \emph{84}, 129216\relax
\mciteBstWouldAddEndPuncttrue
\mciteSetBstMidEndSepPunct{\mcitedefaultmidpunct}
{\mcitedefaultendpunct}{\mcitedefaultseppunct}\relax
\EndOfBibitem
\bibitem[Borodin(2019)]{Borodin:2019}
Borodin,~O. Challenges with prediction of battery electrolyte electrochemical
  stability window and guiding the electrode – electrolyte stabilization.
  \emph{Curr. Opin. Electrochem.} \textbf{2019}, \emph{13}, 86--93\relax
\mciteBstWouldAddEndPuncttrue
\mciteSetBstMidEndSepPunct{\mcitedefaultmidpunct}
{\mcitedefaultendpunct}{\mcitedefaultseppunct}\relax
\EndOfBibitem
\bibitem[Teunissen \latin{et~al.}(2017)Teunissen, {De Proft}, and {De
  Vleeschouwer}]{Teunissen_De_Vleeschouwer:2017}
Teunissen,~J.~L.; {De Proft},~F.; {De Vleeschouwer},~F. Tuning the HOMO–LUMO
  Energy Gap of Small Diamondoids Using Inverse Molecular Design. \emph{J.
  Chem. Theory Comput.} \textbf{2017}, \emph{13}, 1351--1365\relax
\mciteBstWouldAddEndPuncttrue
\mciteSetBstMidEndSepPunct{\mcitedefaultmidpunct}
{\mcitedefaultendpunct}{\mcitedefaultseppunct}\relax
\EndOfBibitem
\bibitem[De_(2020)]{De_Lile_Lee:2020}
Do HOMO-LUMO Energy Levels and Band Gaps Provide Sufficient Understanding of
  Dye-Sensitizer Activity Trends for Water Purification? \emph{ACS Omega}
  \textbf{2020}, \emph{5}, 15052--15062\relax
\mciteBstWouldAddEndPuncttrue
\mciteSetBstMidEndSepPunct{\mcitedefaultmidpunct}
{\mcitedefaultendpunct}{\mcitedefaultseppunct}\relax
\EndOfBibitem
\bibitem[Fromer and Coley(2023)Fromer, and Coley]{Fromer_Coley:2023}
Fromer,~J.~C.; Coley,~C.~W. Computer-aided multi-objective optimization in
  small molecule discovery. \emph{Patterns} \textbf{2023}, \emph{4},
  100678\relax
\mciteBstWouldAddEndPuncttrue
\mciteSetBstMidEndSepPunct{\mcitedefaultmidpunct}
{\mcitedefaultendpunct}{\mcitedefaultseppunct}\relax
\EndOfBibitem
\bibitem[Halgren(1996)]{Halgren:1996_I}
Halgren,~T.~A. Merck molecular force field. I. Basis, form, scope,
  parameterization, and performance of MMFF94. \emph{J. Comput. Chem.}
  \textbf{1996}, \emph{17}, 490--519\relax
\mciteBstWouldAddEndPuncttrue
\mciteSetBstMidEndSepPunct{\mcitedefaultmidpunct}
{\mcitedefaultendpunct}{\mcitedefaultseppunct}\relax
\EndOfBibitem
\bibitem[Halgren(1996)]{Halgren:1996_II}
Halgren,~T.~A. Merck molecular force field. II. MMFF94 van der Waals and
  electrostatic parameters for intermolecular interactions. \emph{J. Comput.
  Chem.} \textbf{1996}, \emph{17}, 520--552\relax
\mciteBstWouldAddEndPuncttrue
\mciteSetBstMidEndSepPunct{\mcitedefaultmidpunct}
{\mcitedefaultendpunct}{\mcitedefaultseppunct}\relax
\EndOfBibitem
\bibitem[Halgren(1996)]{Halgren:1996_III}
Halgren,~T.~A. Merck molecular force field. III. Molecular geometries and
  vibrational frequencies for MMFF94. \emph{J. Comput. Chem.} \textbf{1996},
  \emph{17}, 553--586\relax
\mciteBstWouldAddEndPuncttrue
\mciteSetBstMidEndSepPunct{\mcitedefaultmidpunct}
{\mcitedefaultendpunct}{\mcitedefaultseppunct}\relax
\EndOfBibitem
\bibitem[Halgren and Nachbar(1996)Halgren, and
  Nachbar]{Halgren_Nachbar:1996_IV}
Halgren,~T.~A.; Nachbar,~R.~B. Merck molecular force field. IV. Conformational
  energies and geometries for MMFF94. \emph{J. Comput. Chem.} \textbf{1996},
  \emph{17}, 587--615\relax
\mciteBstWouldAddEndPuncttrue
\mciteSetBstMidEndSepPunct{\mcitedefaultmidpunct}
{\mcitedefaultendpunct}{\mcitedefaultseppunct}\relax
\EndOfBibitem
\bibitem[Halgren(1996)]{Halgren:1996_V}
Halgren,~T.~A. Merck molecular force field. V. Extension of MMFF94 using
  experimental data, additional computational data, and empirical rules.
  \emph{J. Comput. Chem.} \textbf{1996}, \emph{17}, 616--641\relax
\mciteBstWouldAddEndPuncttrue
\mciteSetBstMidEndSepPunct{\mcitedefaultmidpunct}
{\mcitedefaultendpunct}{\mcitedefaultseppunct}\relax
\EndOfBibitem
\bibitem[Halgren(1999)]{Halgren:1999_VI}
Halgren,~T.~A. MMFF VI. MMFF94s option for energy minimization studies.
  \emph{J. Comput. Chem.} \textbf{1999}, \emph{20}, 720--729\relax
\mciteBstWouldAddEndPuncttrue
\mciteSetBstMidEndSepPunct{\mcitedefaultmidpunct}
{\mcitedefaultendpunct}{\mcitedefaultseppunct}\relax
\EndOfBibitem
\bibitem[Halgren(1999)]{Halgren:1999_VII}
Halgren,~T.~A. MMFF VII. Characterization of MMFF94, MMFF94s, and other widely
  available force fields for conformational energies and for
  intermolecular-interaction energies and geometries. \emph{J. Comput. Chem.}
  \textbf{1999}, \emph{20}, 730--748\relax
\mciteBstWouldAddEndPuncttrue
\mciteSetBstMidEndSepPunct{\mcitedefaultmidpunct}
{\mcitedefaultendpunct}{\mcitedefaultseppunct}\relax
\EndOfBibitem
\bibitem[Bannwarth \latin{et~al.}(2019)Bannwarth, Ehlert, and
  Grimme]{Bannwarth_Grimme:2019}
Bannwarth,~C.; Ehlert,~S.; Grimme,~S. GFN2-xTB-An Accurate and Broadly
  Parametrized Self-Consistent Tight-Binding Quantum Chemical Method with
  Multipole Electrostatics and Density-Dependent Dispersion Contributions.
  \emph{J. Chem. Theory Comput.} \textbf{2019}, \emph{15}, 1652–1671\relax
\mciteBstWouldAddEndPuncttrue
\mciteSetBstMidEndSepPunct{\mcitedefaultmidpunct}
{\mcitedefaultendpunct}{\mcitedefaultseppunct}\relax
\EndOfBibitem
\bibitem[Ehlert \latin{et~al.}(2021)Ehlert, Stahn, Spicher, and
  Grimme]{Ehlert_Grimme:2021}
Ehlert,~S.; Stahn,~M.; Spicher,~S.; Grimme,~S. Robust and Efficient Implicit
  Solvation Model for Fast Semiempirical Methods. \emph{J. Chem. Theory
  Comput.} \textbf{2021}, \emph{17}, 4250–4261\relax
\mciteBstWouldAddEndPuncttrue
\mciteSetBstMidEndSepPunct{\mcitedefaultmidpunct}
{\mcitedefaultendpunct}{\mcitedefaultseppunct}\relax
\EndOfBibitem
\bibitem[Ruddigkeit \latin{et~al.}(2012)Ruddigkeit, {van Deursen}, Blum, and
  Reymond]{Ruddigkeit_Reymond:2012}
Ruddigkeit,~L.; {van Deursen},~R.; Blum,~L.~C.; Reymond,~J.-L. Enumeration of
  166 Billion Organic Small Molecules in the Chemical Universe Database GDB-17.
  \emph{J. Chem. Inf. Model.} \textbf{2012}, \emph{52}, 2864–2875\relax
\mciteBstWouldAddEndPuncttrue
\mciteSetBstMidEndSepPunct{\mcitedefaultmidpunct}
{\mcitedefaultendpunct}{\mcitedefaultseppunct}\relax
\EndOfBibitem
\bibitem[Ramakrishnan \latin{et~al.}(2014)Ramakrishnan, Dral, Rupp, and {von
  Lilienfeld}]{Ramakrishnan_Lilienfeld:2014}
Ramakrishnan,~R.; Dral,~P.~O.; Rupp,~M.; {von Lilienfeld},~O.~A. Quantum
  chemistry structures and properties of 134 kilo molecules. \emph{Sci. Data}
  \textbf{2014}, \emph{1}, 140022\relax
\mciteBstWouldAddEndPuncttrue
\mciteSetBstMidEndSepPunct{\mcitedefaultmidpunct}
{\mcitedefaultendpunct}{\mcitedefaultseppunct}\relax
\EndOfBibitem
\bibitem[Jain \latin{et~al.}(2013)Jain, Ong, Hautier, Chen, Richards, Dacek,
  Cholia, Gunter, Skinner, Ceder, and Persson]{Jain_Persson:2013}
Jain,~A.; Ong,~S.~P.; Hautier,~G.; Chen,~W.; Richards,~W.~D.; Dacek,~S.;
  Cholia,~S.; Gunter,~D.; Skinner,~D.; Ceder,~G.; Persson,~K.~A. Commentary:
  The Materials Project: A materials genome approach to accelerating materials
  innovation. \emph{APL Mater.} \textbf{2013}, \emph{1}, 011002\relax
\mciteBstWouldAddEndPuncttrue
\mciteSetBstMidEndSepPunct{\mcitedefaultmidpunct}
{\mcitedefaultendpunct}{\mcitedefaultseppunct}\relax
\EndOfBibitem
\bibitem[Clayden \latin{et~al.}(2012)Clayden, Geeves, and
  Warren]{Clayden_Warren:2012}
Clayden,~J.; Geeves,~N.; Warren,~S. \emph{Organic Chemistry}, 2nd ed.; Oxford
  University Press, 2012\relax
\mciteBstWouldAddEndPuncttrue
\mciteSetBstMidEndSepPunct{\mcitedefaultmidpunct}
{\mcitedefaultendpunct}{\mcitedefaultseppunct}\relax
\EndOfBibitem
\bibitem[Rathore \latin{et~al.}(2005)Rathore, Chopra, and {J. de
  Pablo}]{Rathore_Pablo:2005}
Rathore,~N.; Chopra,~M.; {J. de Pablo},~J. Optimal allocation of replicas in
  parallel tempering simulations. \emph{J. Chem. Phys.} \textbf{2005},
  \emph{122}, 024111\relax
\mciteBstWouldAddEndPuncttrue
\mciteSetBstMidEndSepPunct{\mcitedefaultmidpunct}
{\mcitedefaultendpunct}{\mcitedefaultseppunct}\relax
\EndOfBibitem
\bibitem[Kone and Kofke(2005)Kone, and Kofke]{Kone_Kofke:2005}
Kone,~A.; Kofke,~D.~A. Selection of temperature intervals for
  parallel-tempering simulations. \emph{J. Chem. Phys.} \textbf{2005},
  \emph{122}, 206101\relax
\mciteBstWouldAddEndPuncttrue
\mciteSetBstMidEndSepPunct{\mcitedefaultmidpunct}
{\mcitedefaultendpunct}{\mcitedefaultseppunct}\relax
\EndOfBibitem
\bibitem[Gastegger \latin{et~al.}(2017)Gastegger, Behler, and
  Marquetand]{Gastegger_Marquetand:2017}
Gastegger,~M.; Behler,~J.; Marquetand,~P. Machine learning molecular dynamics
  for the simulation of infrared spectra. \emph{Chem. Sci.} \textbf{2017},
  \emph{8}, 6924–6935\relax
\mciteBstWouldAddEndPuncttrue
\mciteSetBstMidEndSepPunct{\mcitedefaultmidpunct}
{\mcitedefaultendpunct}{\mcitedefaultseppunct}\relax
\EndOfBibitem
\bibitem[Podryabinkin and Shapeev(2017)Podryabinkin, and
  Shapeev]{Podryabinkin_Shapeev:2017}
Podryabinkin,~E.~V.; Shapeev,~A.~V. Active learning of linearly parametrized
  interatomic potentials. \emph{Comput. Mater. Sci.} \textbf{2017}, \emph{140},
  171--180\relax
\mciteBstWouldAddEndPuncttrue
\mciteSetBstMidEndSepPunct{\mcitedefaultmidpunct}
{\mcitedefaultendpunct}{\mcitedefaultseppunct}\relax
\EndOfBibitem
\bibitem[Schaaf \latin{et~al.}(2023)Schaaf, Fako, De, Sch\"{a}fer, and
  Cs\'{a}nyi]{Schaaf_Csanyi:2023}
Schaaf,~L.; Fako,~E.; De,~S.; Sch\"{a}fer,~A.; Cs\'{a}nyi,~G. Accurate Reaction
  Barriers for Catalytic Pathways: An Automatic Training Protocol for Machine
  Learning Force Fields. \emph{arXiv} \textbf{2023}, 2301.09931\relax
\mciteBstWouldAddEndPuncttrue
\mciteSetBstMidEndSepPunct{\mcitedefaultmidpunct}
{\mcitedefaultendpunct}{\mcitedefaultseppunct}\relax
\EndOfBibitem
\bibitem[Vandermause \latin{et~al.}(2022)Vandermause, Xie, Lim, Owen, and
  Kozinsky]{Vandermause_Kozinsky:2023}
Vandermause,~J.; Xie,~Y.; Lim,~J.~S.; Owen,~C.~J.; Kozinsky,~B. Active learning
  of reactive Bayesian force fields applied to heterogeneous catalysis dynamics
  of H/Pt. \emph{Nat. Commun.} \textbf{2022}, \emph{13}, 5183\relax
\mciteBstWouldAddEndPuncttrue
\mciteSetBstMidEndSepPunct{\mcitedefaultmidpunct}
{\mcitedefaultendpunct}{\mcitedefaultseppunct}\relax
\EndOfBibitem
\bibitem[Hern\'{a}ndez-Lobato \latin{et~al.}(2017)Hern\'{a}ndez-Lobato,
  Requeima, Pyzer-Knapp, and Aspuru-Guzik]{Hernandez-Lobato_Aspuru-Guzik:2017}
Hern\'{a}ndez-Lobato,~J.~M.; Requeima,~J.; Pyzer-Knapp,~E.~O.; Aspuru-Guzik,~A.
  Parallel and Distributed Thompson Sampling for Large-scale Accelerated
  Exploration of Chemical Space. \emph{arXiv} \textbf{2017}, 1706.01825\relax
\mciteBstWouldAddEndPuncttrue
\mciteSetBstMidEndSepPunct{\mcitedefaultmidpunct}
{\mcitedefaultendpunct}{\mcitedefaultseppunct}\relax
\EndOfBibitem
\bibitem[Smith \latin{et~al.}(2018)Smith, Nebgen, Lubbers, Isayev, and
  Roitberg]{Smith_Roitberg:2018}
Smith,~J.~S.; Nebgen,~B.; Lubbers,~N.; Isayev,~O.; Roitberg,~A.~E. Less is
  more: Sampling chemical space with active learning. \emph{J. Chem. Phys.}
  \textbf{2018}, \emph{148}, 241733\relax
\mciteBstWouldAddEndPuncttrue
\mciteSetBstMidEndSepPunct{\mcitedefaultmidpunct}
{\mcitedefaultendpunct}{\mcitedefaultseppunct}\relax
\EndOfBibitem
\bibitem[Vousden \latin{et~al.}(2005)Vousden, Farr, and
  Mandel]{Vousden_Mandel:2015}
Vousden,~W.~D.; Farr,~W.~M.; Mandel,~I. Dynamic temperature selection for
  parallel tempering in Markov chain Monte Carlo simulations. \emph{Mon. Not.
  R. Astron. Soc.} \textbf{2005}, \emph{455}, 1919–1937\relax
\mciteBstWouldAddEndPuncttrue
\mciteSetBstMidEndSepPunct{\mcitedefaultmidpunct}
{\mcitedefaultendpunct}{\mcitedefaultseppunct}\relax
\EndOfBibitem
\bibitem[Huang and von Lilienfeld(2020)Huang, and von
  Lilienfeld]{Huang_Lilienfeld:2020}
Huang,~B.; von Lilienfeld,~O.~A. Quantum machine learning using
  atom-in-molecule-based fragments selected on the fly. \emph{Nat. Chem.}
  \textbf{2020}, \emph{12}, 945–951\relax
\mciteBstWouldAddEndPuncttrue
\mciteSetBstMidEndSepPunct{\mcitedefaultmidpunct}
{\mcitedefaultendpunct}{\mcitedefaultseppunct}\relax
\EndOfBibitem
\bibitem[Huang \latin{et~al.}(2023)Huang, von Lilienfeld, Krogel, and
  Benali]{Huang_Benali:2023}
Huang,~B.; von Lilienfeld,~O.~A.; Krogel,~J.~T.; Benali,~A. Toward DMC Accuracy
  Across Chemical Space with Scalable $\Delta$-QML. \emph{J. Chem. Theory
  Comput.} \textbf{2023}, \emph{19}, 1711–1721\relax
\mciteBstWouldAddEndPuncttrue
\mciteSetBstMidEndSepPunct{\mcitedefaultmidpunct}
{\mcitedefaultendpunct}{\mcitedefaultseppunct}\relax
\EndOfBibitem
\end{mcitethebibliography}


\providecommand{\latin}[1]{#1}
\makeatletter
\providecommand{\doi}
  {\begingroup\let\do\@makeother\dospecials
  \catcode`\{=1 \catcode`\}=2 \doi@aux}
\providecommand{\doi@aux}[1]{\endgroup\texttt{#1}}
\makeatother
\providecommand*\mcitethebibliography{\thebibliography}
\csname @ifundefined\endcsname{endmcitethebibliography}
  {\let\endmcitethebibliography\endthebibliography}{}
\begin{mcitethebibliography}{18}
\providecommand*\natexlab[1]{#1}
\providecommand*\mciteSetBstSublistMode[1]{}
\providecommand*\mciteSetBstMaxWidthForm[2]{}
\providecommand*\mciteBstWouldAddEndPuncttrue
  {\def\EndOfBibitem{\unskip.}}
\providecommand*\mciteBstWouldAddEndPunctfalse
  {\let\EndOfBibitem\relax}
\providecommand*\mciteSetBstMidEndSepPunct[3]{}
\providecommand*\mciteSetBstSublistLabelBeginEnd[3]{}
\providecommand*\EndOfBibitem{}
\mciteSetBstSublistMode{f}
\mciteSetBstMaxWidthForm{subitem}{(\alph{mcitesubitemcount})}
\mciteSetBstSublistLabelBeginEnd
  {\mcitemaxwidthsubitemform\space}
  {\relax}
  {\relax}

\bibitem[Ebejer \latin{et~al.}()Ebejer, Morris, and Deane]{Ebejer_Deane:2012}
Ebejer,~J.-P.; Morris,~G.~M.; Deane,~C.~M. Freely Available Conformer
  Generation Methods: How Good Are They? \emph{J. Chem. Inf. Model.} \emph{52},
  1146–1158\relax
\mciteBstWouldAddEndPuncttrue
\mciteSetBstMidEndSepPunct{\mcitedefaultmidpunct}
{\mcitedefaultendpunct}{\mcitedefaultseppunct}\relax
\EndOfBibitem
\bibitem[sof()]{software:Morfeus}
https://kjelljorner.github.io/morfeus\relax
\mciteBstWouldAddEndPuncttrue
\mciteSetBstMidEndSepPunct{\mcitedefaultmidpunct}
{\mcitedefaultendpunct}{\mcitedefaultseppunct}\relax
\EndOfBibitem
\bibitem[Halgren(1996)]{Halgren:1996_I}
Halgren,~T.~A. Merck molecular force field. I. Basis, form, scope,
  parameterization, and performance of MMFF94. \emph{J. Comput. Chem.}
  \textbf{1996}, \emph{17}, 490--519\relax
\mciteBstWouldAddEndPuncttrue
\mciteSetBstMidEndSepPunct{\mcitedefaultmidpunct}
{\mcitedefaultendpunct}{\mcitedefaultseppunct}\relax
\EndOfBibitem
\bibitem[Halgren(1996)]{Halgren:1996_II}
Halgren,~T.~A. Merck molecular force field. II. MMFF94 van der Waals and
  electrostatic parameters for intermolecular interactions. \emph{J. Comput.
  Chem.} \textbf{1996}, \emph{17}, 520--552\relax
\mciteBstWouldAddEndPuncttrue
\mciteSetBstMidEndSepPunct{\mcitedefaultmidpunct}
{\mcitedefaultendpunct}{\mcitedefaultseppunct}\relax
\EndOfBibitem
\bibitem[Halgren(1996)]{Halgren:1996_III}
Halgren,~T.~A. Merck molecular force field. III. Molecular geometries and
  vibrational frequencies for MMFF94. \emph{J. Comput. Chem.} \textbf{1996},
  \emph{17}, 553--586\relax
\mciteBstWouldAddEndPuncttrue
\mciteSetBstMidEndSepPunct{\mcitedefaultmidpunct}
{\mcitedefaultendpunct}{\mcitedefaultseppunct}\relax
\EndOfBibitem
\bibitem[Halgren and Nachbar(1996)Halgren, and
  Nachbar]{Halgren_Nachbar:1996_IV}
Halgren,~T.~A.; Nachbar,~R.~B. Merck molecular force field. IV. Conformational
  energies and geometries for MMFF94. \emph{J. Comput. Chem.} \textbf{1996},
  \emph{17}, 587--615\relax
\mciteBstWouldAddEndPuncttrue
\mciteSetBstMidEndSepPunct{\mcitedefaultmidpunct}
{\mcitedefaultendpunct}{\mcitedefaultseppunct}\relax
\EndOfBibitem
\bibitem[Halgren(1996)]{Halgren:1996_V}
Halgren,~T.~A. Merck molecular force field. V. Extension of MMFF94 using
  experimental data, additional computational data, and empirical rules.
  \emph{J. Comput. Chem.} \textbf{1996}, \emph{17}, 616--641\relax
\mciteBstWouldAddEndPuncttrue
\mciteSetBstMidEndSepPunct{\mcitedefaultmidpunct}
{\mcitedefaultendpunct}{\mcitedefaultseppunct}\relax
\EndOfBibitem
\bibitem[Halgren(1999)]{Halgren:1999_VI}
Halgren,~T.~A. MMFF VI. MMFF94s option for energy minimization studies.
  \emph{J. Comput. Chem.} \textbf{1999}, \emph{20}, 720--729\relax
\mciteBstWouldAddEndPuncttrue
\mciteSetBstMidEndSepPunct{\mcitedefaultmidpunct}
{\mcitedefaultendpunct}{\mcitedefaultseppunct}\relax
\EndOfBibitem
\bibitem[Halgren(1999)]{Halgren:1999_VII}
Halgren,~T.~A. MMFF VII. Characterization of MMFF94, MMFF94s, and other widely
  available force fields for conformational energies and for
  intermolecular-interaction energies and geometries. \emph{J. Comput. Chem.}
  \textbf{1999}, \emph{20}, 730--748\relax
\mciteBstWouldAddEndPuncttrue
\mciteSetBstMidEndSepPunct{\mcitedefaultmidpunct}
{\mcitedefaultendpunct}{\mcitedefaultseppunct}\relax
\EndOfBibitem
\bibitem[Tosco \latin{et~al.}(2014)Tosco, Stiefl, and
  Landrum]{Tosco_Landrum:2014}
Tosco,~P.; Stiefl,~N.; Landrum,~G. Bringing the MMFF force field to the RDKit:
  implementation and validation. \emph{J. Cheminform.} \textbf{2014}, \emph{6},
  37\relax
\mciteBstWouldAddEndPuncttrue
\mciteSetBstMidEndSepPunct{\mcitedefaultmidpunct}
{\mcitedefaultendpunct}{\mcitedefaultseppunct}\relax
\EndOfBibitem
\bibitem[sof()]{software:RDKit}
RDKit: Open-source cheminformatics. https://www.rdkit.org\relax
\mciteBstWouldAddEndPuncttrue
\mciteSetBstMidEndSepPunct{\mcitedefaultmidpunct}
{\mcitedefaultendpunct}{\mcitedefaultseppunct}\relax
\EndOfBibitem
\bibitem[Kim and Kim(2015)Kim, and Kim]{Kim_Kim:2015}
Kim,~Y.; Kim,~W.~Y. Universal Structure Conversion Method for Organic
  Molecules: From Atomic Connectivity to Three-Dimensional Geometry.
  \emph{BKCS} \textbf{2015}, \emph{36}, 1769--1777\relax
\mciteBstWouldAddEndPuncttrue
\mciteSetBstMidEndSepPunct{\mcitedefaultmidpunct}
{\mcitedefaultendpunct}{\mcitedefaultseppunct}\relax
\EndOfBibitem
\bibitem[sof()]{software:xyz2mol}
https://github.com/jensengroup/xyz2mol\relax
\mciteBstWouldAddEndPuncttrue
\mciteSetBstMidEndSepPunct{\mcitedefaultmidpunct}
{\mcitedefaultendpunct}{\mcitedefaultseppunct}\relax
\EndOfBibitem
\bibitem[Bannwarth \latin{et~al.}(2021)Bannwarth, Caldeweyher, Ehlert, Hansen,
  Pracht, Seibert, Spicher, and Grimme]{Bannwarth_Grimme:2021}
Bannwarth,~C.; Caldeweyher,~E.; Ehlert,~S.; Hansen,~A.; Pracht,~P.;
  Seibert,~J.; Spicher,~S.; Grimme,~S. Extended tight-binding quantum chemistry
  methods. \emph{WIREs Computational Molecular Science} \textbf{2021},
  \emph{11}, e1493\relax
\mciteBstWouldAddEndPuncttrue
\mciteSetBstMidEndSepPunct{\mcitedefaultmidpunct}
{\mcitedefaultendpunct}{\mcitedefaultseppunct}\relax
\EndOfBibitem
\bibitem[Bannwarth \latin{et~al.}(2019)Bannwarth, Ehlert, and
  Grimme]{Bannwarth_Grimme:2019}
Bannwarth,~C.; Ehlert,~S.; Grimme,~S. GFN2-xTB-An Accurate and Broadly
  Parametrized Self-Consistent Tight-Binding Quantum Chemical Method with
  Multipole Electrostatics and Density-Dependent Dispersion Contributions.
  \emph{J. Chem. Theory Comput.} \textbf{2019}, \emph{15}, 1652–1671\relax
\mciteBstWouldAddEndPuncttrue
\mciteSetBstMidEndSepPunct{\mcitedefaultmidpunct}
{\mcitedefaultendpunct}{\mcitedefaultseppunct}\relax
\EndOfBibitem
\bibitem[Ehlert \latin{et~al.}(2021)Ehlert, Stahn, Spicher, and
  Grimme]{Ehlert_Grimme:2021}
Ehlert,~S.; Stahn,~M.; Spicher,~S.; Grimme,~S. Robust and Efficient Implicit
  Solvation Model for Fast Semiempirical Methods. \emph{J. Chem. Theory
  Comput.} \textbf{2021}, \emph{17}, 4250–4261\relax
\mciteBstWouldAddEndPuncttrue
\mciteSetBstMidEndSepPunct{\mcitedefaultmidpunct}
{\mcitedefaultendpunct}{\mcitedefaultseppunct}\relax
\EndOfBibitem
\bibitem[Chung \latin{et~al.}(2020)Chung, Gillis, and Green]{Chung_Green:2020}
Chung,~Y.; Gillis,~R.~J.; Green,~W.~H. Temperature-dependent vapor–liquid
  equilibria and solvation free energy estimation from minimal data.
  \emph{AIChE J.} \textbf{2020}, \emph{66}, e16976\relax
\mciteBstWouldAddEndPuncttrue
\mciteSetBstMidEndSepPunct{\mcitedefaultmidpunct}
{\mcitedefaultendpunct}{\mcitedefaultseppunct}\relax
\EndOfBibitem
\end{mcitethebibliography}

\end{document}


\maketitle

\section{Details of elementary mutation and crossover move implementations}
\label{section:implementation_details}

\begin{figure}
          \centering           
\begin{tikzpicture}
        \node[anchor=south west,inner sep=0] (image) at (0,0.25) {\includegraphics[width=0.22\textwidth]{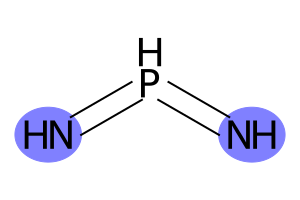}};
        \node[anchor=south west,inner sep=0] (image) at (4.8,0.25) {\includegraphics[width=0.22\textwidth]{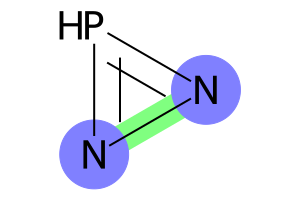}};
        \mutarr{3.5,1.6}{5.0,1.6};
        \node[anchor=south west,inner sep=0] (image) at (0,-3.75) {\includegraphics[width=0.22\textwidth]{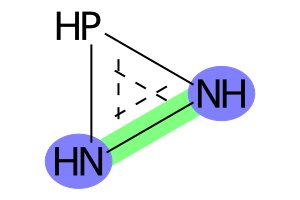}};
        \node[anchor=south west,inner sep=0] (image) at (4.8,-3.75) {\includegraphics[width=0.22\textwidth]{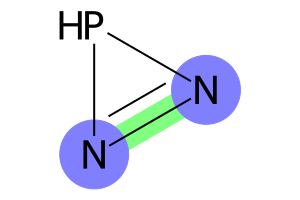}};
        \mutarr{5.0,-2.3}{3.5,-2.3};
        \resstructarroneway{6.6,0.6}{6.6,-1.2};
        \resstructarroneway{1.8,-1.2}{1.8,0.6};
        \draw (1.8,-0.3) node[cross out, draw=black, fill=none, line width=0.5mm, minimum size=5mm] {};
        \node at (0.4,2.5) {\textbf{A}};
        \node at (5.2,2.5) {\textbf{B}};
        \node at (5.2,-1.5) {\textbf{C}};
        \node at (0.4,-1.5) {\textbf{D}};
\end{tikzpicture}
          \caption{Example of an elementary change that is irreversible due to our definition of valid valence of an atom. Applying elementary change \elementarychange{2} to chemical graph \textbf{A} yields chemical graph \textbf{B}, which is actually an incorrectly written chemical graph \textbf{C}. Applying the inverse of \textbf{A}$\rightarrow$\textbf{B} to \textbf{C} yields \textbf{D}, which differs from \textbf{A}.}
     \label{fig:valence_invalid_move} 
 \end{figure}

While in Subsec.~2.3 of the main text it was convenient to speak about invertible elementary mutations, the corresponding procedures were implemented in terms of \emph{elementary changes} listed in the left column of Table~\ref{tab:random_change_choice}, which differ from elementary mutations in treating procedures for creating and destroying nodes separately rather than part of unified mutations changing the number of nodes in a chemical graph. To apply a random mutation to a chemical graph we enumerate all sets of possible elementary change parameters, which is also when we observe constraints on allowed types of covalent bonds and number of nodes. We then randomly choose change parameters among those enumerated, by first choosing the elementary change, and then all other parameters in the order listed in the right column of Table~\ref{tab:random_change_choice}. Lastly, we check that straightforward application of the change yields a molecule without nodes with invalid valences and if that is not the case reject the move automatically. The last step's necessity is illustrated with an example in Figure~\ref{fig:valence_invalid_move}.

In Table~\ref{tab:random_change_choice}, whenever we mention choosing a node or a pair of nodes we mean choosing them among a list of all nodes or pairs of nodes such that none of them are equivalent to any other in the chemical graph. This prevents MOSAiCS from considering several mutations that yield the same chemical graph $X_{2}$ given an initial chemical graph $X_{1}$, making it straightforward to define the unique inverse change parameters that yield $X_{1}$ when applied to $X_{2}$. Enumerating all possible changes for $X_{2}$ analogously to $X_{1}$ allows straightforward calculation of the proposition probability ratio in acceptance probability expression [Eq.~(4) in the main text]. Choosing resonance structure for \elementarychange{2} and \elementarychange{6} is necessary in situations when decreasing bond order, depending on the resonance structure, can yield two distinct molecules corresponding to cases when the changed covalent bond is conserved or broken; such situations are illustrated in Figure~\ref{fig:res_struct_dep_examples}.

 \begin{table}
 \centering

\begin{tabular}{p{0.25\textwidth}|p{0.7\textwidth}}
\toprule
\vspace{-2.0ex}\begin{enumerate}[noitemsep,topsep=0pt,labelwidth=0pt,leftmargin=0pt]\item[] Elementary change \end{enumerate}

\phantom{0}\vspace{-3.0ex}  & \vspace{-2.0ex}\begin{enumerate}[noitemsep,topsep=0pt]\item[] Choice order \end{enumerate}

\phantom{0}\vspace{-3.0ex} \\
 \midrule \vspace{-2.0ex}\begin{enumerate}[noitemsep,topsep=0pt,labelwidth=0pt,leftmargin=0pt] \item[] add node (\elementarychange{1}a)\end{enumerate}

\phantom{0}\vspace{-3.0ex} & \vspace{-2.0ex}\begin{enumerate}[noitemsep,topsep=0pt]\item added node's heavy atom element \item node to which the new node will be connected with a covalent bond \item order of the new covalent bond \end{enumerate}

\phantom{0}\vspace{-3.0ex}\\
 \midrule \vspace{-2.0ex}\begin{enumerate}[noitemsep,topsep=0pt,labelwidth=0pt,leftmargin=0pt] \item[] remove node (\elementarychange{1}b)\end{enumerate}

\phantom{0}\vspace{-3.0ex} & \vspace{-2.0ex}\begin{enumerate}[noitemsep,topsep=0pt]\item removed node's heavy atom element \item removed node \end{enumerate}

\phantom{0}\vspace{-3.0ex}\\
 \midrule \vspace{-2.0ex}\begin{enumerate}[noitemsep,topsep=0pt,labelwidth=0pt,leftmargin=0pt] \item[] change bond order (\elementarychange{2})\end{enumerate}

\phantom{0}\vspace{-3.0ex} & \vspace{-2.0ex}\begin{enumerate}[noitemsep,topsep=0pt]\item step by which bond order is changed \item altered pair of nodes \item resonance structure \end{enumerate}

\phantom{0}\vspace{-3.0ex}\\
 \midrule \vspace{-2.0ex}\begin{enumerate}[noitemsep,topsep=0pt,labelwidth=0pt,leftmargin=0pt] \item[] replace heavy atom (\elementarychange{3})\end{enumerate}

\phantom{0}\vspace{-3.0ex} & \vspace{-2.0ex}\begin{enumerate}[noitemsep,topsep=0pt]\item node's new heavy atom element \item changed node \end{enumerate}

\phantom{0}\vspace{-3.0ex}\\
 \midrule \vspace{-2.0ex}\begin{enumerate}[noitemsep,topsep=0pt,labelwidth=0pt,leftmargin=0pt] \item[] change valence / change hydrogen number (\elementarychange{4})\end{enumerate}

\phantom{0}\vspace{-3.0ex} & \vspace{-2.0ex}\begin{enumerate}[noitemsep,topsep=0pt]\item node whose valence is changed \item new number of hydrogens connected to the node \end{enumerate}

\phantom{0}\vspace{-3.0ex}\\
 \midrule \vspace{-2.0ex}\begin{enumerate}[noitemsep,topsep=0pt,labelwidth=0pt,leftmargin=0pt] \item[] change valence / add heavy atoms (\elementarychange{5}a)\end{enumerate}

\phantom{0}\vspace{-3.0ex} & \vspace{-2.0ex}\begin{enumerate}[noitemsep,topsep=0pt]\item created nodes' heavy atom element \item node to which the created nodes will be connected via covalent bonds \item order of the new covalent bonds (also automatically defines the number of added nodes) \end{enumerate}

\phantom{0}\vspace{-3.0ex}\\
 \midrule \vspace{-2.0ex}\begin{enumerate}[noitemsep,topsep=0pt,labelwidth=0pt,leftmargin=0pt] \item[] change valence / remove heavy atoms (\elementarychange{5}b)\end{enumerate}

\phantom{0}\vspace{-3.0ex} & \vspace{-2.0ex}\begin{enumerate}[noitemsep,topsep=0pt]\item removed nodes' heavy atom element \item node whose neighbors will be removed \item order of covalent bonds connecting removed nodes to the molecule (also automatically defines the number of removed nodes) \end{enumerate}

\phantom{0}\vspace{-3.0ex}\\
 \midrule \vspace{-2.0ex}\begin{enumerate}[noitemsep,topsep=0pt,labelwidth=0pt,leftmargin=0pt] \item[] change valence / change bond order (\elementarychange{6})\end{enumerate}

\phantom{0}\vspace{-3.0ex} & \vspace{-2.0ex}\begin{enumerate}[noitemsep,topsep=0pt]\item step by which bond order is changed \item altered pair of nodes \item resonance structure \end{enumerate}

\phantom{0}\vspace{-3.0ex}\\
 \bottomrule \end{tabular} 
\caption{Elementary changes used in this work along with the corresponding order in which random change parameters were chosen.}
\label{tab:random_change_choice}
\end{table}

 \begin{figure}
          \centering           

\begin{tabularx}{\textwidth}{cc}

\mutationexampleblock{1_28}{(\elementarychange{2})}
&
\mutationexampleblock{1_29}{}
\\
\mutationexampleblock{5_28}{(\elementarychange{6})}
&
\mutationexampleblock{5_29}{}

\end{tabularx}
          \caption{Examples of \elementarychange{2} and \elementarychange{6} elementary changes that decrease order of the covalent bond between the same pairs of nodes by the same values, but yield different results depending on the resonance structure in which the molecules are initialized.}
     \label{fig:res_struct_dep_examples} 
 \end{figure}

\begin{figure}
          \centering           
\begin{tabularx}{\textwidth}{c}
\begin{tikzpicture}
        \node[anchor=south west,inner sep=0] (image) at (0.5,0) {\includegraphics[width=0.2\textwidth]{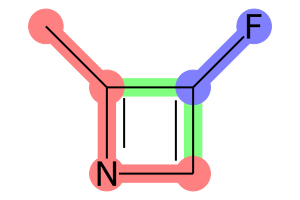}};
        \node[anchor=south west,inner sep=0] (image) at (4.0,0) {\includegraphics[width=0.2\textwidth]{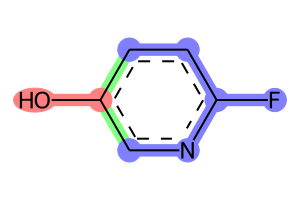}};
        \node[anchor=south west,inner sep=0] (image) at (9,0) {\includegraphics[width=0.2\textwidth]{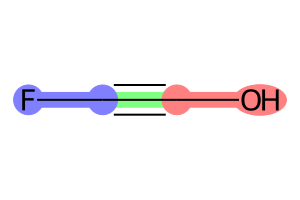}};
        \node[anchor=south west,inner sep=0] (image) at (12.5,0) {\includegraphics[width=0.2\textwidth]{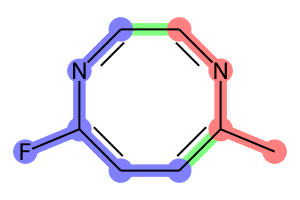}};
        \mutarr{7.5,1.1}{9.0,1.1};
        \node at (3.85,1.1) {\Large{\textbf{+}}};
        \node at (12.45,1.1) {\Large{\textbf{+}}};
\end{tikzpicture}
\end{tabularx}
          \caption{Example of how not checking conservation of number of green bonds during a crossover move results in a non-invertible procedure.}
     \label{fig:irrev_cross_coupling_example} 
 \end{figure}

Choosing the parameters of a crossover move starts with assigning the ``blue fragments'' (as illustrated in Figure~4) which are defined as the set of nodes separated by a given number of covalent bonds from an ``origin'' node. Origin nodes in the two molecules are chosen randomly; then we enumerate all possible blue fragments such that:
\begin{itemize}
\item a pair of resonance structures chosen for both molecules allows a one-on-one mapping between green (see Figure~4) covalent bonds of the same order;
\item exchanging the blue fragments yields chemical graphs satisfying constraints on the number of nodes in a chemical graph;
\item at least one blue fragment and at least one red (see Figure~4) fragment contain more than one node, ensuring that the crossover move is not a simple combination of two \elementarychange{3} elementary changes;
\item the number of green bonds is not larger than a certain threshold (set 3 in this work); this is done to limit the number of chemical graph pairs enumerated during the next step.
\end{itemize}
With the blue fragments chosen we enumerate all pairs of chemical graphs obtained by exchanging them, \emph{i.e.} taking pairs of green bonds from molecules and exchanging nodes between them, thus connecting each molecule's blue fragment to the red fragment of the other molecule. For each thus generated pair we additionally check that the valences in the exchanged blue and red fragments are valid in the new molecules (to prevent situations analogous to the one illustrated in Figure~\ref{fig:valence_invalid_move}), that none of the newly created green bonds connect the same pair of nodes (to prevent irreversible changes illustrated in Figure~\ref{fig:irrev_cross_coupling_example}), and that none of the newly created bonds violate constraints on which types of heavy atoms can be covalently connected.

Of the resulting pairs of chemical graphs one pair, consisting of chemical graphs $\tilde{X}^{\prime}$ and $\tilde{X}^{\prime\prime}$, is randomly chosen. The chemical graphs are assigned to replicas $i^{\prime}$ and $i^{\prime\prime}$ according to the value of sampled probability $P$ [Eq.~(1) in the main text] corresponding to the ordering. In other words, if both replicas $i^{\prime}$ and $i^{\prime\prime}$ are greedy the assignment is random; if exactly one is greedy it is assigned the chemical graph corresponding to the smaller value of $F$; if both are exploration replicas the probability of assigning $\tilde{X}^{\prime}$ and $\tilde{X}^{\prime\prime}$ to replicas $i^{\prime}$ and $i^{\prime\prime}$ is proportional to
\begin{equation}
    P_{\mathrm{order}}(\tilde{X}^{\prime},
    \tilde{X}^{\prime\prime},i^{\prime},i^{\prime\prime})=\exp\left\{-\beta^{(i^{\prime})}\left[F(\tilde{X}^{\prime})+\Vb^{(i^{\prime})}(\tilde{X}^{\prime})\right]
    -\beta^{(i^{\prime\prime})}\left[F(\tilde{X}^{\prime\prime})+\Vb^{(i^{\prime\prime})}(\tilde{X}^{\prime\prime})\right]
    \right\}.\label{eq:trial_ordering_probability}
\end{equation}
Note that if $i^{\prime}$ and $i^{\prime\prime}$ are a greedy and an exploration replicas and $F(X^{(i^{\prime})})>F(X^{(i^{\prime\prime})})$ then the crossover move has no inverse; we avoided such situations by preceding and following a crossover move with attempting a tempering swap move on replicas $i^{\prime}$ and $i^{\prime\prime}$. The inverse crossover parameters required to calculate acceptance probability $\Pacc$ [Eq.~(4) in the main text] correspond to the same blue fragments being exchanged.

\section{Experimental details}

\subsection{Estimating quantities of interest}
\label{subsec:quant_eval}

To estimate free energy of solvation $\dEsolv$, dipole $\dipole$, and HOMO-LUMO gap $\gap$ we used the recipe from Ref.~\citenum{Ebejer_Deane:2012}, as implemented in the Morfeus package,\cite{software:Morfeus} to generate the molecule's conformers and Boltzmann weights $\wb^{(k)}$ (where $k$ is conformer index) at $T=298.15\,\mathrm{K}$ with the MMFF94 forcefield\cite{Halgren:1996_I,*Halgren:1996_II,*Halgren:1996_III,*Halgren_Nachbar:1996_IV,*Halgren:1996_V,*Halgren:1999_VI,*Halgren:1999_VII} as implemented\cite{Tosco_Landrum:2014} in RDKit.\cite{software:RDKit} If converting the lower energy conformer's geometry back to a chemical graph using the xyz2mol code\cite{Kim_Kim:2015,software:xyz2mol} proved impossible or yielded a chemical graph different from the initial one we considered conformer generation to have failed. To decrease the number of conformers considered, we introduced ``cut'' Boltzmann weights $\wbc^{(k)}$ obtained by solving w.r.t. $\wbc^{(k)}$ and $\rho_{-}$
    \begin{align}
        \forall k : \wbc^{(k)}&=\max\left(\wb^{(k)}-\rho_{-}, 0\right),\\
        \frac{\sum_{k}\wbc^{(k)}}{\sum_{k}\wb^{(k)}}&=1-\rcut,
    \end{align}
where $\rcut$ is a user-defined parameter. Note that $\wbc^{(k)}$ is exactly zero for higher-energy conformers of the molecule while being a smooth function of $\wb^{(k)}$. Then for each conformer with non-zero $\wbc^{(k)}$ we used xtb code\cite{Bannwarth_Grimme:2021} to run GFN2-xTB\cite{Bannwarth_Grimme:2019} calculations with analytical linearized Poisson-Boltzmann (ALPB) model\cite{Ehlert_Grimme:2021} used to simulate presence of water. For $\dipole$ and $\gap$ we took weighted averages of the resulting dipole and HOMO-LUMO gap estimates with $\wbc^{(k)}$ as weights; for $\dEsolv$ the averaged quantity was the difference between GFN2-xTB electronic energies obtained with and without ALPB. Since there is a randomness to generating conformers with Morfeus, the procedure was repeated $\Nrepeat$ times with the mean taken as final result and the standard deviation used to estimate the latter's root mean square error (RMSE). A molecule was considered invalid if one of the $\Nrepeat$ generations of MMFF94 coordinates failed or if one of the GFN2-xTB calculations done for the molecule did not converge. This \emph{a priori} limited us to molecules that are not geometrically strained, but we did not consider this restriction important for our end goal applications since geometrically strained molecules are in general less thermodynamically stable, making them less desirable as battery electrolyte components. Lastly, we had to restart some EGP* simulations due to our code's failure to handle some xtb code exceptions. Since we encountered such exceptions during less than one in $10^{7}$ of evaluation of EGP* molecules this issue had a negligible impact on the results presented in this work.

The two different sets of calculation parameters mentioned in Subsec.~2.4 of the main text are $\Nmorfconf=32$ and $\Nrepeat=16$ for ``converged,'' with $\Nmorfconf=8$ and $\Nrepeat=1$ for  ``cheap''; $\rcut=0.1$ was chosen for both cases. Quantity estimates obtained with the former exhibited low RMSE's over QM9 and EGP datasets, but sometimes required relatively large computational time to be calculated, hence the cheap set of parameters were used during our simulations.

Properties of $\dEsolv$, $\dipole$, and $\gap$ over molecules in intersections of QM9 with QM9* and EGP with EGP* satisfying $\gap$ constraints are summarized in Table~\ref{tab:dataset_reference}. Note that RMSE's of properties of interest are reasonable over all such molecules. The table also presents reference values used to calculate relative improvements presented in the main text for candidate molecules.

 \begin{table}
 \centering
\begin{tabular}{lll}
\toprule
              $\gap$ constr. &                         $\phantom{-(}$weak &                    $\phantom{-(}$strong \\
\midrule
\multicolumn{3}{c}{QM9 and QM9* intersection}\\\midrule tot. num. mol. & $\phantom{-(}89015$ & $ \phantom{-(}\_$\\                   num. mol. &                        $\phantom{-(}$67826 &                     $\phantom{-(}$20325 \\
      min. $\dEsolv$, kJ/mol &           $(-9.483 \pm 0.007)\cdot 10^{1}$ &        $(-5.636 \pm 0.037)\cdot 10^{1}$ \\
       min. $\dEsolv$ SMILES &           $\phantom{-(}$NC1=NC(=O)N=C(N)N1 &           $\phantom{-(}$NC(=O)CCNC(N)=O \\
max. RMSE($\dEsolv$), kJ/mol &                        $\phantom{-(}1.807$ &        $\phantom{-(}6.589\cdot 10^{-1}$ \\
 max. RMSE($\dEsolv$) SMILES &           $\phantom{-(}$N=C1NC(N)=C(C=O)O1 &             $\phantom{-(}$C1C2OC1C21CC1 \\
       STD $\dEsolv$, kJ/mol &                        $\phantom{-(}9.977$ &                     $\phantom{-(}7.100$ \\
       max. $\dipole$, debye & $\phantom{-}(1.338 \pm 0.000)\cdot 10^{1}$ & $\phantom{-}\phantom{(}8.532 \pm 0.041$ \\
       max. $\dipole$ SMILES &             $\phantom{-(}$CN1C=NC(=O)N=C1N &           $\phantom{-(}$O=C1NCCNC(=O)N1 \\
 max. RMSE($\dipole$), debye &           $\phantom{-(}9.450\cdot 10^{-1}$ &        $\phantom{-(}3.325\cdot 10^{-1}$ \\
 max. RMSE($\dipole$) SMILES &               $\phantom{-(}$N=C1OCC=C1NC=O &         $\phantom{-(}$C1C2CN3C1C1OC1C23 \\
        STD $\dipole$, debye &                        $\phantom{-(}1.831$ &                     $\phantom{-(}1.355$ \\
       max. RMSE($\gap$), eV &                        $\phantom{-(}1.067$ &                     $\phantom{-(}1.067$ \\
    max. RMSE($\gap$) SMILES &              $\phantom{-(}$CCC1CC12C1CC2C1 &           $\phantom{-(}$CCC1CC12C1CC2C1 \\
\midrule\multicolumn{3}{c}{EGP and EGP* intersection}\\\midrule tot. num. mol. & $\phantom{-(}9437$ & $ \phantom{-(}\_$\\                   num. mol. &                         $\phantom{-(}$6190 &                       $\phantom{-(}$863 \\
      min. $\dEsolv$, kJ/mol &           $(-9.474 \pm 0.006)\cdot 10^{1}$ &        $(-7.814 \pm 0.019)\cdot 10^{1}$ \\
       min. $\dEsolv$ SMILES &           $\phantom{-(}$NC1=NC(=O)N=C(N)N1 & $\phantom{-(}$NS(=O)(=O)CCCCCS(N)(=O)=O \\
max. RMSE($\dEsolv$), kJ/mol &           $\phantom{-(}6.960\cdot 10^{-1}$ &        $\phantom{-(}3.500\cdot 10^{-1}$ \\
 max. RMSE($\dEsolv$) SMILES &             $\phantom{-(}$N=C1NC=CN1C(N)=O &               $\phantom{-(}$ClCC1CCCCO1 \\
       STD $\dEsolv$, kJ/mol &                        $\phantom{-(}9.320$ &                     $\phantom{-(}8.267$ \\
       max. $\dipole$, debye & $\phantom{-}(1.326 \pm 0.000)\cdot 10^{1}$ & $\phantom{-}\phantom{(}8.050 \pm 0.048$ \\
       max. $\dipole$ SMILES &           $\phantom{-(}$NC1=NC(=O)N=C(N)N1 &    $\phantom{-(}$O=S(=O)(F)CCS(=O)(=O)F \\
 max. RMSE($\dipole$), debye &           $\phantom{-(}6.943\cdot 10^{-1}$ &        $\phantom{-(}3.179\cdot 10^{-1}$ \\
 max. RMSE($\dipole$) SMILES &             $\phantom{-(}$N=CNC1=NC=C(N)O1 &        $\phantom{-(}$N\#CC1CCC(C\#N)CC1 \\
        STD $\dipole$, debye &                        $\phantom{-(}1.881$ &                     $\phantom{-(}1.492$ \\
       max. RMSE($\gap$), eV &           $\phantom{-(}2.916\cdot 10^{-1}$ &        $\phantom{-(}2.916\cdot 10^{-1}$ \\
    max. RMSE($\gap$) SMILES &                      $\phantom{-(}$C=C1CC1 &                   $\phantom{-(}$C=C1CC1 \\
\bottomrule
\end{tabular}

\caption{Properties of sets of molecules at the intersection of QM9 and QM9* or EGP and EGP* that satisfy weak or strong constraint on HOMO-LUMO gap $\gap$, namely their number (num. mol.), standard deviations (STD) of optimized quantities (free energy of solvation $\dEsolv$ and dipole $\dipole$) over these molecules, the maximum root mean square errors (RMSEs) observed $\dEsolv$, $\dipole$, and $\gap$, minimum of $\dEsolv$ and maximum of $\dipole$, along with SMILES of the molecules for which these extrema values were observed. The total number of molecules (tot. num. mol.) is the number of molecules whose chemical graphs, if extractable from QM9 or EGP coordinates with xyz2mol code, satisfied constraints of QM9* or EGP* sets.}
\label{tab:dataset_reference}
\end{table}

\subsection{Accessibility of chemical space}

\begin{figure}
          \centering           
          \includegraphics[width=0.3\linewidth]{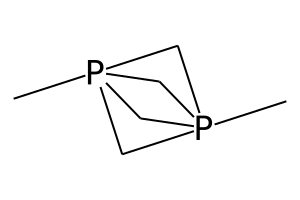}
          \caption{An EGP* molecule that cannot be transformed to or from methane through EGP* molecules using elementary mutations alone.}
     \label{fig:connectivity_counterexample} 
 \end{figure}

To guarantee (at least for $\biasprop=0$ when the sampling is Markovian) that MOSAiCS eventually finds the minimum of loss function over the set of chemical graphs of interest the procedures used to propose trial replica configurations should have \emph{connectivity}, \emph{i.e.} it should be possible to morph each pair of members of the set from one into another using these procedures alone and with only other molecules in this set as intermediates. Mutations \mutationlabel{1}-\mutationlabel{4} are enough to guarantee connectivity in a large variety of relevant sets of chemical graphs, for example those obeying an upper bound on the number of heavy atoms and a constraint on which types of non-carbon heavy atoms can share a covalent bond. Connectivity of this kind of sets is easily seen by ``constructing'' and ``deconstructing'' each set member to and from methane. The same argument holds for QM9* but, unfortunately, not for EGP* due to P and S atoms being forbidden to share a covalent bond with a hydrogen atom, making it impossible to use mutation \mutationlabel{4}. Introducing additional mutations \mutationlabel{5} and \mutationlabel{6} alleviates the problem only partially, with an example of an EGP* molecule that cannot be transformed to and from methane using elementary mutations presented in Figure~\ref{fig:connectivity_counterexample}. This molecule can be considered by the simulations used in this work thanks to crossover moves, but we cannot guarantee this is the case for all such examples. However, we note that all such molecules are geometrically strained, which makes them not interesting for our applications due to decreased thermodynamic stability.

\section{Additional data obtained during optimization in QM9* and EGP*}
\label{sec:add_data}

\begin{figure}
          \centering           
\hspace{-2ex}
\begin{tikzpicture}
        \node[anchor=south west,inner sep=0] (image) at (0,0) {\includegraphics[width=0.15\textwidth]{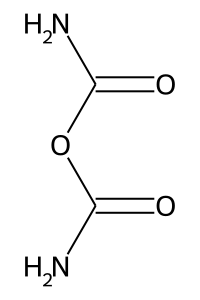}};
        \node[anchor=south west,inner sep=0] (image) at (2.75,0) {\includegraphics[width=0.15\textwidth]{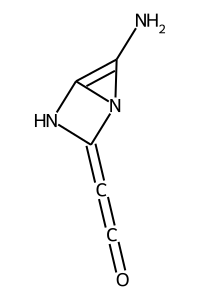}};
        \node[anchor=south west,inner sep=0] (image) at (5.5,0) {\includegraphics[width=0.15\textwidth]{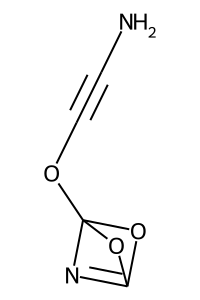}};
        \node[anchor=south west,inner sep=0] (image) at (0,-4) {\includegraphics[width=0.15\textwidth]{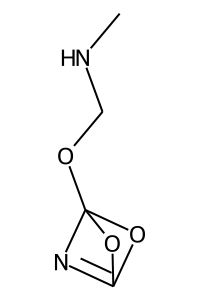}};
        \node[anchor=south west,inner sep=0] (image) at (2.75,-4) {\includegraphics[width=0.15\textwidth]{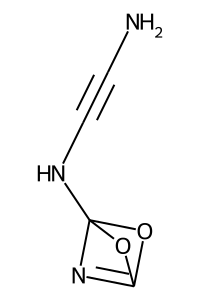}};
        \node[anchor=south west,inner sep=0] (image) at (5.5,-4) {\includegraphics[width=0.15\textwidth]{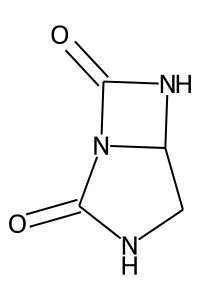}};
        \node at (0.5,2.8) {\candidate{3}{QM9}};
        \node at (3.3,3.3) {\candidate{5}{QM9}};
        \node at (6.3,3.3) {\candidate{7}{QM9}};
        \node at (0.5,-0.66) {\candidate{8}{QM9}};
        \node at (3.4,-0.8) {\candidate{9}{QM9}};
        \node at (6,-1.4) {\candidate{10}{QM9}};
\end{tikzpicture}
          \caption{QM9* candidates that did not exhibit best target property values.}
     \label{fig:qm9_runner_ups} 
 \end{figure}

\begin{figure}
          \centering           
          \includegraphics[width=1.0\linewidth]{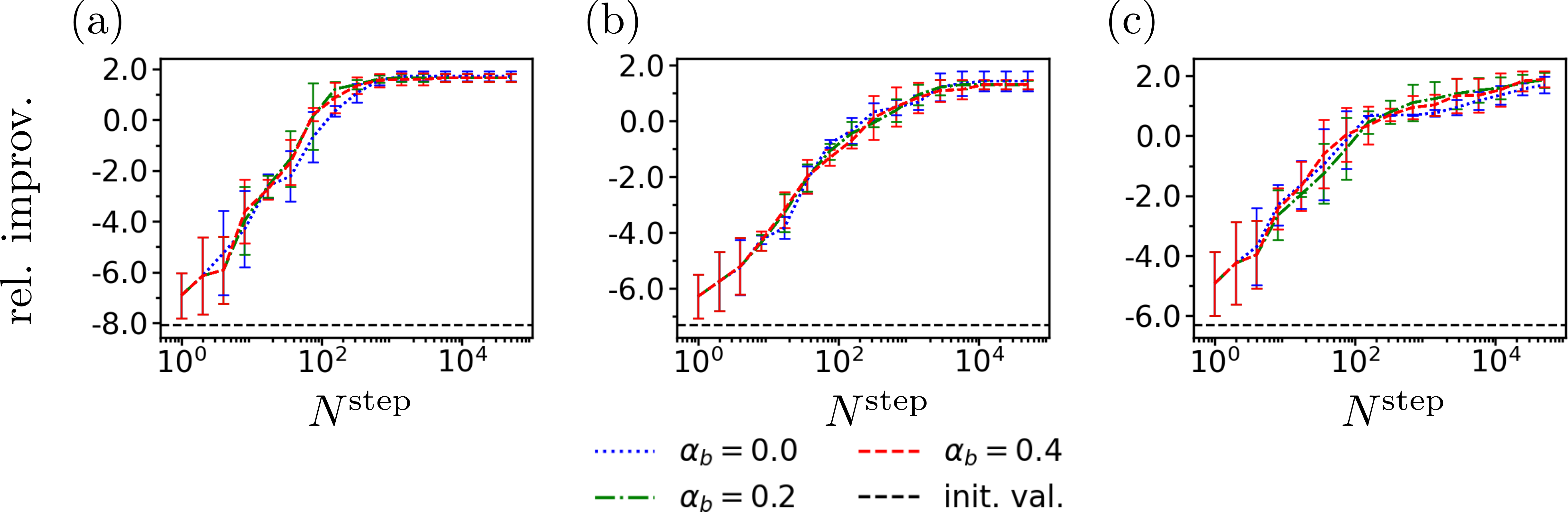}
          \caption{Relative improvement (as defined in Sec.~3 of the main text and estimated from $\dEsolv^{\mathrm{cheap}}$ or $\dipole^{\mathrm{cheap}}$) observed at $N^{\mathrm{step}}$ global Monte Carlo steps for QM9* simulations optimizing (a) $\dipole$ with weak $\gap$ constraint, (b) $\dEsolv$ with strong $\gap$ constraint, and (c) $\dipole$ with strong $\gap$ constraint. For each value of the biasing potential $\biasprop$ we plot the mean value over different random generator seeds, the error bars corresponding to standard deviation. "init. val." is the relative improvement of the simulations' starting molecule, \emph{i.e.} methane.}
     \label{fig:qm9_SI_opt_logs} 
 \end{figure}

\begin{figure}
          \centering           
          \includegraphics[width=1.0\linewidth]{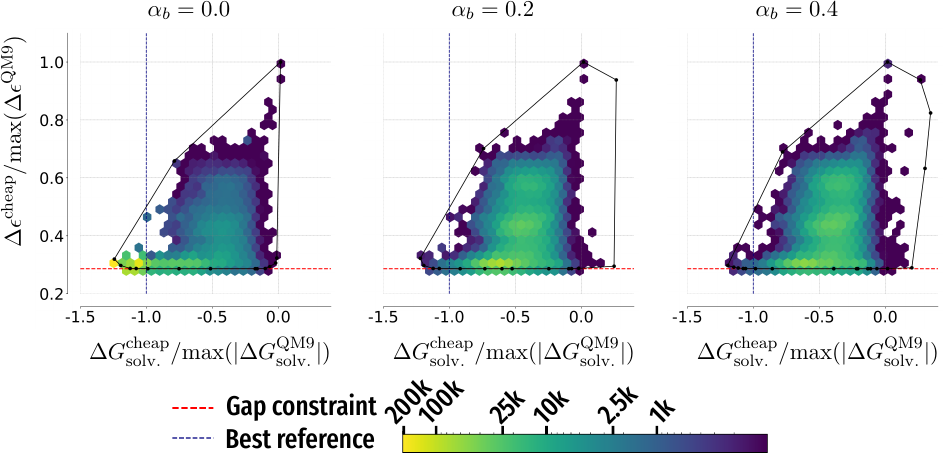}
          \caption{Densities of molecules encountered by example simulations minimizing free energy of solvation $\dEsolv$ with strong HOMO-LUMO gap $\gap$ constraint in QM9* that produced the candidate with smallest $\dEsolv$ for a given value of bias proportionality coefficient $\biasprop$.}
     \label{fig:qm9_pareto_front_solvation_strong} 
 \end{figure}

\begin{figure}
          \centering           
          \includegraphics[width=1.0\linewidth]{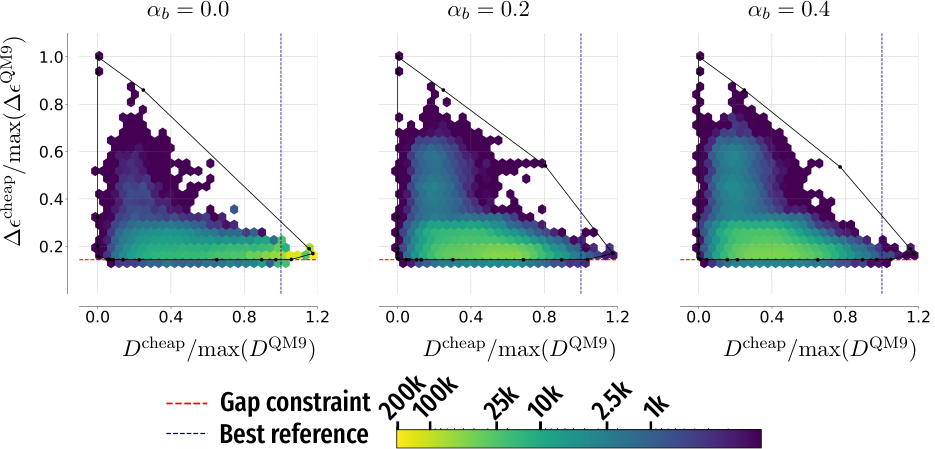}
          \caption{Densities of molecules encountered by example simulations maximizing dipole $\dipole$ with weak HOMO-LUMO gap $\gap$ constraint in QM9* that produced the candidate with largest $\dipole$ for a given value of bias proportionality coefficient $\biasprop$.}
     \label{fig:qm9_pareto_front_dipole_weak} 
 \end{figure}

\begin{figure}
          \centering           
          \includegraphics[width=1.0\linewidth]{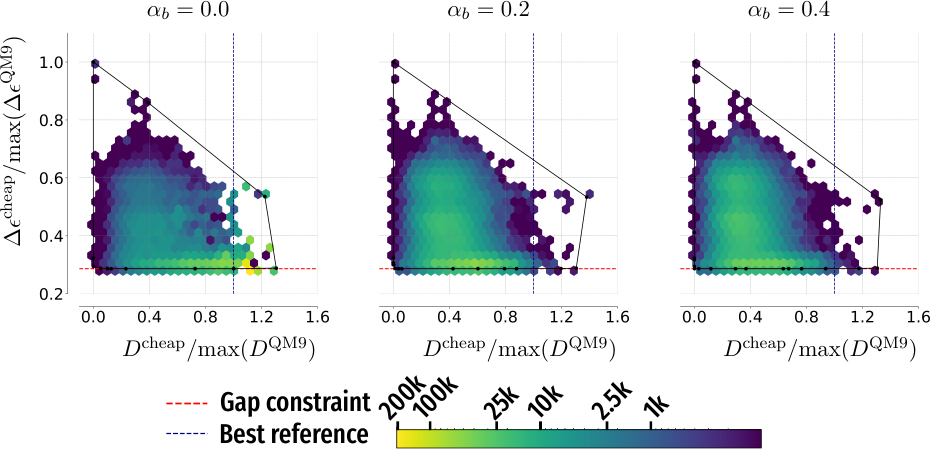}
          \caption{Densities of molecules encountered by example simulations maximizing dipole $\dipole$ with strong HOMO-LUMO gap $\gap$ constraint in QM9* that produced the candidate with largest $\dipole$ for a given value of bias proportionality coefficient $\biasprop$.}
     \label{fig:qm9_pareto_front_dipole_strong} 
 \end{figure}

Full information about QM9* candidates, including the corresponding values of optimized quantities and how often they were proposed by our simulations, is presented in Table~\ref{tab:qm9_candidates}; we see that for all optimization problems but for maximizing $\dipole$ with strong $\gap$ constraint the agreement between simulations in terms of proposed candidates is close to unanimous. Figure~\ref{fig:qm9_runner_ups} presents QM9* candidates that were not plotted in Figure~5; while we once again see MOSAiCS' ability to generate unconventional molecular structures, candidates \candidate{7}{QM9}, \candidate{8}{QM9}, and \candidate{9}{QM9} demonstrate that our criterion for considering a molecule stable was not always realistic. To compare how fast MOSAiCS found candidates for different optimization problems and $\biasprop$ values, we also calculated numbers of chemical graphs considered during a simulation before the candidate was encountered $\tpreq$, total number of chemical graphs considered during a simulation $\tottp$, and the number of global steps (as defined in Subsec.~2.5 of the main text) taken by a simulation before the candidate was encountered $\beststepfound$. For QM9* simulations mean and standard deviation values of these quantities are summarized in Table~\ref{tab:qm9_step_comp}. Changing the bias proportionality coefficient $\biasprop$ did not significantly affect the number of global steps required on average to find a candidate molecule, but did lead to a more thorough (but also costly) exploration of chemical space, as demonstrated by significant increase of $\overline{\tpreq}$ and $\overline{\tottp}$. Another interesting observation is that for a given $\gap$ constraint $\beststepfound$ values are significantly higher for optimization of $\dipole$ than optimization of $\dEsolv$. This is likely due to $\dEsolv$ being easier to optimize since free energy of solvation can be accurately decomposed as the sum of scalar contributions from different functional groups in the molecule,\cite{Chung_Green:2020} while for the dipole moment such contributions would take vector form whose sum would depend on the groups' placement in Cartesian space. Also note that for all optimization problems but for maximizing $\dipole$ with a strong $\gap$ constraint the $\beststepfound$ was typically significantly smaller than the simulation length, which is an indicator of the simulations being converged in these cases. Relative improvement progress plots for optimization problems not shown in the main text are presented in Figure~\ref{fig:qm9_SI_opt_logs}; while they look largely analogous to Figure~6, our simulations look less converged for minimization under strong $\gap$ constraint than under weak $\gap$ constraint. Plots of densities of QM9* molecules analogous to Figure~7 are found in Figures~\ref{fig:qm9_pareto_front_solvation_strong}-\ref{fig:qm9_pareto_front_dipole_strong}; analogously to Figure~7 we observe an increase in diversity of considered molecules largely achieved by increasing exploration of chemical space regions less valuable in terms of properties of interest.

\begin{table}
 \centering
\begin{tabular}{llllll}
\toprule
 \multirow{2}{*}{molecule} & \multirow{2}{*}{SMILES} & \multirow{2}{*}{$\phantom{-(}$opt. quant.} & \multicolumn{3}{l}{prop. with $\biasprop$} \\
 \cline{4-6}\phantom{1} & \phantom{1} & \phantom{1} & 0.0 & 0.2 & 0.4 \\
\midrule
 \multicolumn{6}{c}{min. $\dEsolv$ (weak $\gap$ constraint)}           \\ \midrule
 \candidate{1}{QM9} & NC1=NC(=O)N=C(N)N1 & $(-9.479 \pm 0.006)\cdot 10^{1}$ & $8$ & $8$ & $8$ \\
\midrule\multicolumn{6}{c}{min. $\dEsolv$ (strong $\gap$ constraint)}           \\ \midrule
 \candidate{2}{QM9} & NC(=O)NCNC(N)=O & $(-6.819 \pm 0.060)\cdot 10^{1}$ & $7$ & $8$ & $8$ \\
 \candidate{3}{QM9} & NC(=O)OC(N)=O & $(-6.727 \pm 0.000)\cdot 10^{1}$ & $1$ & $0$ & $0$ \\
 \midrule\multicolumn{6}{c}{max. $\dipole$ (weak $\gap$ constraint)}           \\ \midrule
 \candidate{4}{QM9} & NC1=C2OC(=CC=O)N12 & $\phantom{-}(1.573 \pm 0.010)\cdot 10^{1}$ & $8$ & $7$ & $8$ \\
 \candidate{5}{QM9} & NC1=C2NC(=C=C=O)N12 & $\phantom{-}(1.523 \pm 0.000)\cdot 10^{1}$ & $0$ & $1$ & $0$ \\
\midrule\multicolumn{6}{c}{max. $\dipole$ (strong $\gap$ constraint)}           \\ \midrule
 \candidate{6}{QM9} & FC(F)=C=C=C1OCO1 & $\phantom{-}(1.114 \pm 0.000)\cdot 10^{1}$ & $1$ & $4$ & $1$ \\
 \candidate{7}{QM9} & NC\#COC12N=C(O1)O2 & $\phantom{-}(1.083 \pm 0.003)\cdot 10^{1}$ & $4$ & $2$ & $2$ \\
 \candidate{8}{QM9} & CNCOC12N=C(O1)O2 & $\phantom{-}(1.051 \pm 0.020)\cdot 10^{1}$ & $0$ & $1$ & $4$ \\
 \candidate{9}{QM9} & NC\#CNC12N=C(O1)O2 & $\phantom{-}(1.036 \pm 0.007)\cdot 10^{1}$ & $1$ & $1$ & $1$ \\
 \candidate{10}{QM9} & O=C1NCC2NC(=O)N12 & $\phantom{-}(1.000 \pm 0.000)\cdot 10^{1}$ & $2$ & $0$ & $0$ \\
\bottomrule
\end{tabular}

\caption{QM9* candidates along with their corresponding optimized quantity values (opt. quant.) and the number of times the molecule was proposed as a candidate by simulations run with different values of the bias proportionality coefficient $\biasprop$ (prop. with $\biasprop$). $\dEsolv$ and $\dipole$ values are in kJ/mol and debye.}
\label{tab:qm9_candidates}
\end{table}

 \begin{table}
 \centering
\begin{tabular}{llllllll}
\toprule
                            $\gap$ constr. & \multicolumn{3}{c}{weak} & \phantom{\_} & \multicolumn{3}{c}{strong} \\
         \cline{2-4}\cline{6-8}$\biasprop$ &                 0.0 &                 0.2 &                 0.4 & \phantom{\_} &                 0.0 &                 0.2 &                 0.4 \\
\midrule
        \multicolumn{8}{c}{min. $\dEsolv$}                      \\
              \midrule $\overline{\tpreq}$ & $4.072\cdot 10^{3}$ & $5.164\cdot 10^{3}$ & $1.449\cdot 10^{4}$ & \phantom{\_} & $3.854\cdot 10^{3}$ & $8.391\cdot 10^{3}$ & $8.544\cdot 10^{3}$ \\
                          $\sigma(\tpreq)$ & $1.672\cdot 10^{3}$ & $2.979\cdot 10^{3}$ & $1.053\cdot 10^{4}$ & \phantom{\_} & $2.347\cdot 10^{3}$ & $4.339\cdot 10^{3}$ & $6.707\cdot 10^{3}$ \\
                       $\overline{\tottp}$ & $2.352\cdot 10^{5}$ & $6.229\cdot 10^{5}$ & $6.862\cdot 10^{5}$ & \phantom{\_} & $1.394\cdot 10^{5}$ & $4.304\cdot 10^{5}$ & $4.559\cdot 10^{5}$ \\
                          $\sigma(\tottp)$ & $3.447\cdot 10^{3}$ & $1.588\cdot 10^{3}$ & $2.171\cdot 10^{3}$ & \phantom{\_} & $3.188\cdot 10^{3}$ & $1.395\cdot 10^{3}$ & $1.677\cdot 10^{3}$ \\
               $\overline{\beststepfound}$ & $4.234\cdot 10^{2}$ & $3.231\cdot 10^{2}$ & $7.781\cdot 10^{2}$ & \phantom{\_} & $6.704\cdot 10^{2}$ & $5.584\cdot 10^{2}$ & $5.181\cdot 10^{2}$ \\
                  $\sigma(\beststepfound)$ & $1.786\cdot 10^{2}$ & $1.823\cdot 10^{2}$ & $5.627\cdot 10^{2}$ & \phantom{\_} & $3.729\cdot 10^{2}$ & $2.785\cdot 10^{2}$ & $4.025\cdot 10^{2}$ \\
\midrule\multicolumn{8}{c}{max. $\dipole$}                      \\
              \midrule $\overline{\tpreq}$ & $2.126\cdot 10^{4}$ & $3.131\cdot 10^{4}$ & $5.182\cdot 10^{4}$ & \phantom{\_} & $1.164\cdot 10^{5}$ & $2.290\cdot 10^{5}$ & $2.181\cdot 10^{5}$ \\
                          $\sigma(\tpreq)$ & $1.589\cdot 10^{4}$ & $1.762\cdot 10^{4}$ & $4.687\cdot 10^{4}$ & \phantom{\_} & $3.757\cdot 10^{4}$ & $1.424\cdot 10^{5}$ & $1.285\cdot 10^{5}$ \\
                       $\overline{\tottp}$ & $2.426\cdot 10^{5}$ & $6.591\cdot 10^{5}$ & $7.169\cdot 10^{5}$ & \phantom{\_} & $2.022\cdot 10^{5}$ & $4.765\cdot 10^{5}$ & $4.842\cdot 10^{5}$ \\
                          $\sigma(\tottp)$ & $4.052\cdot 10^{3}$ & $9.418\cdot 10^{2}$ & $1.312\cdot 10^{3}$ & \phantom{\_} & $7.042\cdot 10^{3}$ & $1.593\cdot 10^{3}$ & $1.150\cdot 10^{3}$ \\
               $\overline{\beststepfound}$ & $2.779\cdot 10^{3}$ & $1.896\cdot 10^{3}$ & $2.886\cdot 10^{3}$ & \phantom{\_} & $2.482\cdot 10^{4}$ & $2.063\cdot 10^{4}$ & $1.851\cdot 10^{4}$ \\
                  $\sigma(\beststepfound)$ & $2.357\cdot 10^{3}$ & $1.099\cdot 10^{3}$ & $2.661\cdot 10^{3}$ & \phantom{\_} & $9.271\cdot 10^{3}$ & $1.474\cdot 10^{4}$ & $1.314\cdot 10^{4}$ \\
\bottomrule
\end{tabular}

\caption{Mean and standard deviation of number of chemical graphs considered by a simulation before its candidate was encountered [$\overline{\tpreq}$ and $\sigma(\tpreq)$], total number of chemical graphs considered by a simulation [$\overline{\tottp}$ and $\sigma(\tottp)$], and number of global steps taken by a simulation before the candidate was encountered [$\overline{\beststepfound}$ and $\sigma(\beststepfound)$] observed for trajectories in QM9* for different optimization problems (minimize free energy of solvation $\dEsolv$ or maximize dipole $\dipole$ under weak or strong constraint on HOMO-LUMO gap $\gap$) and different values of the bias proportionality coefficient $\biasprop$.}
\label{tab:qm9_step_comp}
\end{table}

\begin{figure}
          \centering           
          \includegraphics[width=1.0\linewidth]{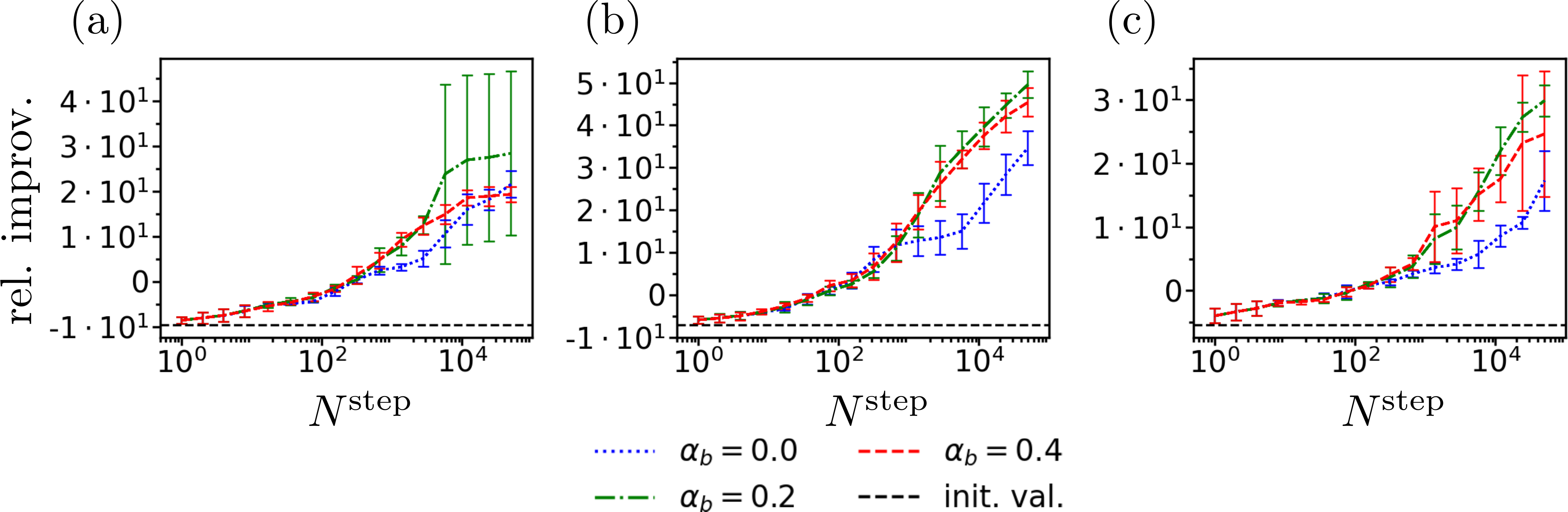}
          \caption{Relative improvement observed at $N^{\mathrm{step}}$ global Monte Carlo steps for EGP* simulations optimizing (a) $\dipole$ with weak $\gap$ constraint, (b) $\dEsolv$ with strong $\gap$ constraint, and (c) $\dipole$ with strong $\gap$ constraint. Results are organized analogously to Figure~\ref{fig:qm9_SI_opt_logs}.}
     \label{fig:egp_SI_opt_logs} 
 \end{figure}

\begin{figure}
          \centering           
          \includegraphics[width=1.0\linewidth]{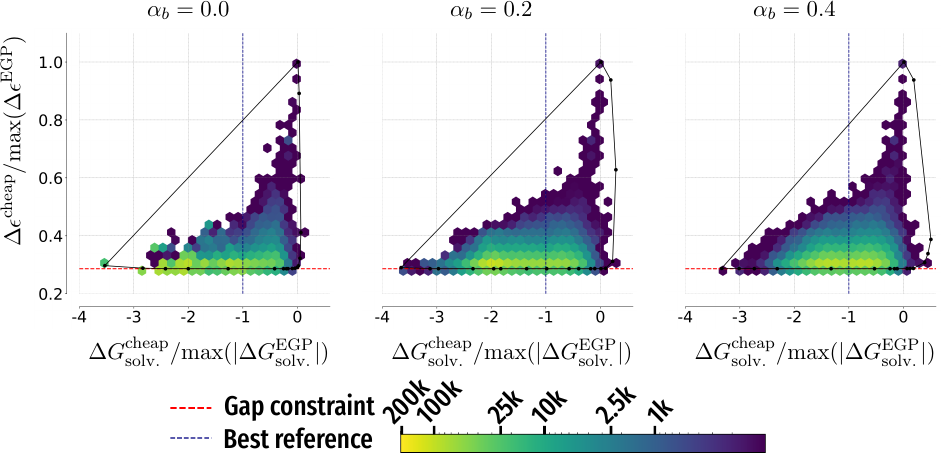}
          \caption{Densities of molecules encountered by simulations minimizing free energy of solvation $\dEsolv$ with strong HOMO-LUMO gap $\gap$ constraint in EGP* that produced the candidate with smallest $\dEsolv$ for a given value of bias proportionality coefficient $\biasprop$.}
     \label{fig:egp_pareto_front_solvation_strong} 
 \end{figure}

\begin{figure}
          \centering           
          \includegraphics[width=1.0\linewidth]{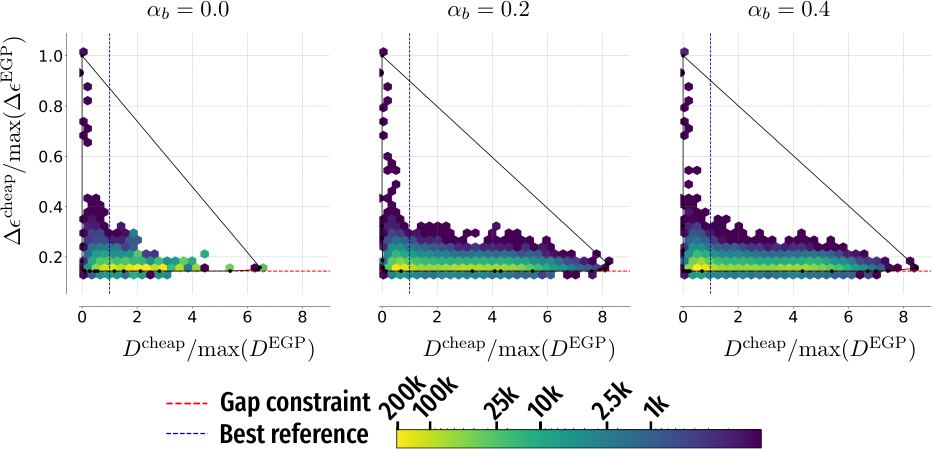}
          \caption{Densities of molecules encountered by simulations maximizing dipole $\dipole$ with weak HOMO-LUMO gap $\gap$ constraint in EGP* that produced the candidate with largest $\dipole$ for a given value of bias proportionality coefficient $\biasprop$.}
     \label{fig:egp_pareto_front_dipole_weak} 
 \end{figure}

\begin{figure}
          \centering           
          \includegraphics[width=1.0\linewidth]{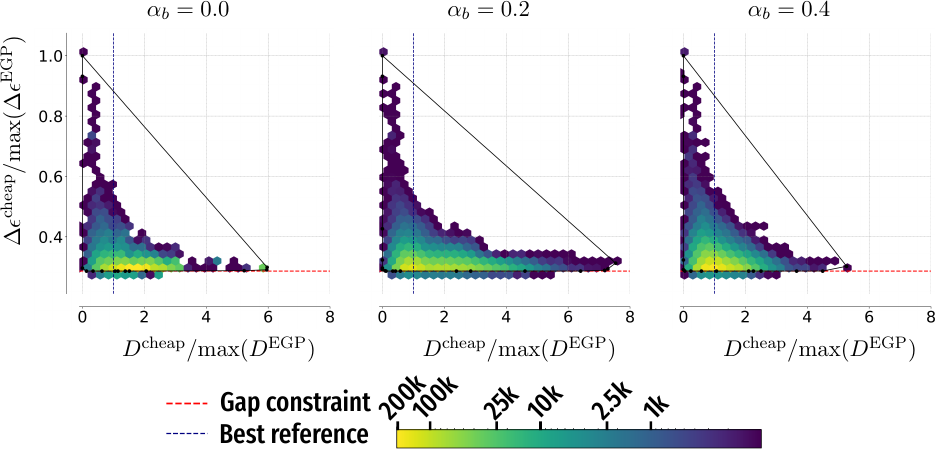}
          \caption{Densities of molecules encountered by simulations maximizing dipole $\dipole$ with strong HOMO-LUMO gap $\gap$ constraint in EGP* that produced the candidate with largest $\dipole$ for a given value of bias proportionality coefficient $\biasprop$.}
     \label{fig:egp_pareto_front_dipole_strong} 
 \end{figure}

Due to the large number of candidates found in EGP* their full lists are left for Subsec.~\ref{subsec:egp_full_candidate_lists}. We summarize how well these molecules were optimized for different optimization problems and $\biasprop$ values in Table~\ref{tab:egp_step_comp}; we observe from the behavior of average optimized property values of candidates $\overline{\dipole^{\mathrm{best}}}$ and $\overline{\dEsolv^{\mathrm{best}}}$ that adding biasing potential at worst did not significantly affect optimization results and at best provided significant improvements. Table~\ref{tab:egp_step_comp} summarizes the behavior of $\tpreq$, $\tottp$, and $\beststepfound$ for our EGP* simulations. Increasing $\biasprop$ consistently increased the number of molecules explored by a trajectory; the $\beststepfound$ values comparable to the total simulation length is an indicator of these simulations not being converged. Analogously to Figure~9, Figure~\ref{fig:egp_SI_opt_logs} shows how using non-zero $\biasprop$ could accelerate finding better candidates w.r.t the number of global Monte Carlo steps. However, we need to underline that, as indicated by $\tpreq$ and $\tottp$ values in Table~\ref{tab:egp_step_comp}, a non-zero $\biasprop$ actually increased the computational time spent on proposing a candidate since MOSAiCS considered more chemical graphs during the corresponding simulations. Plots of densities of EGP* molecules analogous to Figure~10 are found in Figures~\ref{fig:egp_pareto_front_solvation_strong}-\ref{fig:egp_pareto_front_dipole_strong}; analogously to Figure~10 we observe that increasing $\biasprop$ tended to help the simulations explore regions of chemical space more valuable in terms of optimized quantities.


\begin{table}
\begin{tabular}{llllllll}
\toprule
                            $\gap$ constr. & \multicolumn{3}{c}{weak} & \phantom{\_} & \multicolumn{3}{c}{strong} \\
         \cline{2-4}\cline{6-8}$\biasprop$ &                 0.0 &                 0.2 &                 0.4 & \phantom{\_} &                 0.0 &                 0.2 &                 0.4 \\
\midrule
        \multicolumn{8}{c}{min. $\dEsolv$}                      \\
              \midrule $\overline{\tpreq}$ & $1.533\cdot 10^{5}$ & $5.509\cdot 10^{5}$ & $4.916\cdot 10^{5}$ & \phantom{\_} & $2.518\cdot 10^{5}$ & $4.992\cdot 10^{5}$ & $4.442\cdot 10^{5}$ \\
                          $\sigma(\tpreq)$ & $3.857\cdot 10^{4}$ & $1.428\cdot 10^{5}$ & $1.658\cdot 10^{5}$ & \phantom{\_} & $3.899\cdot 10^{4}$ & $2.052\cdot 10^{5}$ & $2.479\cdot 10^{5}$ \\
                       $\overline{\tottp}$ & $1.824\cdot 10^{5}$ & $6.895\cdot 10^{5}$ & $7.736\cdot 10^{5}$ & \phantom{\_} & $3.085\cdot 10^{5}$ & $7.555\cdot 10^{5}$ & $8.969\cdot 10^{5}$ \\
                          $\sigma(\tottp)$ & $1.236\cdot 10^{4}$ & $8.907\cdot 10^{3}$ & $6.908\cdot 10^{3}$ & \phantom{\_} & $1.047\cdot 10^{4}$ & $6.802\cdot 10^{3}$ & $4.660\cdot 10^{3}$ \\
               $\overline{\beststepfound}$ & $3.841\cdot 10^{4}$ & $3.963\cdot 10^{4}$ & $3.119\cdot 10^{4}$ & \phantom{\_} & $3.777\cdot 10^{4}$ & $3.228\cdot 10^{4}$ & $2.428\cdot 10^{4}$ \\
                  $\sigma(\beststepfound)$ & $1.174\cdot 10^{4}$ & $1.091\cdot 10^{4}$ & $1.095\cdot 10^{4}$ & \phantom{\_} & $7.706\cdot 10^{3}$ & $1.406\cdot 10^{4}$ & $1.399\cdot 10^{4}$ \\
\midrule\multicolumn{8}{c}{max. $\dipole$}                      \\
              \midrule $\overline{\tpreq}$ & $2.141\cdot 10^{5}$ & $5.646\cdot 10^{5}$ & $6.574\cdot 10^{5}$ & \phantom{\_} & $2.759\cdot 10^{5}$ & $5.532\cdot 10^{5}$ & $5.779\cdot 10^{5}$ \\
                          $\sigma(\tpreq)$ & $2.881\cdot 10^{4}$ & $1.192\cdot 10^{5}$ & $7.904\cdot 10^{4}$ & \phantom{\_} & $3.755\cdot 10^{4}$ & $1.794\cdot 10^{5}$ & $2.566\cdot 10^{5}$ \\
                       $\overline{\tottp}$ & $2.537\cdot 10^{5}$ & $7.740\cdot 10^{5}$ & $8.427\cdot 10^{5}$ & \phantom{\_} & $3.269\cdot 10^{5}$ & $8.604\cdot 10^{5}$ & $9.540\cdot 10^{5}$ \\
                          $\sigma(\tottp)$ & $1.296\cdot 10^{4}$ & $1.401\cdot 10^{4}$ & $1.016\cdot 10^{4}$ & \phantom{\_} & $1.532\cdot 10^{4}$ & $7.360\cdot 10^{3}$ & $2.848\cdot 10^{3}$ \\
               $\overline{\beststepfound}$ & $3.905\cdot 10^{4}$ & $3.617\cdot 10^{4}$ & $3.882\cdot 10^{4}$ & \phantom{\_} & $4.033\cdot 10^{4}$ & $3.150\cdot 10^{4}$ & $3.001\cdot 10^{4}$ \\
                  $\sigma(\beststepfound)$ & $6.732\cdot 10^{3}$ & $8.028\cdot 10^{3}$ & $4.668\cdot 10^{3}$ & \phantom{\_} & $5.221\cdot 10^{3}$ & $1.073\cdot 10^{4}$ & $1.360\cdot 10^{4}$ \\
\bottomrule
\end{tabular}

\caption{Numbers of chemical graphs considered by a trajectory before the candidate molecule was encountered $\tpreq$, total numbers of chemical graphs considered by a trajectory $\tottp$, and numbers of global steps after which the candidate was encountered [$\overline{\beststepfound}$ and $\sigma(\beststepfound)$] observed for EGP* simulations; results are labeled analogously to Table~\ref{tab:qm9_step_comp}.}
\label{tab:egp_step_comp}
\end{table}

Lastly, the difference between cheap and converged estimates of $\dipole$ and $\dEsolv$ is very small for rigid molecules that only have one local minimum in configuration space (\emph{e.g.} benzene), but it can become significant for larger non-rigid molecules with many valid conformers. To compare how much this difference affected candidate molecules obtained in this work we introduced ``numerical error measure'' $\cheapquantnoise$ defined as the ratio of mean absolute error of the cheap estimate relative to the converged one divided by standard deviation of the converged estimate over the pre-final (as defined in Subsec.~2.5 of the main text) molecules. Its values for QM9* and EGP* simulations are presented in Table~\ref{tab:cheap_quant_noise}. For QM9*, we note that maximizing $\dipole$ under strong $\gap$ constraint was apparently much more affected by using $\dipole^{\mathrm{cheap}}$ in the minimization function, which may be the reason simulations for that optimization problem agreed on candidate molecules much less frequently than for the other three. We also observe that for a given optimization problem simulations in EGP* were much more affected by cheap/converged difference than the ones in QM9*, likely due to EGP* consisting of larger molecules with more conformers to be accounted for. This added to EGP* being the more challenging set of molecules to search through compared to QM9*.

\begin{table}
 \centering
\begin{tabular}{llll}
\toprule
    \multirow{2}{*}{Optimization problem} & \multicolumn{3}{c}{$\cheapquantnoise$} \\
                              \cline{2-4} &      $\biasprop=0.0$ &      $\biasprop=0.2$ &      $\biasprop=0.4$ \\
\midrule \multicolumn{4}{c}{QM9*}\\ \midrule
  min. $\dEsolv$ (weak $\gap$ constraint) & $2.009\cdot 10^{-1}$ & $2.013\cdot 10^{-1}$ & $2.010\cdot 10^{-1}$ \\
min. $\dEsolv$ (strong $\gap$ constraint) & $1.827\cdot 10^{-1}$ & $1.814\cdot 10^{-1}$ & $1.834\cdot 10^{-1}$ \\
  max. $\dipole$ (weak $\gap$ constraint) & $2.635\cdot 10^{-1}$ & $2.658\cdot 10^{-1}$ & $2.714\cdot 10^{-1}$ \\
max. $\dipole$ (strong $\gap$ constraint) & $4.108\cdot 10^{-1}$ & $4.189\cdot 10^{-1}$ & $4.270\cdot 10^{-1}$ \\
\midrule \multicolumn{4}{c}{EGP*}\\ \midrule
  min. $\dEsolv$ (weak $\gap$ constraint) & $2.889\cdot 10^{-1}$ & $8.493\cdot 10^{-1}$ & $7.993\cdot 10^{-1}$ \\
min. $\dEsolv$ (strong $\gap$ constraint) & $4.145\cdot 10^{-1}$ & $5.359\cdot 10^{-1}$ & $4.409\cdot 10^{-1}$ \\
  max. $\dipole$ (weak $\gap$ constraint) & $5.687\cdot 10^{-1}$ & $8.752\cdot 10^{-1}$ & $9.565\cdot 10^{-1}$ \\
max. $\dipole$ (strong $\gap$ constraint) & $4.722\cdot 10^{-1}$ & $8.612\cdot 10^{-1}$ & $7.068\cdot 10^{-1}$ \\
\bottomrule
\end{tabular}

\caption{Measure of error caused by using cheap version of optimized quantity during the simulation $\cheapquantnoise$ (see Sec.~\ref{sec:add_data} for definition) observed during simulations in QM9* or EGP*, while maximizing dipole $\dipole$ or minimizing free energy of solvation $\dEsolv$, under a weak or strong constraint on the HOMO-LUMO gap $\gap$, and with different values of the bias proportionality coefficient $\biasprop$.}
\label{tab:cheap_quant_noise}
\end{table}

\subsection{Effect of initial replica positions on optimization results}
\label{subsec:random_init_mols_results}

\begin{figure}
          \centering           
          \includegraphics[width=0.6\textwidth]{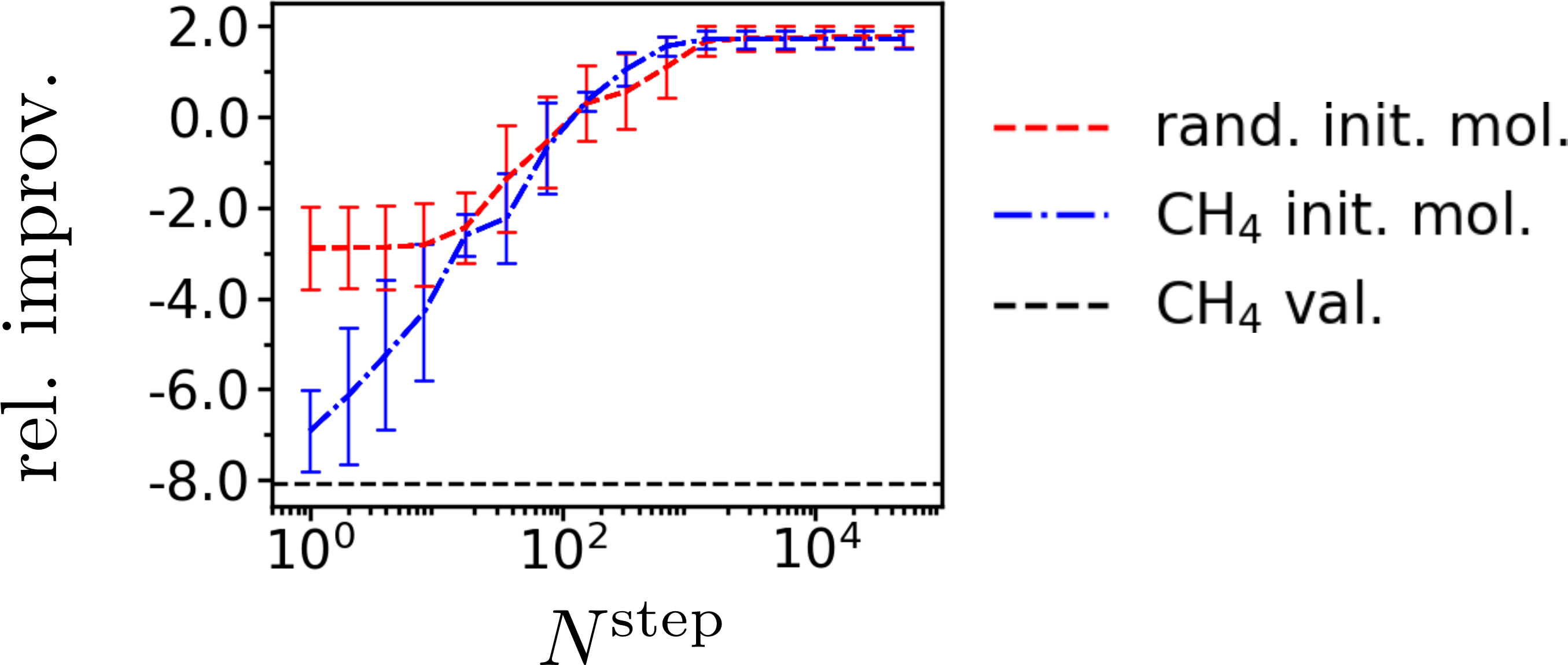}
          \caption{Relative improvement observed at $N^{\mathrm{step}}$ global Monte Carlo steps for QM9* simulations optimizing $\dEsolv$ with strong $\gap$ constraint if the molecules initially occupied by replicas are chosen randomly from QM9 dataset (``rand. init. mol.'') with bias proportionality coefficient $\biasprop$ set to 0.0. Plotted for comparison are corresponding values when all replicas are initialized at methane (``$\mathrm{CH}_{4}$ init. mol.'', also appearing in Figure~\ref{fig:qm9_SI_opt_logs}a) as well as the relative improvement value for methane (``$\mathrm{CH}_{4}$ val.'').}
     \label{fig:init_cond_opt_logs} 
 \end{figure}

It is natural to assume that increasing diversity of molecules initially occupied by replicas should make it easier for MOSAiCS to propose better candidate molecules. We tested this hypothesis on minimizing $\dEsolv$ with strong $\gap$ constraint over QM9*, since it is a problem for which converged results proved easy to obtain and for which improvement over QM9 was observed. For that optimization problem we launched 8 simulations with $\biasprop=0.0$ which only differed from the others presented in this work by the choice of initial molecules: each was chosen with uniform probability from molecules in the intersection of QM9 and QM9* that satisfied strong constraint on $\gap^{\mathrm{cheap}}$, then initial molecules were assigned to replicas in such a way that molecules with larger $\dEsolv^{\mathrm{cheap}}$ were occupied by replicas with smaller $\beta^{(i)}$. The resulting mean optimization progress is plotted in Figure~\ref{fig:init_cond_opt_logs}, with the corresponding values observed when all replicas were initialized in methane added for comparison. Surprisingly, while simulations with random starting molecules had significantly larger starting values of relative improvement, they failed to go beyond those values significantly earlier than simulations initialized with methane, and they all proposed previously mentioned \candidate{2}{QM9} as the candidate. While there might be better ways to choose initial molecules that would make search for candidates tangibly faster, finding them was beyond the scope of this work.

\subsection{EGP* full candidate list}
\label{subsec:egp_full_candidate_lists}

\begin{figure}
          \centering           
\hspace{-2ex}
\begin{tikzpicture}
        \node[anchor=south west,inner sep=0] (image) at (0,0) {\includegraphics[width=0.15\textwidth]{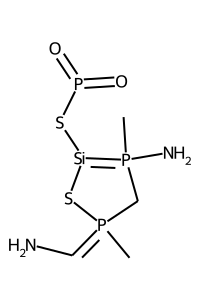}};
        \node[anchor=south west,inner sep=0] (image) at (2.75,0) {\includegraphics[width=0.15\textwidth]{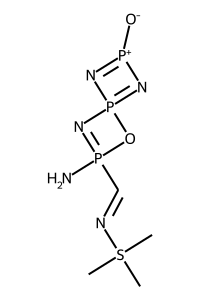}};
        \node[anchor=south west,inner sep=0] (image) at (5.5,0) {\includegraphics[width=0.15\textwidth]{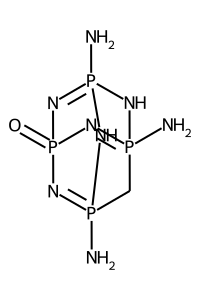}};
        \node[anchor=south west,inner sep=0] (image) at (0,-4) {\includegraphics[width=0.15\textwidth]{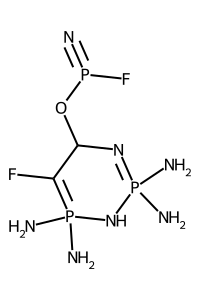}};
        \node[anchor=south west,inner sep=0] (image) at (2.75,-4) {\includegraphics[width=0.15\textwidth]{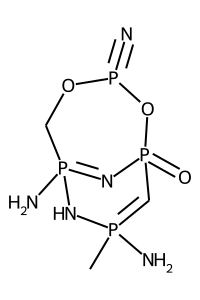}};
        \node[anchor=south west,inner sep=0] (image) at (5.5,-4) {\includegraphics[width=0.15\textwidth]{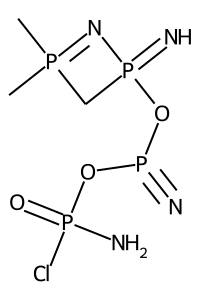}};
        \node at (0.3,2.55) {\candidate{14}{EGP}};
        \node at (3.25,3.2) {\candidate{17}{EGP}};
        \node at (6,3.2) {\candidate{32}{EGP}};
        \node at (0.3,-1.35) {\candidate{38}{EGP}};
        \node at (3.25,-1) {\candidate{40}{EGP}};
        \node at (5.45,-1.0) {\candidate{82}{EGP}};
\end{tikzpicture}
          \caption{Examples of EGP* candidates.}
     \label{fig:egp_candidate_examples} 
 \end{figure}

EGP* candidates proposed by MOSAiCS are listed in Tables~\ref{tab:egp_candidates_solvation_weak}-\ref{tab:egp_candidates_dipole_strong}. Examples of candidates plotted in Figure~\ref{fig:egp_candidate_examples} demonstrate MOSAiCS' ability to produce candidates with complicated structures, including unusual heterocycles (\candidate{14}{EGP}, \candidate{38}{EGP}, and \candidate{82}{EGP}), conjugated heterocycles (\candidate{17}{EGP}, \candidate{40}{EGP}), and caged molecules (\candidate{32}{EGP}).

 \begin{table}
 \centering
\begin{tabular}{llllll}
\toprule
\multirow{2}{*}{molecule} &                     \multirow{2}{*}{SMILES} & \multirow{2}{*}{$\phantom{-(}\dEsolv$, kJ/mol} & \multicolumn{3}{l}{prop. with $\biasprop$} \\
   \cline{4-6}\phantom{1} &                                 \phantom{1} &                                    \phantom{1} &                    0.0 & 0.2 & 0.4 \\
\midrule
       \candidate{1}{EGP} & CS(C)=P(C)(C)[Si](C)(Cl)S[Si]1=P[Si]1(Br)Br &               $(-1.194 \pm 0.007)\cdot 10^{3}$ &                    $0$ & $0$ & $1$ \\
       \candidate{2}{EGP} &               CS(C)(C)CCP1CS1=P(=O)SP(=O)=O &               $(-1.080 \pm 0.001)\cdot 10^{3}$ &                    $0$ & $1$ & $2$ \\
       \candidate{3}{EGP} &               CS(C)(C)CNP1CS1=P(=O)SP(=O)=O &               $(-1.041 \pm 0.001)\cdot 10^{3}$ &                    $0$ & $1$ & $0$ \\
       \candidate{4}{EGP} &         CS(C)(C)CC(Br)S[Si](Br)(Br)SP(=O)=O &               $(-9.896 \pm 0.092)\cdot 10^{2}$ &                    $0$ & $2$ & $1$ \\
       \candidate{5}{EGP} &               CS(C)(C)CNP1NS1=P(=O)SP(=O)=O &               $(-9.868 \pm 0.050)\cdot 10^{2}$ &                    $0$ & $0$ & $1$ \\
       \candidate{6}{EGP} &       CS(C)(C)CS[Si]1(Br)S[Si]1(Br)SP(=O)=O &               $(-9.660 \pm 0.075)\cdot 10^{2}$ &                    $0$ & $0$ & $1$ \\
       \candidate{7}{EGP} &         CS(C)(C)CP(Br)S[Si](Br)(Br)SP(=O)=O &               $(-9.573 \pm 0.127)\cdot 10^{2}$ &                    $0$ & $1$ & $0$ \\
       \candidate{8}{EGP} &            CS1(C)CCS(=P(=O)SP(=O)=O)P(Cl)C1 &               $(-9.556 \pm 0.055)\cdot 10^{2}$ &                    $0$ & $1$ & $0$ \\
       \candidate{9}{EGP} &         CS(C)(C)CC(Cl)S[Si](Br)(Br)SP(=O)=O &               $(-9.405 \pm 0.226)\cdot 10^{2}$ &                    $0$ & $0$ & $1$ \\
      \candidate{10}{EGP} &          CS1(C)CCP(Cl)[Si](Br)(SP(=O)=O)SC1 &               $(-9.244 \pm 0.092)\cdot 10^{2}$ &                    $0$ & $1$ & $0$ \\
      \candidate{11}{EGP} &             CS(=NP(=O)=S=P(=O)Br)NCS(C)(C)C &               $(-8.418 \pm 0.102)\cdot 10^{2}$ &                    $0$ & $1$ & $0$ \\
      \candidate{12}{EGP} &              CS(P1CSP(SP(=O)=O)S1)=P(C)(C)N &               $(-7.724 \pm 0.026)\cdot 10^{2}$ &                    $0$ & $0$ & $1$ \\
      \candidate{13}{EGP} &                CS(C)(C)CCSNSP(=O)=CP(\#N)Br &               $(-7.144 \pm 0.056)\cdot 10^{2}$ &                    $1$ & $0$ & $0$ \\
      \candidate{14}{EGP} &           CP1(=CN)CP(C)(N)=[Si](SP(=O)=O)S1 &               $(-5.121 \pm 0.076)\cdot 10^{2}$ &                    $1$ & $0$ & $0$ \\
      \candidate{15}{EGP} &           CP1(N)=CP(SP(=O)=O)C(N)=P(C)(C)C1 &               $(-4.511 \pm 0.059)\cdot 10^{2}$ &                    $1$ & $0$ & $0$ \\
      \candidate{16}{EGP} &               CP(C)(N)=C(N)NCOC(=N)SP(=O)=O &               $(-4.456 \pm 0.079)\cdot 10^{2}$ &                    $1$ & $0$ & $0$ \\
      \candidate{17}{EGP} &          CS(C)(C)N=CP1(N)=NP2(=NP(=O)=N2)O1 &               $(-4.261 \pm 0.003)\cdot 10^{2}$ &                    $1$ & $0$ & $0$ \\
      \candidate{18}{EGP} &                CP(N)(N)=C1NCCOP(OP(=O)=O)N1 &               $(-4.153 \pm 0.006)\cdot 10^{2}$ &                    $1$ & $0$ & $0$ \\
      \candidate{19}{EGP} &           CS(C)(=NCCS(\#N)=NP(\#N)Br)=C(N)N &               $(-3.825 \pm 0.013)\cdot 10^{2}$ &                    $1$ & $0$ & $0$ \\
      \candidate{20}{EGP} &             CP1(N)=C(N)P(OP(=O)=O)S(=O)CCC1 &               $(-3.822 \pm 0.012)\cdot 10^{2}$ &                    $1$ & $0$ & $0$ \\
\bottomrule
\end{tabular}

\caption{EGP* candidates proposed during minimization of free energy of solvation $\dEsolv$ with weak constraint on the HOMO-LUMO gap $\gap$, the corresponding values of $\dEsolv$, and the number of times a molecule was proposed as a candidate by simulations run with different values of the bias proportionality coefficient $\biasprop$ (prop. with $\biasprop$).}
\label{tab:egp_candidates_solvation_weak}
\end{table}

 \begin{table}
 \centering
\begin{tabular}{llllll}
\toprule
\multirow{2}{*}{molecule} &            \multirow{2}{*}{SMILES} & \multirow{2}{*}{$\phantom{-(}\dEsolv$, kJ/mol} & \multicolumn{3}{l}{prop. with $\biasprop$} \\
   \cline{4-6}\phantom{1} &                        \phantom{1} &                                    \phantom{1} &                    0.0 & 0.2 & 0.4 \\
\midrule
      \candidate{21}{EGP} &       COP(\#N)OP(\#N)NCCP(N)(N)=CN &               $(-2.698 \pm 0.021)\cdot 10^{2}$ &                    $1$ & $0$ & $0$ \\
      \candidate{22}{EGP} &       N\#P(N)OP(\#N)OCNCP(N)(N)=CN &               $(-2.670 \pm 0.043)\cdot 10^{2}$ &                    $1$ & $1$ & $0$ \\
      \candidate{23}{EGP} &     CC(OCOP(\#N)OP(\#N)N)=P(C)(C)N &               $(-2.483 \pm 0.037)\cdot 10^{2}$ &                    $0$ & $1$ & $0$ \\
      \candidate{24}{EGP} &        CP(N)(=CN)NCNP(\#N)OP(\#N)N &               $(-2.480 \pm 0.014)\cdot 10^{2}$ &                    $0$ & $1$ & $0$ \\
      \candidate{25}{EGP} &  N\#P(F)OP(\#N)NP1(=N)NC=P(N)(N)N1 &               $(-2.431 \pm 0.013)\cdot 10^{2}$ &                    $0$ & $1$ & $0$ \\
      \candidate{26}{EGP} &     CC(CP(N)(N)=CN)NP(\#N)OP(\#N)N &               $(-2.405 \pm 0.051)\cdot 10^{2}$ &                    $1$ & $0$ & $0$ \\
      \candidate{27}{EGP} &       CP(C)(N)=COCCOP(\#N)OP(\#N)N &               $(-2.383 \pm 0.042)\cdot 10^{2}$ &                    $1$ & $0$ & $0$ \\
      \candidate{28}{EGP} &   N\#P(F)OP(\#N)OC1CC(=P(N)(N)N)O1 &               $(-2.376 \pm 0.007)\cdot 10^{2}$ &                    $0$ & $1$ & $0$ \\
      \candidate{29}{EGP} &       COP(\#N)NP(\#N)NCCP(N)(N)=CN &               $(-2.368 \pm 0.023)\cdot 10^{2}$ &                    $1$ & $0$ & $0$ \\
      \candidate{30}{EGP} &     CP(C)(C)=COC(N)OP(\#N)OP(\#N)N &               $(-2.366 \pm 0.007)\cdot 10^{2}$ &                    $0$ & $0$ & $1$ \\
      \candidate{31}{EGP} &     CP(C)(N)=C(F)OCOP(\#N)OP(\#N)F &               $(-2.335 \pm 0.020)\cdot 10^{2}$ &                    $0$ & $1$ & $0$ \\
      \candidate{32}{EGP} & NP12=NP3(=O)N=P(N)(C1)NP(N)(=N3)N2 &               $(-2.295 \pm 0.008)\cdot 10^{2}$ &                    $0$ & $1$ & $0$ \\
      \candidate{33}{EGP} & N\#P(F)OP1(=O)C=P(N)(N)CP(N)(N)=N1 &               $(-2.291 \pm 0.010)\cdot 10^{2}$ &                    $0$ & $1$ & $0$ \\
      \candidate{34}{EGP} & CP1(N)=CP(=O)(OP(\#N)N)N=P(N)(N)N1 &               $(-2.261 \pm 0.027)\cdot 10^{2}$ &                    $0$ & $0$ & $2$ \\
      \candidate{35}{EGP} & CP12=NP3(=O)N=P(N)(CP(N)(=N3)N1)N2 &               $(-2.257 \pm 0.007)\cdot 10^{2}$ &                    $0$ & $0$ & $1$ \\
      \candidate{36}{EGP} & NP12=NP3(=O)N=P(N)(N1)NP(N)(=N3)N2 &               $(-2.253 \pm 0.008)\cdot 10^{2}$ &                    $0$ & $0$ & $1$ \\
      \candidate{37}{EGP} &     CP(N)(N)=C1COP(\#N)OP(N)(=O)N1 &               $(-2.232 \pm 0.007)\cdot 10^{2}$ &                    $2$ & $0$ & $1$ \\
      \candidate{38}{EGP} &    N\#P(F)OC1N=P(N)(N)NP(N)(N)=C1F &               $(-2.142 \pm 0.005)\cdot 10^{2}$ &                    $0$ & $0$ & $1$ \\
      \candidate{39}{EGP} & N\#P(N)OP1(=O)C=P(N)(N)OP(N)(N)=N1 &               $(-2.099 \pm 0.019)\cdot 10^{2}$ &                    $0$ & $0$ & $1$ \\
      \candidate{40}{EGP} & CP1(N)=CP2(=O)N=P(N)(COP(\#N)O2)N1 &               $(-2.076 \pm 0.009)\cdot 10^{2}$ &                    $1$ & $0$ & $0$ \\
\bottomrule
\end{tabular}

\caption{EGP* candidates proposed during minimization of free energy of solvation $\dEsolv$ with strong constraint on HOMO-LUMO gap $\gap$; columns are labeled analogously to Table~\ref{tab:egp_candidates_solvation_weak}.}
\label{tab:egp_candidates_solvation_strong}
\end{table}
 
 \begin{table}
 \centering
\begin{tabular}{llllll}
\toprule
\multirow{2}{*}{molecule} &          \multirow{2}{*}{SMILES} & \multirow{2}{*}{$\phantom{-(}\dipole$, debye} & \multicolumn{3}{l}{prop. with $\biasprop$} \\
   \cline{4-6}\phantom{1} &                      \phantom{1} &                                   \phantom{1} &                    0.0 & 0.2 & 0.4 \\
\midrule
      \candidate{41}{EGP} &     CS(C)(C)CCC=PSP(\#N)SP(=O)=O &    $\phantom{-}(1.095 \pm 0.009)\cdot 10^{2}$ &                    $0$ & $1$ & $0$ \\
      \candidate{42}{EGP} &     CS(C)(C)CSC=CSP(\#N)SP(=O)=O &    $\phantom{-}(1.075 \pm 0.009)\cdot 10^{2}$ &                    $0$ & $1$ & $0$ \\
      \candidate{43}{EGP} & O=P(=O)SP(=O)=S1N[Si]1=S1CP1CSCl &    $\phantom{-}(1.063 \pm 0.013)\cdot 10^{2}$ &                    $0$ & $1$ & $0$ \\
      \candidate{44}{EGP} &       CP(SC=CSCS(C)(C)C)SP(=O)=S &    $\phantom{-}(1.029 \pm 0.010)\cdot 10^{2}$ &                    $0$ & $0$ & $1$ \\
      \candidate{45}{EGP} &        CP(NCNCCS(C)(C)C)SP(=O)=O &    $\phantom{-}(1.012 \pm 0.010)\cdot 10^{2}$ &                    $0$ & $1$ & $0$ \\
      \candidate{46}{EGP} &       CP(SC=CSCS(C)(C)C)SP(=O)=O &    $\phantom{-}(9.944 \pm 0.077)\cdot 10^{1}$ &                    $0$ & $1$ & $0$ \\
      \candidate{47}{EGP} &        CS(C)(C)CCCCNP(F)SP(=O)=O &    $\phantom{-}(9.743 \pm 0.127)\cdot 10^{1}$ &                    $0$ & $1$ & $0$ \\
      \candidate{48}{EGP} &  CS(C)(C)CCCS1=[Si](SP(=S)=S)C1N &    $\phantom{-}(9.713 \pm 0.073)\cdot 10^{1}$ &                    $0$ & $0$ & $1$ \\
      \candidate{49}{EGP} &      CS(C)(C)CCNSNP(\#N)SP(=O)=O &    $\phantom{-}(9.640 \pm 0.238)\cdot 10^{1}$ &                    $0$ & $0$ & $1$ \\
      \candidate{50}{EGP} &        CS(C)(C)CCCCSP(F)SP(=O)=O &    $\phantom{-}(9.563 \pm 0.060)\cdot 10^{1}$ &                    $0$ & $1$ & $0$ \\
      \candidate{51}{EGP} &        CS(C)(C)N=PSCSCSN=P(=O)Br &    $\phantom{-}(9.499 \pm 0.044)\cdot 10^{1}$ &                    $0$ & $1$ & $0$ \\
      \candidate{52}{EGP} &     C=S(C)CSCP1CS1=P(=O)SP(=O)=O &    $\phantom{-}(9.420 \pm 0.082)\cdot 10^{1}$ &                    $0$ & $0$ & $1$ \\
      \candidate{53}{EGP} &      CS(C)(C)NC=CSP(N=S)SP(=O)=O &    $\phantom{-}(9.267 \pm 0.026)\cdot 10^{1}$ &                    $0$ & $0$ & $1$ \\
      \candidate{54}{EGP} &        CP(CCCSP(=O)=O)CCS(C)(C)C &    $\phantom{-}(9.264 \pm 0.110)\cdot 10^{1}$ &                    $0$ & $0$ & $1$ \\
      \candidate{55}{EGP} & CS(C)(N)C[Si]1(F)CSP(SP(=O)=S)S1 &    $\phantom{-}(8.639 \pm 0.076)\cdot 10^{1}$ &                    $0$ & $0$ & $1$ \\
      \candidate{56}{EGP} &      C=S(C)(C)=PSC1OP1SCSP(=O)=O &    $\phantom{-}(8.447 \pm 0.059)\cdot 10^{1}$ &                    $0$ & $0$ & $1$ \\
      \candidate{57}{EGP} &      CS(=CN)CCNCSP(\#N)SP(\#N)Br &    $\phantom{-}(7.938 \pm 0.134)\cdot 10^{1}$ &                    $1$ & $0$ & $0$ \\
      \candidate{58}{EGP} &    CS1(CSC=CSP(\#N)SP(\#N)Br)CO1 &    $\phantom{-}(7.709 \pm 0.071)\cdot 10^{1}$ &                    $1$ & $0$ & $0$ \\
      \candidate{59}{EGP} &    C=S(C)(C)=PC=CCSC(=O)SP(=O)=O &    $\phantom{-}(7.685 \pm 0.093)\cdot 10^{1}$ &                    $1$ & $0$ & $0$ \\
      \candidate{60}{EGP} &  CS(C)(=CC=COS(=O)P(=O)=S)=C(N)N &    $\phantom{-}(7.587 \pm 0.020)\cdot 10^{1}$ &                    $1$ & $0$ & $0$ \\
      \candidate{61}{EGP} &   C=S(C)(=CN)CCCSP(\#N)SP(\#N)Br &    $\phantom{-}(7.495 \pm 0.137)\cdot 10^{1}$ &                    $1$ & $0$ & $0$ \\
      \candidate{62}{EGP} &      CP(C)(C)=C(N)NC=CCNCP(=S)=S &    $\phantom{-}(7.457 \pm 0.087)\cdot 10^{1}$ &                    $1$ & $0$ & $0$ \\
      \candidate{63}{EGP} &    CP(C)(NC=CNC\#CP(=S)=S)=C(N)N &    $\phantom{-}(7.453 \pm 0.009)\cdot 10^{1}$ &                    $1$ & $0$ & $0$ \\
      \candidate{64}{EGP} &        CS(C)(=CN)=NSCC=CN=S=NC=O &    $\phantom{-}(5.369 \pm 0.012)\cdot 10^{1}$ &                    $1$ & $0$ & $0$ \\
\bottomrule
\end{tabular}

\caption{EGP* candidates proposed during maximization of dipole $\dipole$ with weak constraint on HOMO-LUMO gap $\gap$; columns are labeled analogously to Table~\ref{tab:egp_candidates_solvation_weak}.}
\label{tab:egp_candidates_dipole_weak}
\end{table}

 \begin{table}
 \centering
\begin{tabular}{llllll}
\toprule
\multirow{2}{*}{molecule} &             \multirow{2}{*}{SMILES} & \multirow{2}{*}{$\phantom{-(}\dipole$, debye} & \multicolumn{3}{l}{prop. with $\biasprop$} \\
   \cline{4-6}\phantom{1} &                         \phantom{1} &                                   \phantom{1} &                    0.0 & 0.2 & 0.4 \\
\midrule
      \candidate{65}{EGP} &           NC=P(N)(N)CCCCNCN=S(=O)=O &    $\phantom{-}(5.981 \pm 0.114)\cdot 10^{1}$ &                    $0$ & $1$ & $0$ \\
      \candidate{66}{EGP} &        CP(C)(C)=CCCCOP(\#N)OP(\#N)F &    $\phantom{-}(4.870 \pm 0.027)\cdot 10^{1}$ &                    $0$ & $5$ & $0$ \\
      \candidate{67}{EGP} &        N\#P(N)OP(\#N)NCNCP(N)(N)=CN &    $\phantom{-}(4.865 \pm 0.073)\cdot 10^{1}$ &                    $0$ & $1$ & $0$ \\
      \candidate{68}{EGP} &        C=P(C)(C)CCNCNP(\#N)OP(\#N)F &    $\phantom{-}(4.587 \pm 0.049)\cdot 10^{1}$ &                    $1$ & $0$ & $0$ \\
      \candidate{69}{EGP} &    CP(C)(N)=C1CC(OP(\#N)OP(\#N)F)C1 &    $\phantom{-}(4.401 \pm 0.084)\cdot 10^{1}$ &                    $0$ & $1$ & $0$ \\
      \candidate{70}{EGP} &         C=P(C)(C)CCCOP(\#N)OP(\#N)F &    $\phantom{-}(4.024 \pm 0.038)\cdot 10^{1}$ &                    $1$ & $0$ & $0$ \\
      \candidate{71}{EGP} &      C=P1(N)CC(OCOP(\#N)OP(\#N)F)C1 &    $\phantom{-}(3.924 \pm 0.057)\cdot 10^{1}$ &                    $0$ & $0$ & $1$ \\
      \candidate{72}{EGP} &         CC=P(C)(N)NCOP(\#N)OP(\#N)F &    $\phantom{-}(3.847 \pm 0.052)\cdot 10^{1}$ &                    $0$ & $0$ & $1$ \\
      \candidate{73}{EGP} &      C=P(N)(N)C(N)CCNP(\#N)OP(\#N)F &    $\phantom{-}(3.842 \pm 0.099)\cdot 10^{1}$ &                    $0$ & $0$ & $1$ \\
      \candidate{74}{EGP} &         C=P(C)(N)CNCOP(\#N)OP(\#N)F &    $\phantom{-}(3.777 \pm 0.035)\cdot 10^{1}$ &                    $0$ & $0$ & $2$ \\
      \candidate{75}{EGP} &      C=P(C)(C)OCC(F)OP(\#N)OP(\#N)F &    $\phantom{-}(3.768 \pm 0.034)\cdot 10^{1}$ &                    $0$ & $0$ & $1$ \\
      \candidate{76}{EGP} &            CP(N)(N)=CCCCCOCNP(\#N)F &    $\phantom{-}(3.674 \pm 0.091)\cdot 10^{1}$ &                    $0$ & $0$ & $1$ \\
      \candidate{77}{EGP} &     CC(CNP(=O)(F)NP(\#N)F)=P(C)(C)N &    $\phantom{-}(3.427 \pm 0.070)\cdot 10^{1}$ &                    $0$ & $0$ & $1$ \\
      \candidate{78}{EGP} &        CP(C)(N)=CCC(F)(F)CCOP(\#N)F &    $\phantom{-}(3.195 \pm 0.048)\cdot 10^{1}$ &                    $1$ & $0$ & $0$ \\
      \candidate{79}{EGP} &        C=P(C)(N)CCOP(\#N)OCOP(\#N)F &    $\phantom{-}(3.134 \pm 0.126)\cdot 10^{1}$ &                    $1$ & $0$ & $0$ \\
      \candidate{80}{EGP} &           CC=P(C)(N)CC(F)CCOP(\#N)F &    $\phantom{-}(2.977 \pm 0.052)\cdot 10^{1}$ &                    $1$ & $0$ & $0$ \\
      \candidate{81}{EGP} &        N\#P(F)NCCC12C=P(N)(NCN1)NC2 &    $\phantom{-}(2.870 \pm 0.041)\cdot 10^{1}$ &                    $1$ & $0$ & $0$ \\
      \candidate{82}{EGP} & CP1(C)=NP(=N)(OP(\#N)OP(N)(=O)Cl)C1 &    $\phantom{-}(2.643 \pm 0.006)\cdot 10^{1}$ &                    $1$ & $0$ & $0$ \\
      \candidate{83}{EGP} &   CP(C)(C)=N[Si]1(N)COC(OP(\#N)F)C1 &    $\phantom{-}(2.588 \pm 0.017)\cdot 10^{1}$ &                    $1$ & $0$ & $0$ \\
\bottomrule
\end{tabular}

\caption{EGP* candidates proposed during maximization of dipole $\dipole$ with strong constraint on HOMO-LUMO gap $\gap$; columns are labeled analogously to Table~\ref{tab:egp_candidates_solvation_weak}.}
\label{tab:egp_candidates_dipole_strong}
\end{table}

\bibliography{references}